\documentclass[a4paper,fleqn,usenatbib]{mnras}
\usepackage{ragged2e} 
\usepackage[T1]{fontenc}
\usepackage{ae,aecompl}
\usepackage{graphicx}	
\usepackage{amsmath}	
\usepackage{amssymb}	
\usepackage{cleveref}   
\usepackage{hyperref}   
\usepackage[countmax]{subfloat}
\usepackage{booktabs}
\newcommand{\RNum}[1]{\uppercase\expandafter{\romannumeral #1\relax}}

\title[]{Investigation of the correlation between optical and $\gamma$-ray flux variation in the blazar Ton 599}

\author[Bhoomika et al.]{
Bhoomika Rajput$^{1,2}$,\thanks{E-mail: bhoomikarjpt2@gmail.com}
Amit Kumar Mandal$^{3}$,
Ashwani Pandey$^{4}$,
C. S. Stalin$^{5}$,
Walter Max-Moerbeck$^{1}$,  
\newauthor
$ $ and Blesson Mathew$^{2}$
\\
$^{1}$Departamento de Astronomi{\'a}, Universidad de Chile, Camino El Observatorio 1515, Las Condes, Santiago, Chile\\
$^{2}$Department of Physics, CHRIST (Deemed to be University), Hosur Road, Bangalore 560 029, India\\
$^{3}$Astronomy Program, Department of Physics and Astronomy, Seoul National University, Seoul 151-742, Republic of Korea\\
$^{4}$Center for Theoretical Physics, Polish Academy of Sciences, Al. Lotnik{\'o}w 32/46, 02-668 Warsaw, Poland\\
$^{5}$Indian Institute of Astrophysics, Block II, Koramangala, Bangalore 560034, India\\
}

\date{Accepted XXX. Received YYY; in original form ZZZ}

\pubyear{2015}

\begin{document}
\label{firstpage}
\pagerange{\pageref{firstpage}--\pageref{lastpage}}
\maketitle

\begin{abstract}
The correlation between optical and $\gamma$-ray flux variations in blazars reveals a complex behaviour. In this study, we present our analysis of the connection between changes in optical and $\gamma$-ray emissions in the blazar Ton 599 over a span of approximately 15 years, from August 2008 to March 2023. Ton 599 reached its highest flux state across the entire electromagnetic spectrum during the second week of January 2023. To investigate the connection between changes in optical and $\gamma$-ray flux, we have designated five specific time periods, labeled as epochs A, B, C, D, and E. During periods B, C, D, and E, the source exhibited optical flares, while it was in its quiescent state during period A. The $\gamma$-ray counterparts to these optical flares are present during periods B, C, and E, however during period D, the $\gamma$-ray counterpart is either weak or absent. We conducted a broadband spectral energy distribution (SED) fitting by employing a one-zone leptonic emission model for these epochs. The SED analysis unveiled that the optical-UV emission primarily emanated from the accretion disk in quiescent period A, whereas synchrotron radiation from the jet dominated during periods B, C, D, and E. Diverse correlated patterns in the variations of optical and $\gamma$-ray emissions, like correlated optical and $\gamma$-ray flares, could be accounted for by changes in factors such as the magnetic field, bulk Lorentz factor, and electron density. On the other hand, an orphan optical flare could result from increased magnetic field and bulk Lorentz factor.
\end{abstract}



\section{Introduction}
Blazars, the peculiar category of active galactic nuclei (AGN) known for emitting non-thermal variable radiation covering the whole range of the electromagnetic spectrum, are among the most luminous objects in the Universe (luminosity $\approx$ 10$^{\rm {42}}$$-$10$^{\rm {48}}$ erg s$^{\rm {-1}}$). These objects emit relativistic jets that are oriented in the same direction as the observer's line of sight \citep{1993ARA&A..31..473A, 1995PASP..107..803U}, resulting in strong Doppler boosting. They reside within elliptical galaxies \citep{2000ApJ...532..816U} and their energy source is derived from the process of material accreting onto a supermassive black hole. (10$^{\rm {6}}$-10$^{\rm {10}}$ M$_{\rm {\odot}}$)\citep{1973A&A....24..337S, 2012ApJ...748...49S}.

Blazars are categorized into two groups: flat spectrum radio quasars (FSRQs) and BL Lacs. This classification relies on the rest-frame equivalent width (EW) of the emission lines observed in their optical spectra. FSRQs exhibit strong emission lines (EW $>$ 5 {\rm \AA}), while BL Lacs exhibit either weak or no emission lines (EW $<$ 5 {\rm \AA}) \citep{1991ApJS...76..813S, 1997A&A...325..109S}. The absence of broad emission lines in the optical spectra of BL Lac objects may be caused by Doppler-boosted non-thermal continuum, which could swamp out the spectral emission lines \citep{1978bllo.conf..328B}. Additionally, the presence of EW $>$ 5 {\rm \AA} observed in BL Lac objects might be the outcome of an especially low state of the beamed continuum \citep{1995ApJ...452L...5V}. Consequently, the EW by itself is not a reliable indicator of the difference between the two categories of blazars. \cite{2011MNRAS.414.2674G} and \cite{2012MNRAS.421.1764S} proposed a categorization scheme for FSRQs and BL Lacs that hinges on the luminosity of the broad-line region (BLR), which is measured in Eddington units. In accordance with this scheme, the luminosity of the broad line region (L$_{\rm BLR}$) to the Eddington luminosity ratio for FSRQs is greater than 5 $\times$ 10$^{\rm {-4}}$ while it is less than 5 $\times$ 10$^{-4}$ for BL Lacs.

The broadband spectral energy distribution (SED) of blazars has a two-hump structure, where one hump peaks at lower energies in the optical/IR/X-ray bands and the second hump peaks at higher energies in the MeV/GeV bands \citep{1998MNRAS.299..433F, 2016ApJS..224...26M}. The low-energy emission is attributed to the synchrotron emission process of the relativistic electrons in the jets \citep{1982ApJ...253...38U, 1988AJ.....95..307I}. \cite{1995ApJ...444..567P} and \cite{2010ApJ...716...30A} proposed an alternative method of categorizing blazars by considering the position of the synchrotron peak frequency ($\nu^{\rm syn}_{\rm peak}$) in their SED. These classes include low-synchrotron peaked blazars (LSP, $\nu^{\rm syn}_{\rm peak}$ $<$ 10$^{14}$ Hz), intermediate-synchrotron peaked blazars (ISP, 10$^{14}$ Hz $<$ $\nu^{\rm syn}_{\rm peak}$ $<$ 10$^{15}$ Hz), and high-synchrotron peaked blazars (HSP, 10$^{15}$ Hz $<$ $\nu^{\rm syn}_{\rm peak}$ $<$ 10$^{17}$ Hz). Most FSRQs are LSP blazars, while BL Lac sources exhibit all three blazar behaviors. Some BL Lacs are extremely high-synchrotron peaked blazars (EHBL or extreme blazars) with $\nu^{\rm syn}_{\rm peak}$ $>$ 10$^{17}$ Hz \citep{2019MNRAS.486.1741F}.

It is renowned that the synchrotron emission process results in low-energy emission. However, the origin of the higher energy peak remains partially understood. To explain this numerous models have been developed. Leptonic model often provides a satisfactory explanation for the observed broadband SED of the majority of blazars. In the leptonic scenario, the high-energy emission is produced by the inverse compton (IC) process \citep{2010ApJ...716...30A}. The seed photons that cause the IC process because of the scattering of the relativistic electron in the jets can originate from a variety of locations. These could originate from within the jets by synchrotron emission process (synchrotron self-Compton or SSC; \citealt{1981ApJ...243..700K, 1985ApJ...298..114M, 1989ApJ...340..181G, 2009A&A...501..879T})  or from outside the jets (external Compton or EC) mainly from the accretion disk \citep{1997A&A...324..395B, 2002ApJ...575..667D}, from the BLR \citep{1996MNRAS.280...67G, 2009ApJ...692...32D} and from the dusty torus \citep{2000ApJ...545..107B, 2008ApJ...675...71S}.

In addition to the leptonic model, the high-energy peak in the broadband SED of blazars can also be accounted for by the hadronic model \citep{2013ApJ...768...54B}. Hadronic model for blazar postulates that protons are accelerated to ultra-relativistic energies and dominate the high-energy emission through proton-synchrotron radiation \citep{2000NewA....5..377A, 2003APh....18..593M} or photo-pion production via proton-proton or proton-photon interaction \citep{1993A&A...269...67M}, which results in the production of ultra-high-energy photons and neutrinos. The hybrid lepto-hadronic model has also been proposed in some cases to explain high-energy emission in blazar \citep{2016ApJ...826...54D, 2016ApJ...817...61P, 2019MNRAS.483L..12C}. 

In addition to the broadband SED, the study of the temporal correlation between the two major components of the broadband emission i.e. optical and $\gamma$-ray emission can give major insights to the high-energy emission in blazars. In the leptonic model of blazar emission, since the relativistic electrons in the jets are responsible for both the optical and $\gamma$-ray emission \citep{2007Ap&SS.309...95B}, a close correlation between the optical and $\gamma$-ray flux changes is anticipated. The launch of the {\it Fermi} Gamma-ray space telescope\footnote{https://fermi.gsfc.nasa.gov/ssc/data/access/lat/} (referred to as {\it Fermi} hereafter; \citealt{2009ApJ...697.1071A}) in 2008 gave us unusual opportunity to characterise the $\gamma$-ray for a large number of sources \citep{2020ApJS..247...33A}. To investigate the connection between the optical and $\gamma$-ray bands, {\it Fermi} data was combined with numerous ground-based optical and IR observations.
Several studies have been conducted on this topic, yielding diverse conclusions. For example, correlations between optical and $\gamma$-ray changes, with or without lag, have been found in blazars \citep{2009ApJ...697L..81B, 2012ApJ...749..191C, 2014ApJ...783...83L, 2018MNRAS.480.5517L}. These findings firmly support the leptonic model of blazar emission. However, many studies disfavour the one-zone leptonic model due to the absence of a correlation between changes in optical and $\gamma$-ray flux. In these studies, either $\gamma$-ray flares are present without their corresponding optical flares (orphan $\gamma$-ray flares) \citep{2013ApJ...779..174D, 2014ApJ...797..137C, 2015ApJ...804..111M, 2020MNRAS.498.5128R} or optical flares are present without their $\gamma$-ray counterparts (orphan optical flares) (\citealt{2013ApJ...763L..11C, 2019MNRAS.486.1781R, 2020MNRAS.498.5128R} and references therein).

The existing observations suggest that the correlation between variations in optical flux and $\gamma$-ray flux is intricate. In this work, we conducted a thorough examination over the long-term duration to understand the diverse behaviour between changes in optical and $\gamma$-ray flux through broadband SED modeling in the blazar Ton 599. This blazar was recently observed at its brightest state \citep{2023ATel15854....1P, 2023ATel15859....1G, 2023ATel15875....1T}. The availability of multi-wavelength data over the long-term period renders this source appropriate candidate for conducting study on the optical-GeV connection. Ton 599, also known as 4C $+$29.45, is named as 4FGL J1159.5$+$2914 in the 4th catalog of {\it Fermi} Large Area Telescope (4FGL; \citealt{2022ApJS..260...53A}). It is located at a redshift z = 0.725 \citep{2010MNRAS.405.2302H}, with RA = 179$^\circ$.882641 and Dec = 29$^\circ$.245507. This source is Classified as an FSRQ source and it falls under the category of LSP source \citep{2010ApJ...710.1271A}. It displays high optical variability and exhibits strong polarization \citep{2006PASJ...58..797F}. In the $\gamma$-ray band this source was first detected by the Energetic Gamma Ray Experiment Telescope (EGRET) \citep{1995ApJS..101..259T} and later by the {\it Fermi} \citep{2010ApJ...715..429A}. Moreover, this source was detected in the very high-energy band (VHE $>$ 100 GeV) by VERITAS \citep{2017ATel11075....1M}. This source is observed with significant variability in the optical and $\gamma$-ray, as well as in other bands of the electromagnetic spectrum \citep{2006PASJ...58..797F, 2018ApJ...866..102P, 2019ApJ...871..101P, 2021Galax...9..118R, 2023MNRAS.tmp..320B}. In 2017, Ton 599 was detected in an unprecedented massive flaring state across the entire electromagnetic spectrum \citep{2017ATel10931....1C, 2017ATel10932....1C, 2017ATel10949....1A}. The optical and GeV connection for this source during its 2017 outburst was investigated by \cite{2019ApJ...871..101P}, and optical and $\gamma$-ray emission regions are found to be co-spatial with the time lag of a few days. This source was modelled using a two-zone leptonic emission model during the same flare by \cite{2020MNRAS.492...72P}, and it was found that the GeV emission was from EC process, where the seed photons originated from the dusty torus. The largest flare ever recorded by {\it Fermi} for this source in a period of around 15 years occurred in January 2023 when Ton 599 was once again found to be flaring, with a flux value of 3.60 $\pm$ 0.27 $\times$ 10$^{-6}$ {\rm ph cm$^{-2}$ s$^{-1}$}. Here, we present the findings of a multi-wavelength analysis performed on the blazar Ton 599 using data collected over a period of about 15 years, from August 2008 to March 2023. The main objective of this analysis is to determine any potential correlation between variations in the optical V-band flux and $\gamma$-ray flux and, as a result, putting constraints on the emission process in blazars. We provide the information about the data utilized in this work in section \ref{sec:mw_data}. Section \ref{sec:analysis} provides a detail of the analysis, and section \ref{sec:discussion} presents the findings and a discussion. The summary of the work is given in the final section.
\begin{figure*}
\hspace*{-3cm}
\includegraphics[width=1.3\textwidth]{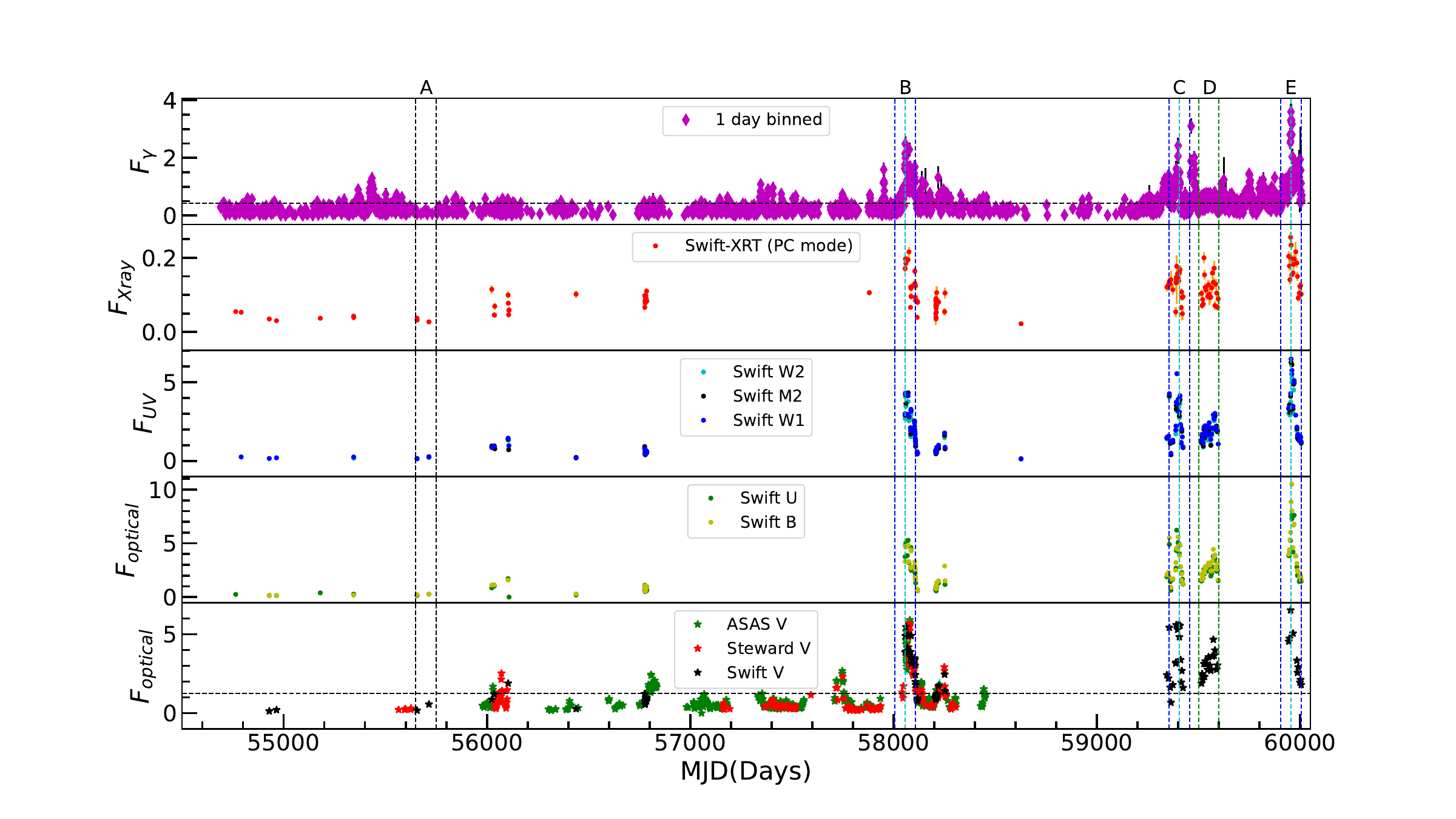}
\caption{Multi-wavelength light curves of the source Ton 599 for the period from MJD 54686 to MJD 60008. The top panel refers to the one day binned $\gamma$-ray (0.1-300 {\rm GeV}) light curve, the second panel refers to the  X-ray (0.3-10 {\rm keV}) light curve, the third panel refers to the UV light curve, the fourth and fifth panels refer to the optical light curves. The $\gamma$-ray fluxes are measured in the units of 10$^{-6}$ {\rm ph cm$^{-2}$ s$^{-1}$} and the UV and optical fluxes are measured in the units of 10$^{-11}$ {\rm erg cm$^{-2}$ s$^{-1}$}. The two vertical black dashed lines represent the quiescent epoch A. The vertical blue dashed lines refer to the flaring epochs B, C and E, when both optical and $\gamma$-ray flares are present. The two vertical green dashed lines represent the flaring epoch D, when optical flare is present without its corresponding $\gamma$-ray counterpart. The peaks of optical and $\gamma$-ray flares are denoted by the vertical cyan dashed lines. The horizontal black dashed lines in the $\gamma$-ray and optical V-band light curves represents the average flux from August 2008 to March 2023.}
\label{figure-lightcurve}
\end{figure*}  
\begin{table*}
\caption{Details of the epochs studied in this work for optical-GeV correlation and broadband SED modelling. In this table the optical flux values are in units of 10$^{-11}$ {\rm erg cm$^{-2}$ s$^{-1}$}, and the $\gamma$-ray flux values are in units of 10$^{-6}$ {\rm ph cm$^{-2}$ s$^{-1}$}.}
\resizebox{1.0\textwidth} {!}{  
\begin{tabular} {lcccccccccr} \hline
& & \multicolumn{2}{c}{Time-period (MJD)}  & \multicolumn{2}{c}{Calender date}   & \multicolumn{2}{c}{Average Flux} & \multicolumn{2}{c}{Peak Flux}  \\
Epoch & Status & Start & End & Start & End & Optical & $\gamma$ & Optical & $\gamma$\\ \hline
A & Quiescent & 55650 & 55750 & 2011-03-30 & 2011-07-08 & 0.37$\pm$0.02 & 0.11$\pm$0.02 & 0.56$\pm$0.03 & 0.21$\pm$0.09  \\
B & Optical flare with $\gamma$-ray counterpart & 58007 & 58107 & 2017-09-11 & 2017-12-20 & 3.89$\pm$0.02 & 0.99$\pm$0.02  & 5.47$\pm$0.13 & 2.49$\pm$0.28   \\
C & Optical flare with $\gamma$-ray counterpart & 59355 & 59455 & 2021-05-21 & 2021-08-29 & 3.62$\pm$0.03 & 0.63$\pm$0.02 & 5.60$\pm$0.14 & 2.42$\pm$0.28   \\
D & Optical flare without $\gamma$-ray counterpart & 59500 & 59600 & 2021-10-13 & 2022-01-21 & 3.03$\pm$0.02 & 0.45$\pm$0.01 & 4.67$\pm$0.12 & 0.81$\pm$0.05  \\
E & Optical flare with $\gamma$-ray counterpart & 59905 & 60005 & 2022-11-22 & 2023-03-02 & 5.55$\pm$0.03 & 1.32$\pm$0.03 & 10.30$\pm$0.14 & 3.60$\pm$0.27   \\
\hline
\end{tabular}}
\label{table-epochs}
\end{table*}
\begin{figure*}
\begin{center}$
\begin{array}{rrr}
\hspace{-1 cm}
\includegraphics[width=68mm,height=90mm]{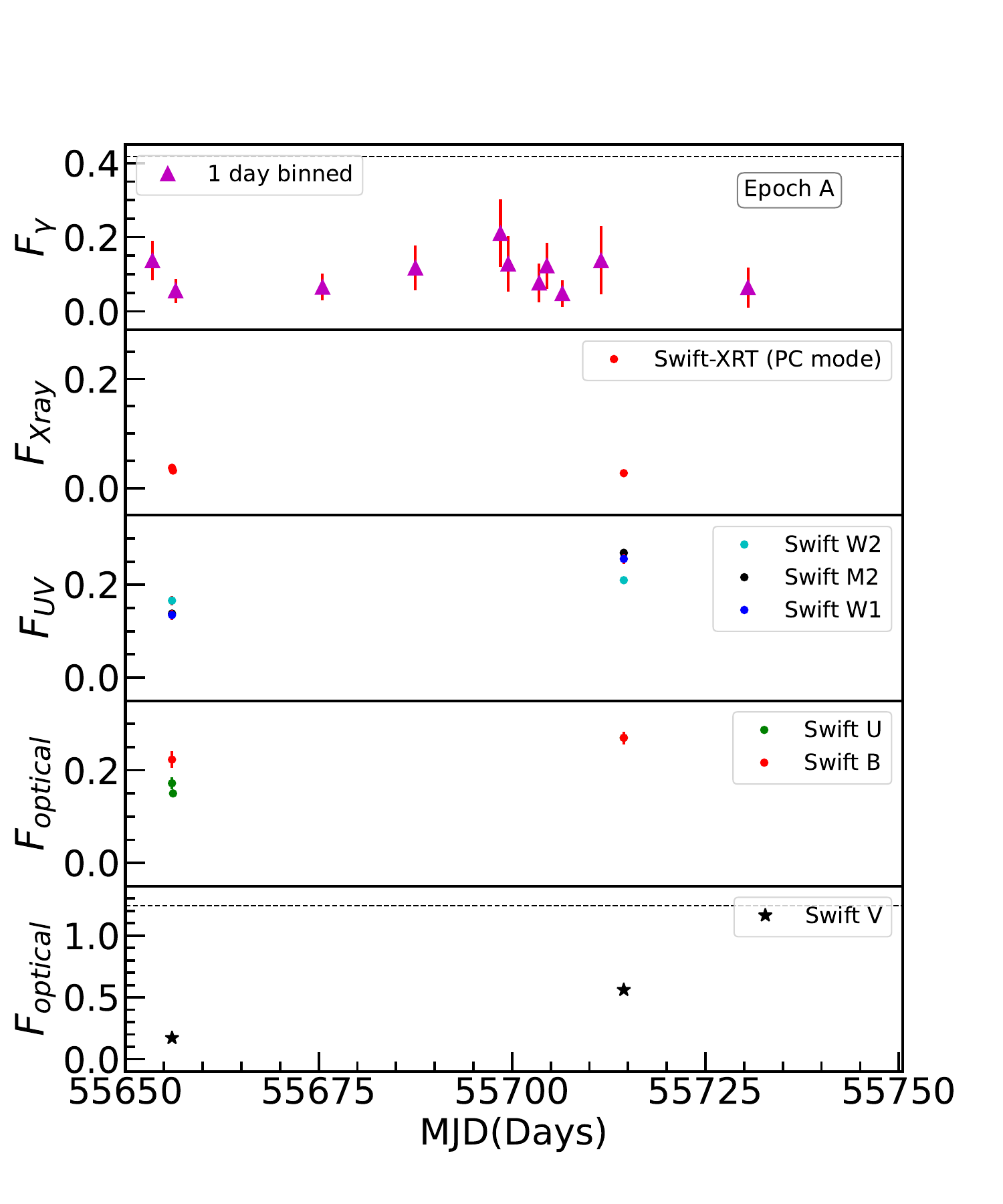}
\includegraphics[width=68mm,height=90mm]{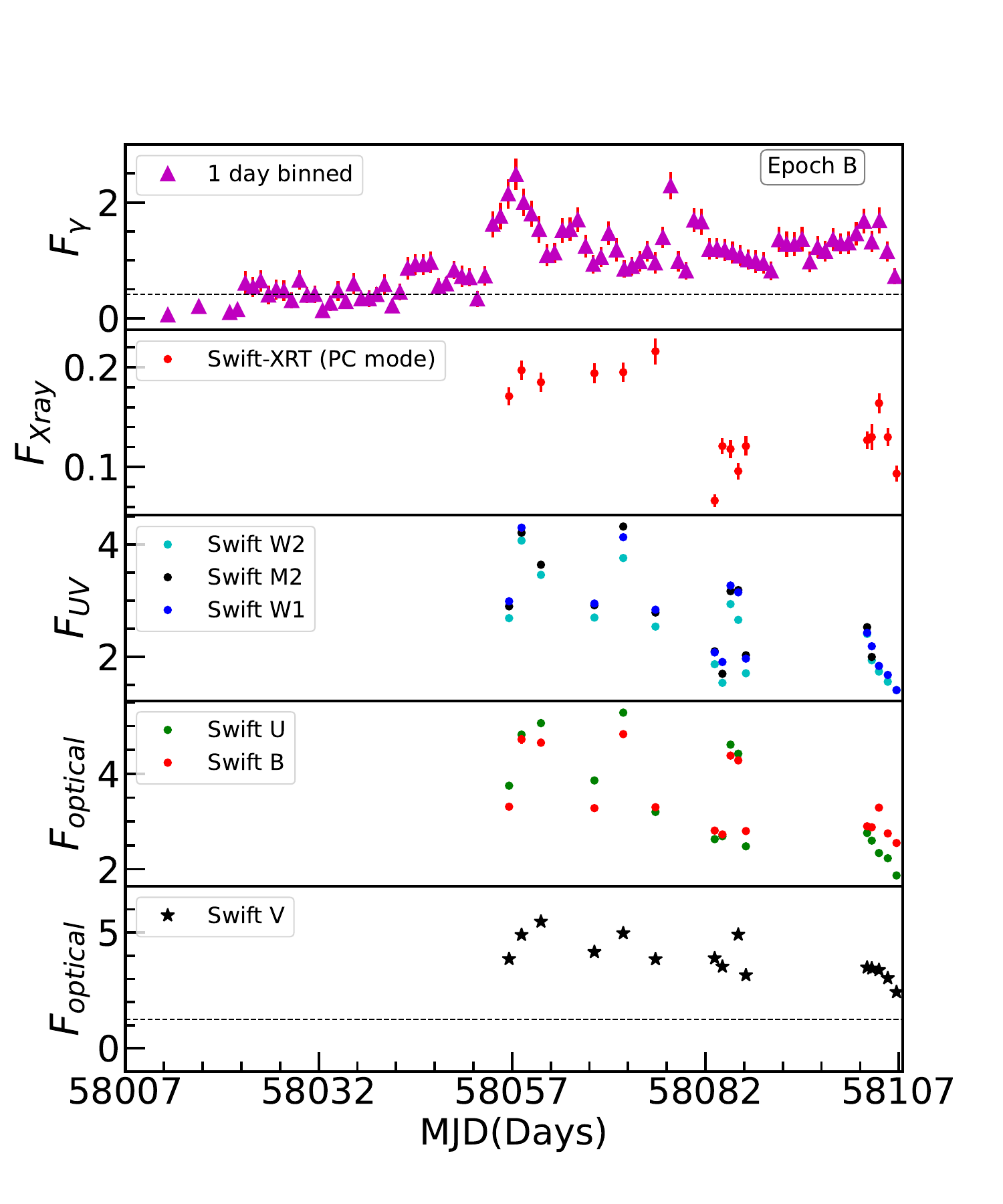}
\includegraphics[width=68mm,height=90mm]{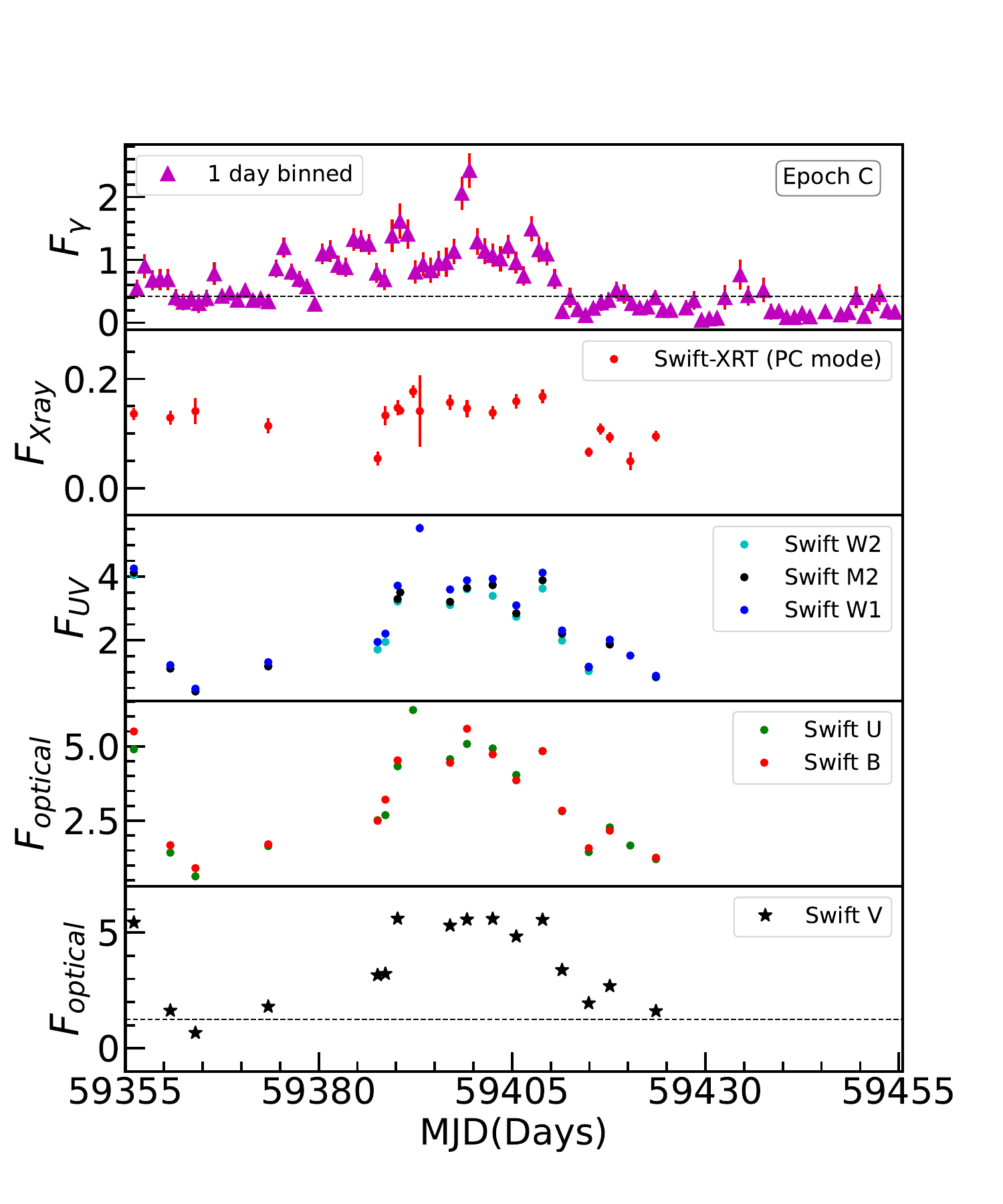}
\end{array}$
\end{center}
\begin{center}$
\begin{array}{rr}
\includegraphics[width=68mm,height=90mm]{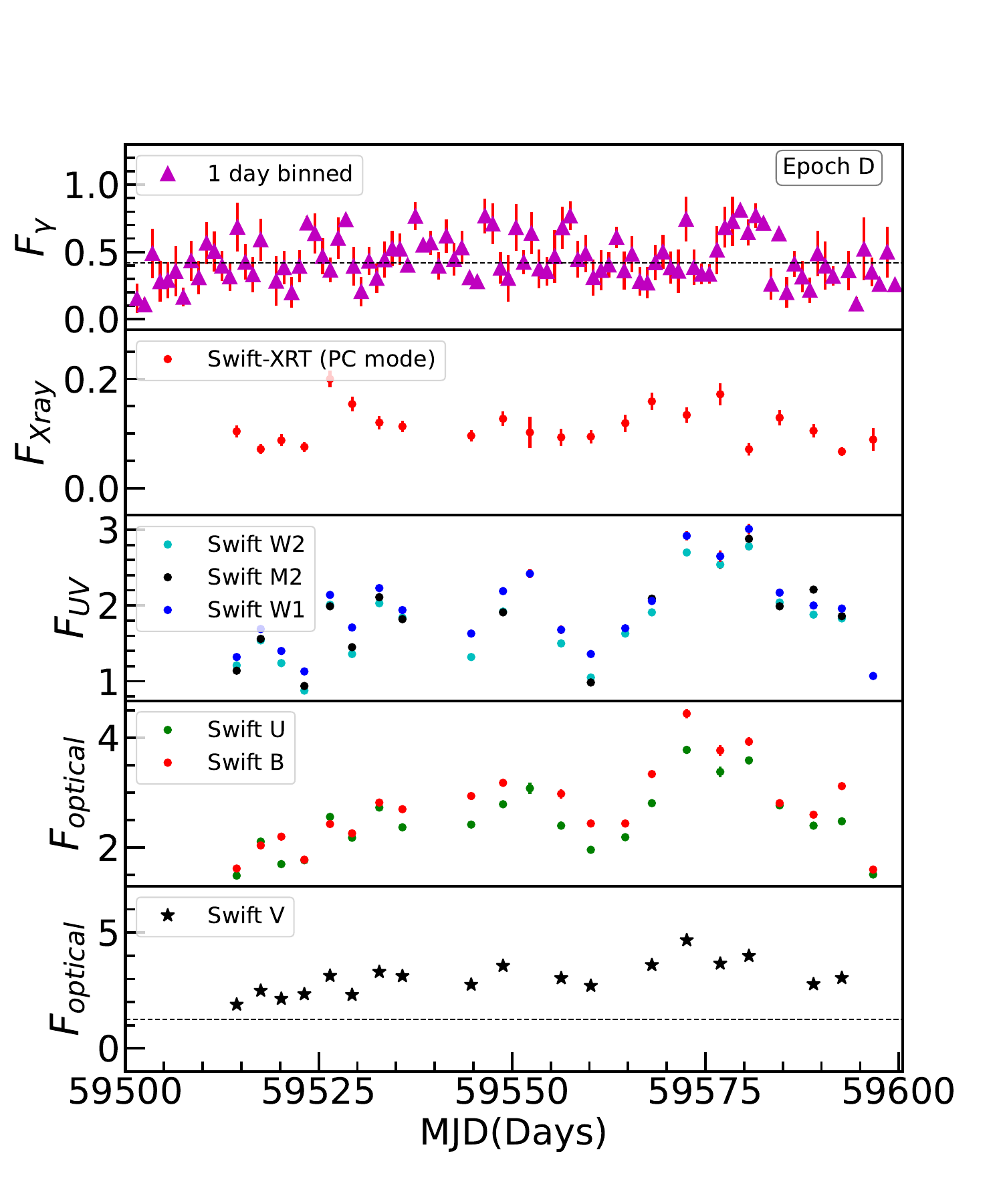}
\includegraphics[width=68mm,height=90mm]{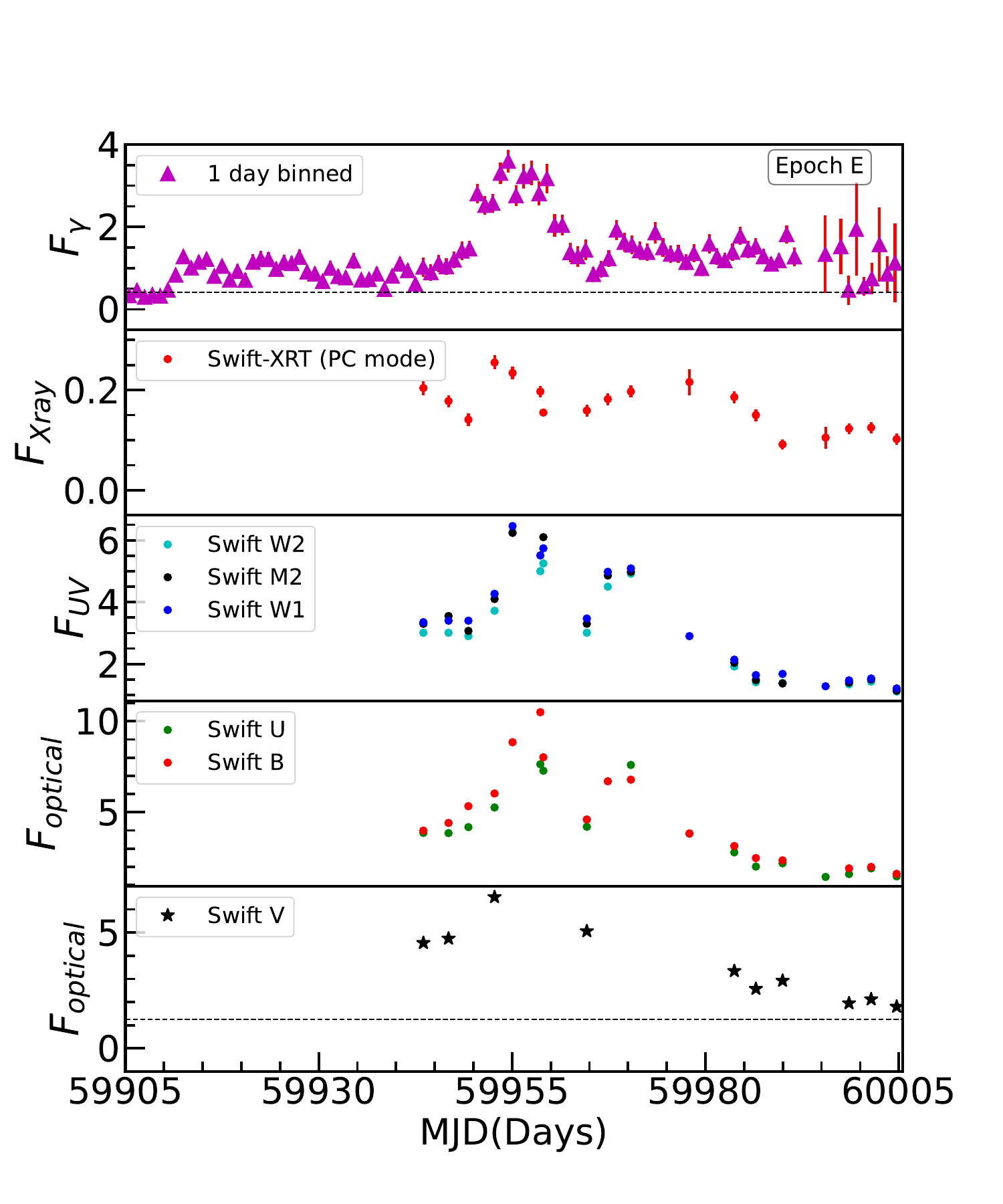}
\end{array}$
\end{center}
\caption{Multi-wavelength light curves for the selected epochs A, B, C, D, and E. The light curves for epochs A, B, and C are displayed in the upper left, middle, and right panels, respectively. The bottom left and right panels represent the light curves for epochs D and E, respectively. The detail of the panels are same as in Figure \ref{figure-lightcurve}.}
\label{figure-lightcurves_epoch1}
\end{figure*}

\section{Multi-wavelength data reduction}\label{sec:mw_data}
To carry out a multi-wavelength analysis of the source Ton 599, we used publicly accessible data in the $\gamma$-ray, X-ray, UV, and optical bands that span around 15 years, from August 2008 to March 2023.

\subsection{$\gamma$-ray data}
We analyzed about 15 years of data collected from the Large Area Telescope (LAT) aboard the {\it Fermi}. The {\it Fermi}-LAT is a pair-conversion telescope and has a very large effective area ($>$ 8000 {\rm cm$^2$}) \citep{2009ApJ...697.1071A}. It usually operates in the scanning mode and scans the entire sky once every $\sim$ 3 hours covering an extensive spectral energy range from 20 MeV to TeV energies. For this study, we performed binned likelihood analysis by using the {\it Fermi}-LAT data in the energy range of 100 MeV-300 GeV from 2008 August 08 to 2023 March 05, a period of approximately 15 years (MJD 54686$-$60008). To analyze the {\it Fermi} data we used the current version of Fermitools 2.2.0. We used PASS 8 data where the photon-like events are categorized as ’evclass=128, evtype=3’. All the events were chosen within a {10$^\circ$} circular region of interest (ROI) centered on the source and a {90$^\circ$} zenith angle cut was made to eliminate background $\gamma$-rays from the Earth's limb. After that, we generated a good time interval using the suggested criteria "(DATA QUAL $>$ 0)\&\&(LAT CONFIG==1)". For this analysis, we used the latest instrument response function (IRF) "P8R3\_SOURCE\_V3", Galactic diffuse emission model "gll\_iem\_v07" and the extragalactic isotropic diffuse emission model "iso\_P8R3\_SOURCE\_V3\_v1". Our source model includes all the point sources from the 4FGL catalog that fall within {10$^\circ$} of the source, as well as the Galactic and the extragalactic isotropic diffuse emission components. The spectral shapes of all the sources are adopted from the 4FGL catalog. For the sources lying between {0$^\circ$} to {5$^\circ$}, the spectral parameters are left free except the scaling factor and for the source lying between {5$^\circ$} to {10$^\circ$}, all the spectral parameters are kept fixed to their 4FGL catalog value. To determine the significance of the detection, we used the maximum likelihood (ML) test statistic, defined as TS = 2$\Delta$ log(L), where L represents the likelihood function comparing models with and without a $\gamma$-ray point source at the source's location. During the likelihood analysis, the sources with TS $<$ 25, which corresponds to a 5$\sigma$ detection \citep{1996ApJ...461..396M}, were removed. This modified model is then employed for the analysis of the light curve and spectra. To generate the one-day binned $\gamma$-ray light curve over the time period of interest, we regarded the source as detectable when the value of TS $>$ 9, equivalent to a 3$\sigma$ detection.

\subsection{X-ray data}

\subsubsection{{\it Swift}-XRT data}
The X-ray telescope (XRT), as detailed in \cite{2005SSRv..120..165B}, mounted on the {\it Swift} satellite \citep{2004ApJ...611.1005G}, operates within the 0.3 to 10 KeV energy range, observed Ton 599 concurrently with {\it Fermi}. The X-ray data utilized in this work was taken from the archives at HEASARC\footnote{https://heasarc.gsfc.nasa.gov/docs/archive.html}. To analyze the data, we adhered the standard procedure given by the instrument pipeline. We used the data obtained only in the photon counting (PC) mode for the light curve and spectrum analysis. We analyzed the data with the {\tt xrtpipeline} task, utilizing the latest CALDB and response files provided in HEASOFT version 6.29. We employed the standard grade selection 0-12. To generate the energy spectra the calibrated and cleaned events files were added. The source spectra were extracted from a circular area with a radius of 60$^{\prime\prime}$, positioned at the center of the source, whereas, the background spectra were extracted from a region with a radius of 80$^{\prime\prime}$, located away from the source. The exposure maps were combined using the {\tt ximage} task, and the ancillary response files (ARFs) were created using the {\tt xrtmkarf} task. The corresponding response matrix files (RMFs) were obtained from CALDB. The RMFs and ARFs files were loaded into the {\tt grppha} tool together with the source and background spectra in order to combine and group them. To perform the fitting of the final X-ray spectra within {\tt XSPEC} \citep{1996ASPC..101...17A}, we employed an absorbed simple power law model with a Galactic neutral hydrogen column density N$_{\rm {H}}$ = 1.63$\times$10$^{20}$ {\rm cm$^{-2}$} \citep{1990ARA&A..28..215D, 2005A&A...440..775K}. 

\subsubsection{{\it NuSTAR} data}
{\it NuSTAR}, the first focusing hard X-ray telescope, was launched in June 2012. It consists of two co-aligned X-ray detectors pair with their corresponding focal plane modules, called FPMA and FPMB and it operates over a wide energy range from 3 to 79 keV \citep{2013ApJ...770..103H}. {\it NuSTAR} recorded two observations of Ton 599 in May 2019 and June 2021. For this study, we used one observation (ObsID 60463037004) during the 2021 flare of Ton 599 with an exposure time of 17.5 ks. The data was reduced using the {\it NuSTAR} Data Analysis Software package {\tt NuSTARDAS} within the HEASOFT\footnote{http://heasarc.gsfc.nasa.gov/docs/nustar/analysis/} version 6.29. The calibrated and cleaned level 2 event files were produced using the {\tt nupipeline} task and the CALDB version 20211202. We extracted the source and background spectra using a circular
region of 30$^{\prime\prime}$ with the {\tt nuproducts} script. We grouped the spectra using {\tt grppha} with at least 20 counts per bin and further processed it into {\tt XSPEC}. 

\subsection{UV and optical data}
We obtained the data in the UV and optical bands from {\it Swift}-UV-Optical telescope ({\it Swift}-UVOT). We analyzed this data using the online tool\footnote{https://www.ssdc.asi.it/cgi-bin/swiftuvarchint}. To generate the light curves in the UV and optical bands, the magnitudes obtained using the online tool were converted into fluxes using the zero point method described in \cite{2011AIPC.1358..373B} without correcting for galactic reddening. However, to examine the spectral characteristics of the source, the magnitudes were corrected for galactic extinction (A$_{\rm \lambda}$), whose values were obtained from NED\footnote{https://ned.ipac.caltech.edu/}. In addition to using optical data obtained from {\it Swift}-UVOT, we also incorporated optical V-band data from the ground-based survey ASAS-SN\footnote{https://asas-sn.osu.edu/} and from the Steward observatory (SO)\footnote{http://james.as.arizona.edu/$\sim$psmith/Fermi/}. The ASAS-SN ground-based survey operates with five stations positioned in both the northern and southern hemispheres. It has the capability to conduct daily observations of the entire visible sky, reaching a depth of g = 18.5 mag \citep{2014ApJ...788...48S, 2017PASP..129j4502K}. The Steward observatory, which is located in Arizona, employs the 2.3 m Bok Telescope and the 1.54 m Kuiper Telescope for photometric observations.

\section{Analysis}\label{sec:analysis}

\subsection{Multi-wavelength light curves}\label{sec:analysis1}
We have generated multi-wavelength light curves of the source Ton 599 for a period of about 15 years from August 08, 2008 to March 05, 2023 (MJD 54686$-$60008). The multi-wavelength light curves, which include 1-day binned $\gamma$-ray, X-ray (PC mode), UV and optical light curves, are shown in Figure \ref{figure-lightcurve}. From Figure \ref{figure-lightcurve} it is conspicuous that Ton 599 has numerous quiescent and flaring stages during this period. We have identified specific time periods within this 15-year duration. These include one quiescent period labeled as epoch A and four flaring periods (when an optical or $\gamma$-ray flare is present), marked as epochs B, C, D, and E. These quiescent and flaring epochs have a 100-day time span and flaring epochs are centered on the peak of either the $\gamma$-ray or optical light curves. The time periods of these epochs are given in Table \ref{table-epochs}. These epochs were chosen based on the following criteria: 1) The accessibility of multi-wavelength data 2) optical and $\gamma$-ray light-curve data during the quiescent phase should be below the average level 3) optical or $\gamma$-ray flare should gradually increase by more than two times from the average level. The values of average optical and $\gamma$-ray fluxes are 1.24$\pm$0.003 $\times$ $10^{-11}$ {\rm erg cm$^{-2}$ s$^{-1}$} and 0.42$\pm$0.003 $\times$ $10^{-6}$ {\rm ph cm$^{-2}$ s$^{-1}$}, respectively, over the period from August 2008 to March 2023. The average and peak values of optical and $\gamma$-ray flux for each epoch are given in Table \ref{table-epochs}. The details of each of the epochs are given below.

\subsubsection{Epoch A (MJD 55650$-$55750)}
The source was in a quiescent state in the $\gamma$-ray, X-ray, UV and  optical bands during this epoch. The multi-wavelength light curve of this epoch is shown in the upper left panel of Figure \ref{figure-lightcurves_epoch1}. 

\subsubsection{Epoch B (MJD 58007$-$58107)}
During this epoch, the source showed major variation in $\gamma$-ray flux. The $\gamma$-ray flux increased by about six times from the average level during this epoch. Along with the large changes in the $\gamma$-ray flux, changes in the X-ray, UV and optical fluxes were also observed (refer to upper middle panel of Figure \ref{figure-lightcurves_epoch1}). The optical flux also increased by nearly 5 times from the average level. During this epoch, two short-term intense $\gamma$-ray flares were observed superimposed on the large $\gamma$-ray flare at around MJD 58057 and MJD 58077. Visual inspection of the Figure \ref{figure-lightcurves_epoch1} upper middle panel) also gives the hint of corresponding short-term optical flare at around MJD 58057 but there is no hint of an optical flare at around MJD 58077 due to lack of data points. We deduce that during this time period, we noted the same patterns of flux variations in both the optical and GeV energy ranges.

\subsubsection{Epoch C (MJD 59355$-$59455)}
The source experienced a significant $\gamma$-ray flare during this epoch, this one being around six times greater than the $\gamma$-ray flux on average. This substantial flaring event was subsequently accompanied by flares in the X-ray, UV, and optical bands. Notably, the optical flux experienced an increase of approximately five times compared to its average level. Thus, in this epoch too, we found that the variability pattern in the optical and $\gamma$-ray bands appeared to be correlated. The multi-wavelength light curve for this epoch is shown in the upper right panel of Figure \ref{figure-lightcurves_epoch1}. 

\subsubsection{Epoch D (MJD 59500$-$59600)}
During this epoch, the $\gamma$-ray flux variations are moderate but the changes in the X-ray, UV and optical flux are significant (see the lower left panel of the Figure \ref{figure-lightcurves_epoch1}). In comparison to the average flux level, the $\gamma$-ray flux increased by less than two times, whereas the optical flux increased by about four times compared to the average flux level. Since there was no clear large amplitude variation in the $\gamma$-ray light curve at this epoch, we can infer that in this epoch we observed an orphan optical flare.

\subsubsection{Epoch E (MJD 59905$-$60005)}
The source underwent the major $\gamma$-ray flare during this epoch, which is the largest $\gamma$-ray flare over the whole time period examined in this work. Such large flux variations were also noticed in the X-ray, UV and optical bands. The multi-wavelength light curve for this epoch is shown in lower right panel of Figure \ref{figure-lightcurves_epoch1}. During this epoch the $\gamma$-ray flux increased about nine times from the average flux level and the optical flux increased nearly eight times compared to the average flux. We conclude that in this epoch the optical and $\gamma$-ray flux changes are structurally correlated.

\begin{figure*}
\begin{center}$
\begin{array}{rr}
\includegraphics[width=75mm,height=75mm]{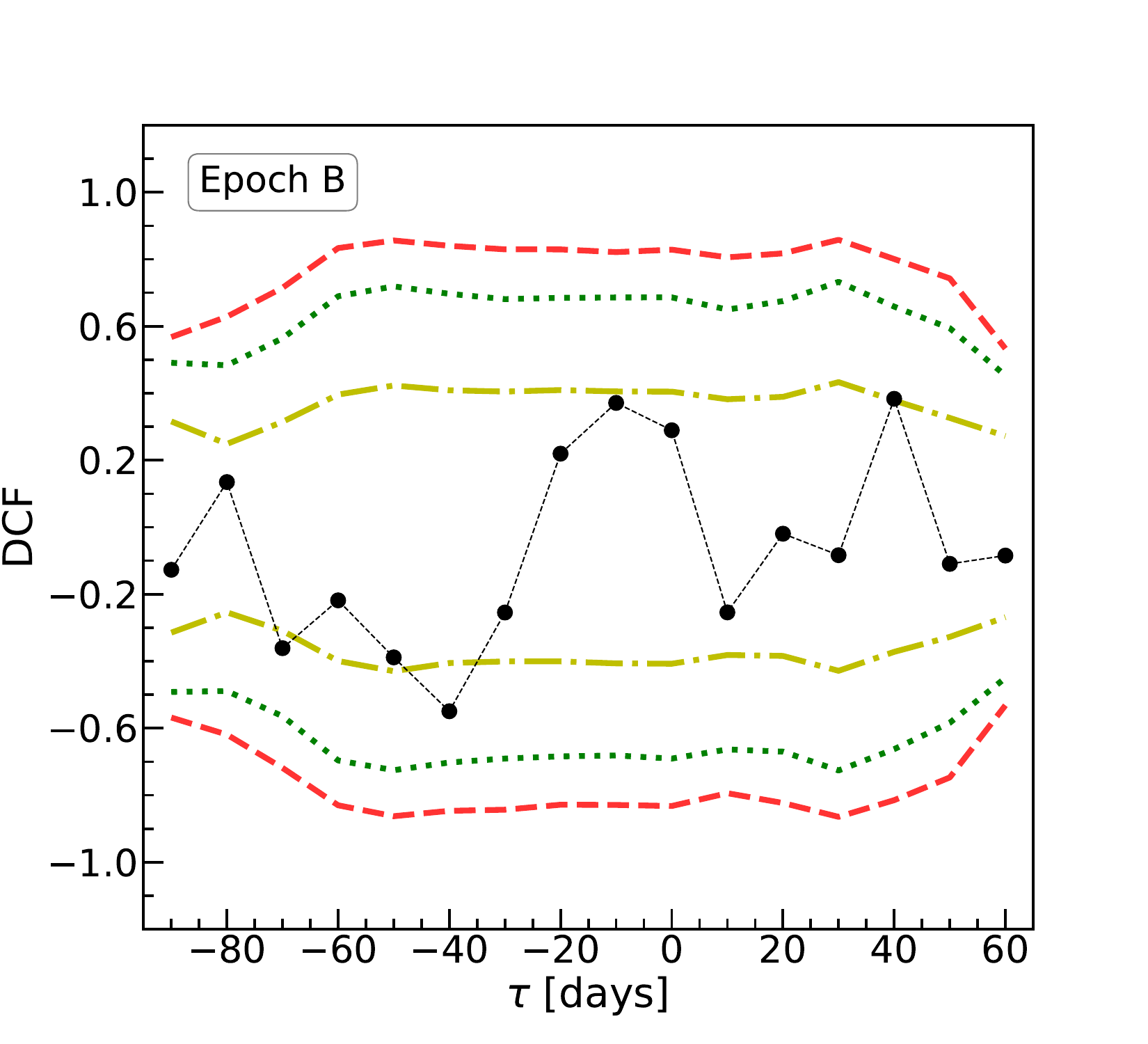}
\includegraphics[width=75mm,height=75mm]{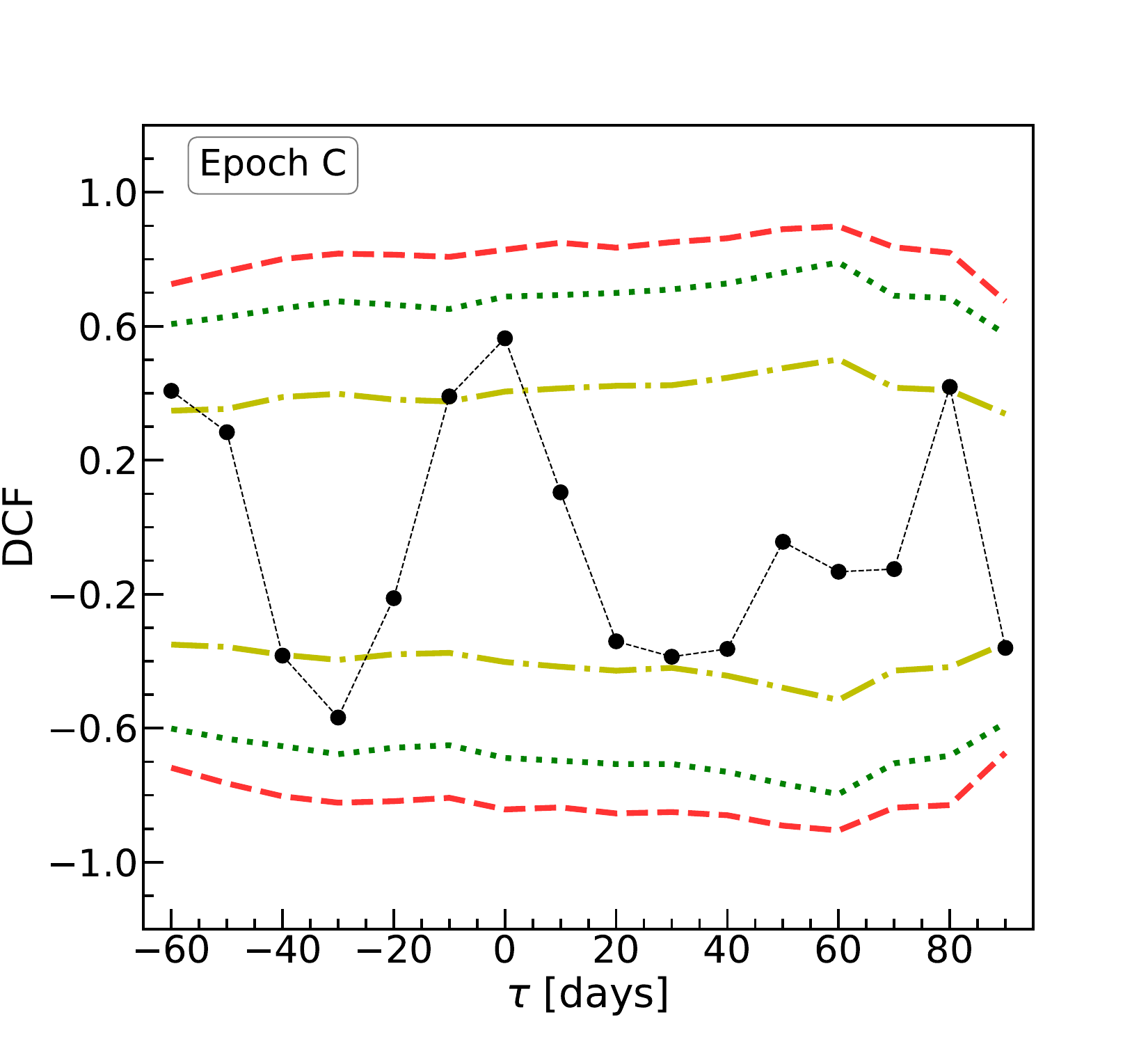}
\end{array}$
\end{center}
\begin{center}$
\begin{array}{rr}
\includegraphics[width=75mm,height=75mm]{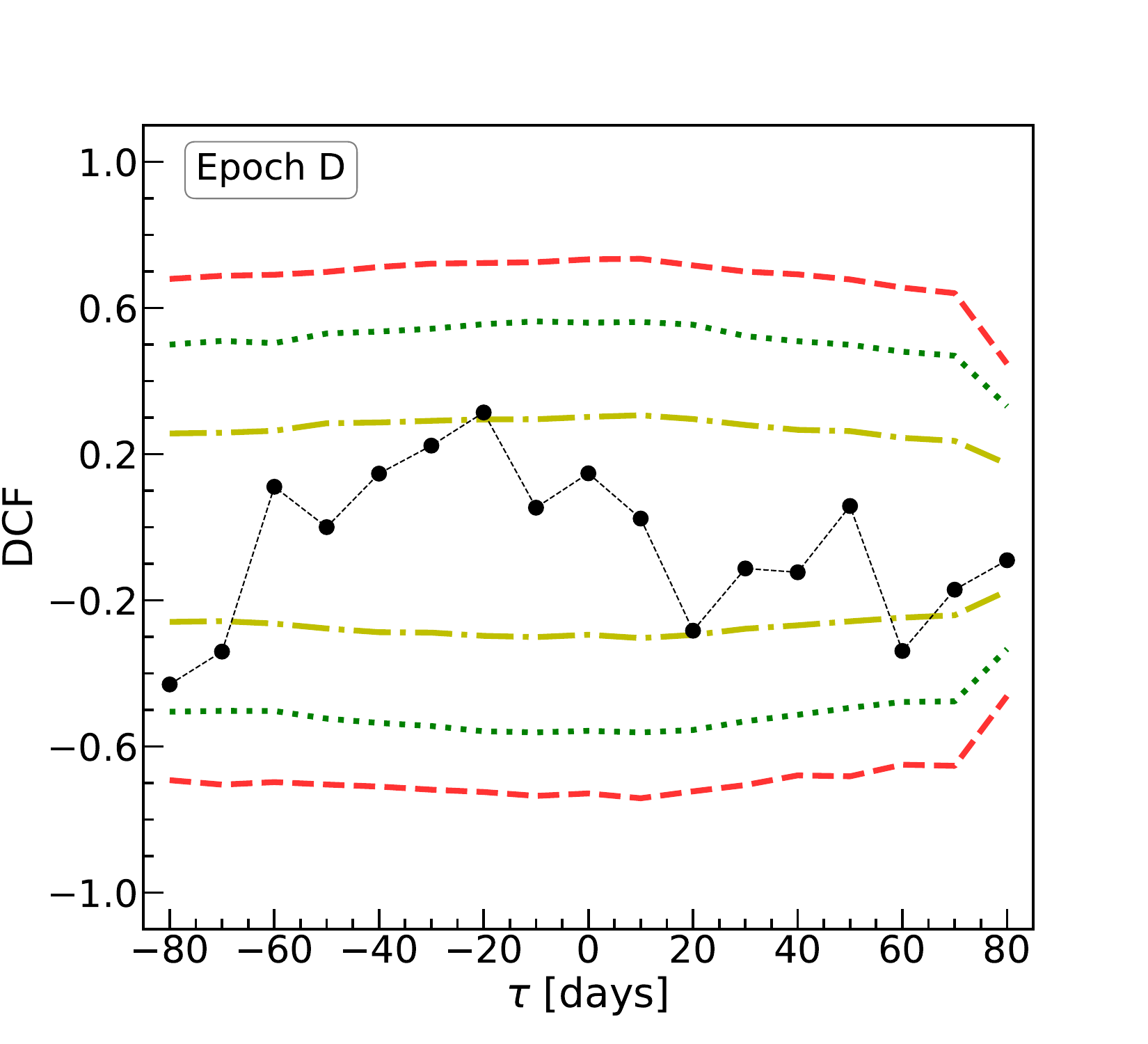}
\includegraphics[width=75mm,height=75mm]{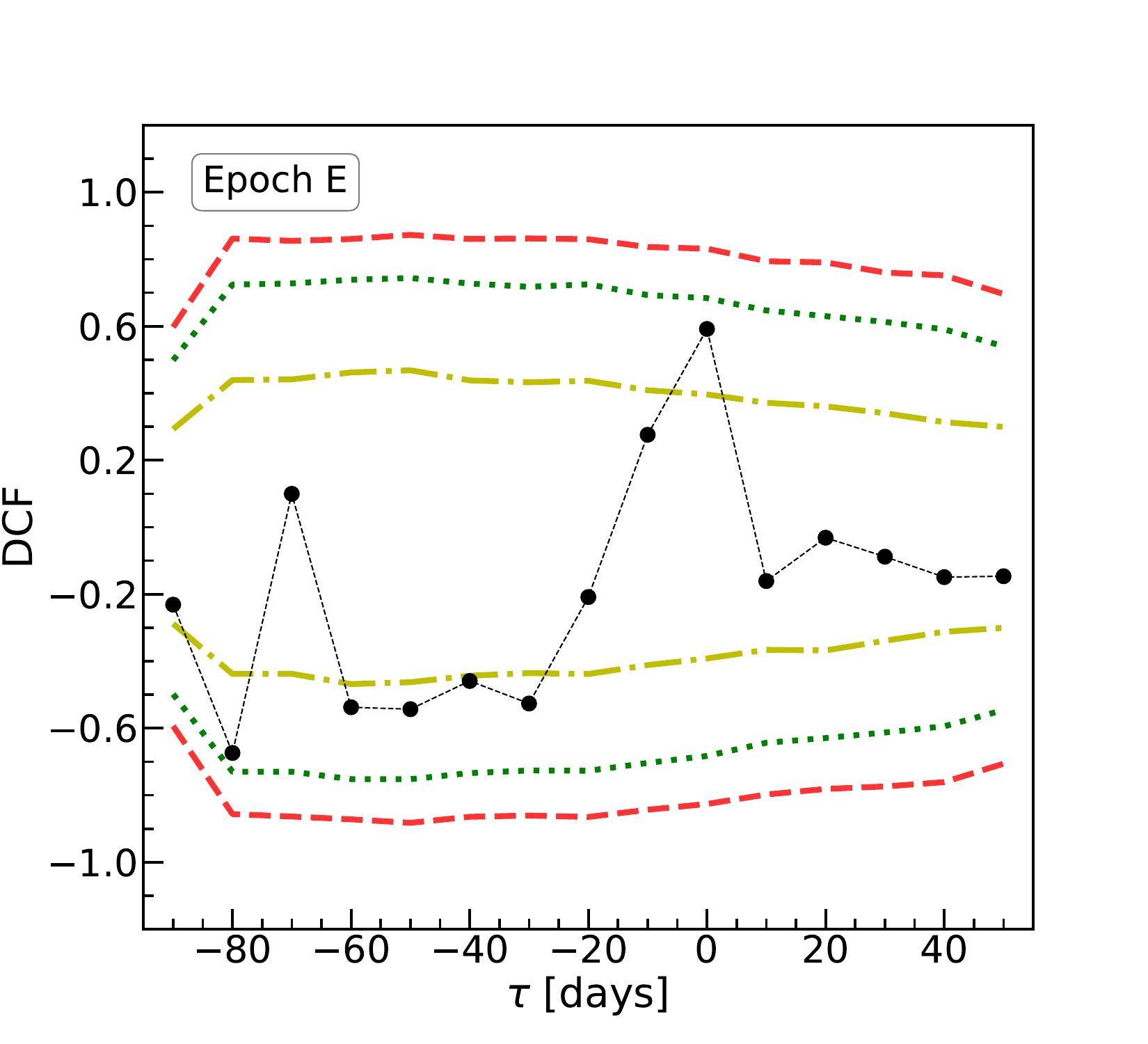}
\end{array}$
\end{center}
\caption{The outcomes of the discrete cross-correlation analysis conducted on the optical V-band and $\gamma$-ray light curves during epochs B, C, D, and E. The DCF is indicated by the black dashed line. The significance levels are represented by the yellow dashed-dotted line (68.27\%, 1$\sigma$), green dotted line (95.45\%, 2$\sigma$), and red dashed line (99.73\%, 3$\sigma$). A negative lag indicates the $\gamma$-ray flux leads the optical flux variations.}
\label{figure-cross_cor}
\end{figure*}

\subsection{Cross-correlation analysis} 
The initial investigation described in Section \ref{sec:analysis1} involved visually examining the correlation between changes in optical and $\gamma$-ray emissions over different time periods. Furthermore, we explored this correlation in depth by employing advanced cross-correlation analysis methods. Analyzing the cross-correlation between variations in flux across various energy bands allows for the determination of whether emission in different bands are co-spatial or if they emanate from different regions in the jet. In this study, we conduct a cross-correlation analysis using the discrete cross-correlation function (DCF) as outlined in \cite{1988ApJ...333..646E}) to measure the time lag between unevenly sampled optical (V-band) light curve and $\gamma$-ray light curve binned over one day. The significance of the cross-correlation is evaluated through Monte Carlo simulations, with the assumption of a simple power-law power spectral density (PSD) model. This model is defined as P($\nu$) $\propto$ 1/$\nu^{\beta}$, where $\nu$ represents the temporal frequency, and $\beta$ denotes the slope. A comprehensive explanation of this methodology is provided by \cite{2014MNRAS.445..437M}. We have adopted the value of $\beta_\gamma$ = 1.5 \citep{2014MNRAS.445..428M} and $\beta_{\rm {optical}}$ = 2 \citep{2021ApJ...909...39G}. The DCF between optical and $\gamma$-ray light curves for epochs B, C, D, and E are shown in Figure \ref{figure-cross_cor}. During these epochs, we  detected no significant correlations between variations in optical and $\gamma$-ray flux. The results of the DCF analysis are provided in Table \ref{table-cross-cor}.

\begin{table}
\caption{The outcomes of the DCF analysis between optical flux and $\gamma$-ray flux variations during epochs B, C, D, and E. Here $\tau$ represents the time lag, with negative values indicating that the $\gamma$-ray is leading the optical flux variations. The DCF values provide estimates of the cross-correlation coefficient, the p-value represents the likelihood of observing the correlation under the assumption of the null hypothesis.}
\begin{tabular} {lcccr} \hline
Epoch & $\tau$ (days) & DCF & p-value  \\ \hline
B & -10.0$\pm$6.1 & 0.37 & 0.365  \\
C & 0.0$\pm$3.7 & 0.56 & 0.139   \\
D & -20.0$\pm$14.8 & 0.31 & 0.287\\
E & 0.0$\pm$3.7 & 0.59 & 0.103 \\
\hline
\end{tabular}
\label{table-cross-cor}
\end{table}

\subsection{Optical spectral variations} 
\begin{figure*}
\begin{center}$
\begin{array}{rr}
\includegraphics[width=75mm,height=65mm]{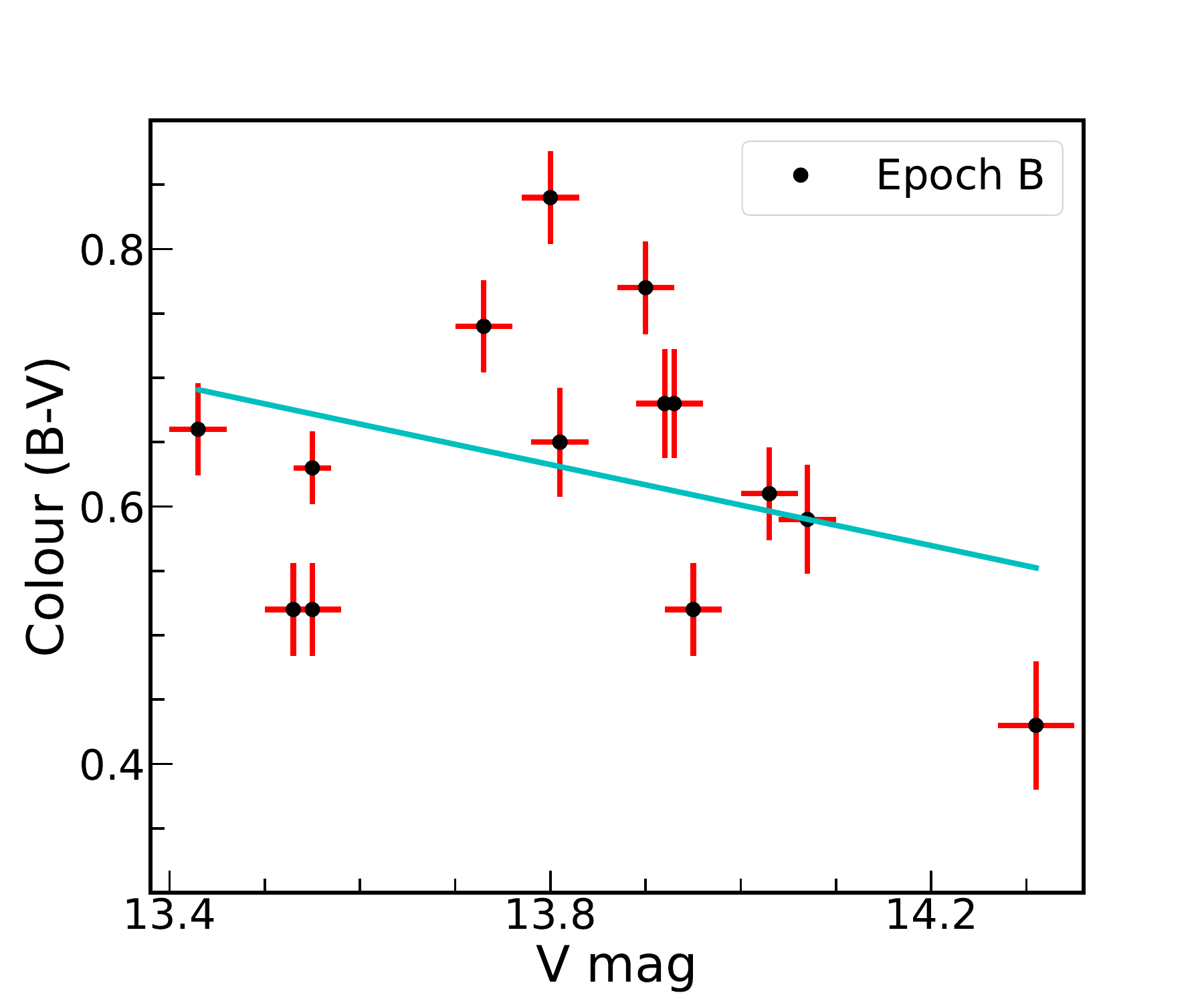}
\includegraphics[width=75mm,height=65mm]{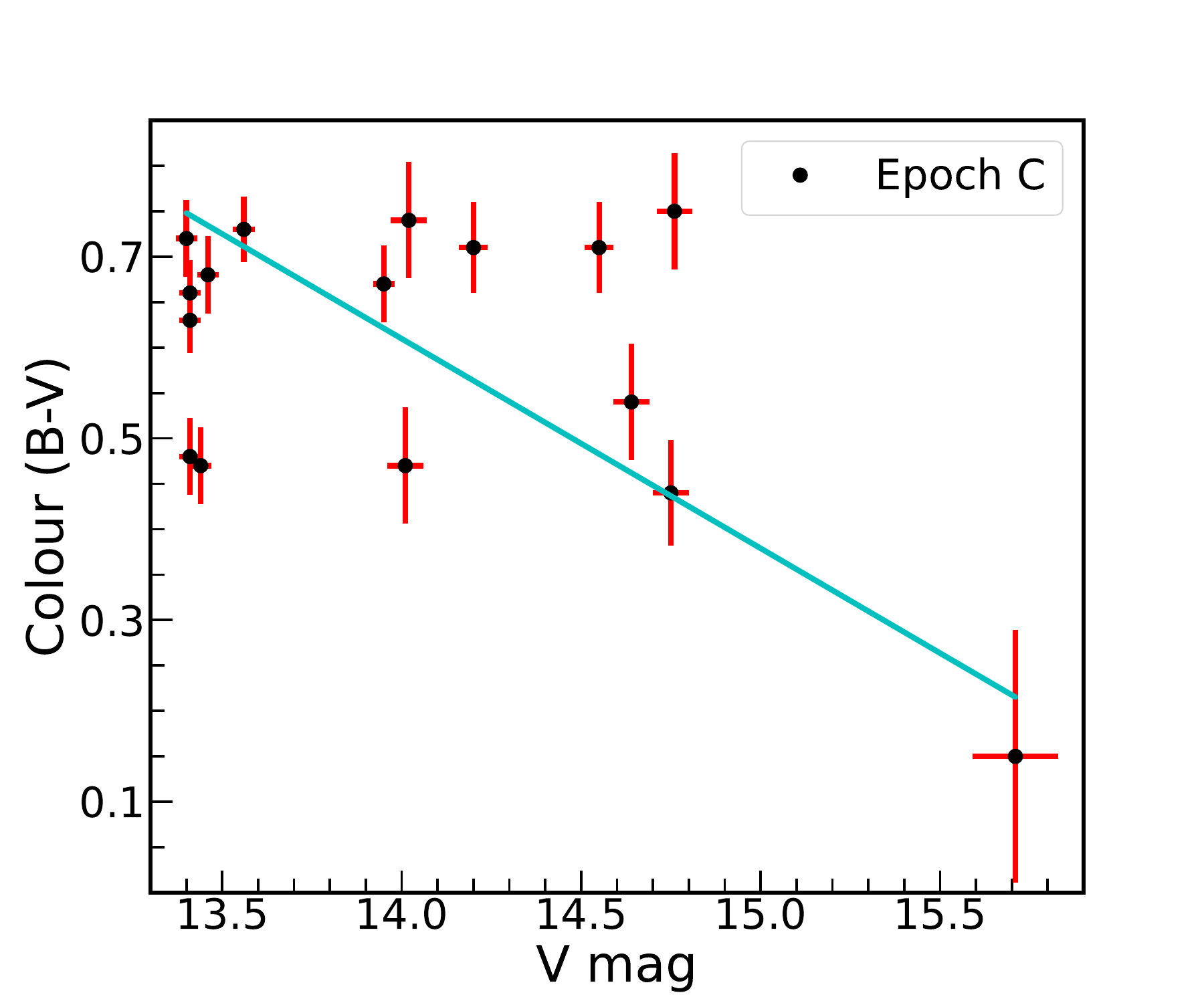}
\end{array}$
\end{center}
\begin{center}$
\begin{array}{rr}
\includegraphics[width=75mm,height=65mm]{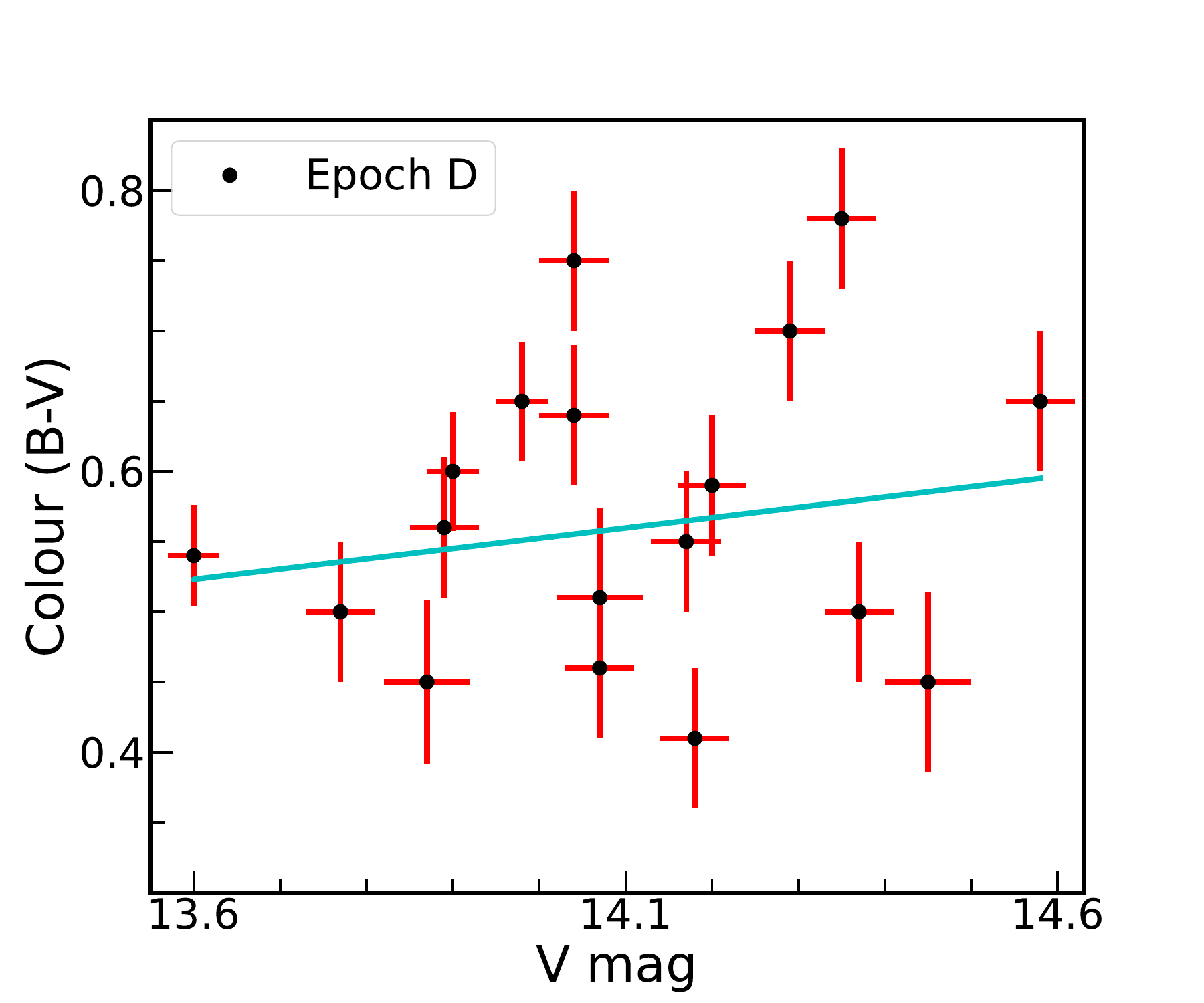}
\includegraphics[width=75mm,height=65mm]{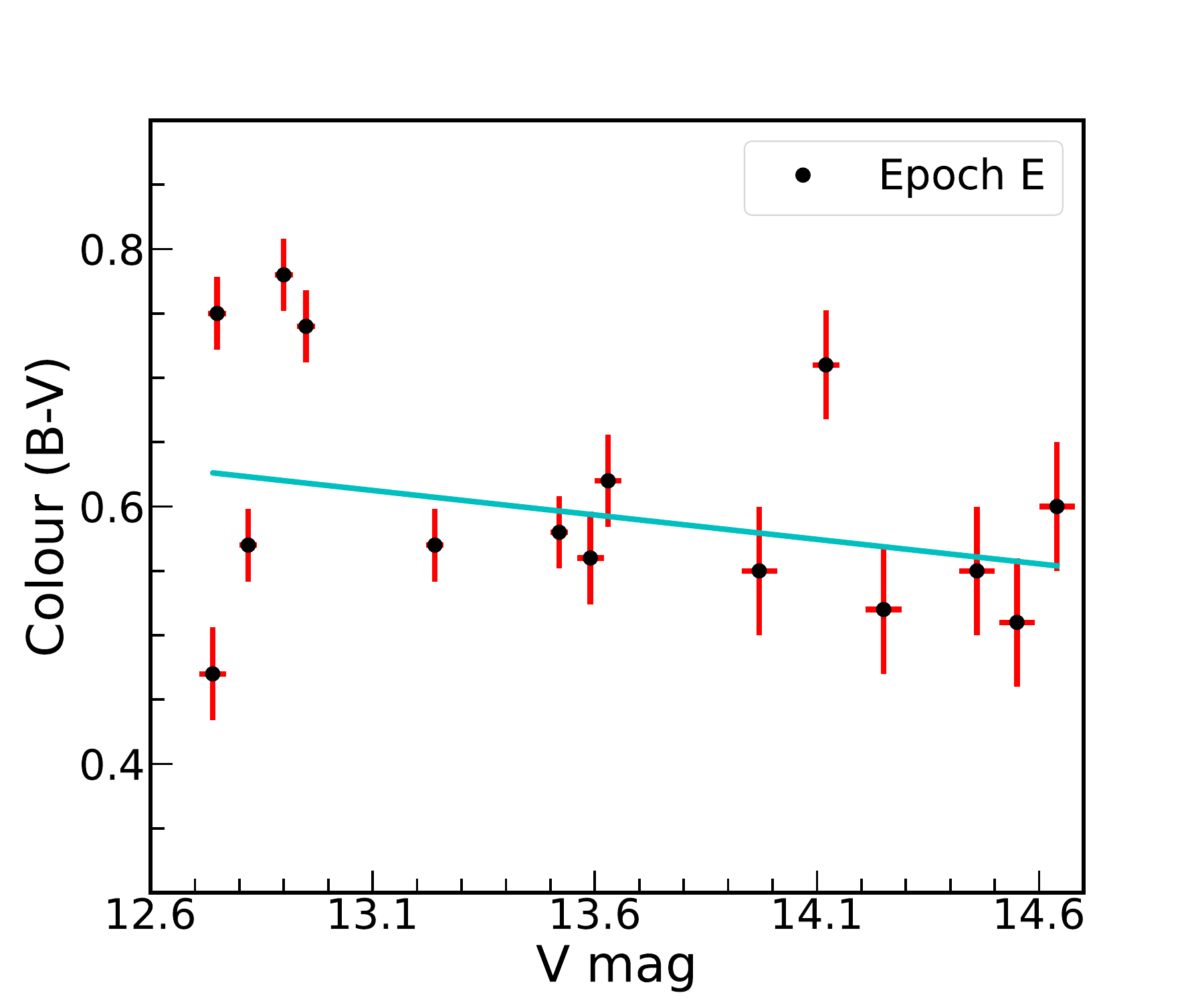}
\end{array}$
\end{center}
\caption{Colour-magnitude diagram for Epochs B, C, D and E. The cyan line depicts the weighted linear least squares fit to the data.}
\label{figure-col-mag}
\end{figure*}

Apart from changes in flux, blazars are also known for exhibiting variations in their spectral characteristics. In general, FSRQs exhibit redder-when-brighter (RWB) trend \citep{2006A&A...450...39G, 2012ApJ...756...13B, 2015RAA....15.1784Z, 2019MNRAS.486.1781R, 2020MNRAS.498.3578S, 2022MNRAS.510.1791N}. However bluer-when-brighter (BWB) pattern is also observed in FSRQ sources \citep{2011MNRAS.418.1640W, 2019MNRAS.486.1781R, 2020MNRAS.498.3578S, 2020MNRAS.498.5128R}. The optical and UV spectral regions (V, B, U, W1, M2, and W2 bands) investigated in this paper include contributions from both non-thermal emission from the relativistic jet and thermal emission from the accretion disk. Hence, by examining spectral variations, one can identify the many factors causing the observed flux variations. To describe the spectral behaviour of the epochs studied in this work we looked for the variations in the B-V colour against the V-band magnitude. The colour-magnitude diagrams of these epochs are shown in Figure \ref{figure-col-mag}. Due to lack of data, we have not included the color-magnitude diagram for epoch A. The trend of spectral behaviour during epochs B, C, D and E was identified by weighted linear least squares fit, which take into account errors in both colour and V-band magnitude. We determined the Spearman rank correlation coefficient (R) and probability of obtaining this correlation under the assumption of the null hypothesis (p-value). If R $>$ 0.5 or R $<$ $-$0.5 with p-value $<$ 0.05, the source is considered to have displayed any significant colour trend over the epochs. We found no spectral variations during the epochs considered in this work. The results of the weighted linear least squares fitting are given in Table \ref{table-col_mag}.

\begin{table}
\caption{Results of the weighted linear least squares fit and Spearman rank correlation test applied to the colour-magnitude diagram during epochs B, C, D, and E.  None of the correlations are significant for a chosen significance level of 0.05.}
\begin{tabular} {lcccr} \hline
Epoch & Intercept & Slope & R & p-value \\ \hline
B & 2.80$\pm$1.64 & -0.15$\pm$0.12 & -0.18 & 0.52  \\
C & 3.84$\pm$0.63 & -0.23$\pm$0.04 & -0.08 & 0.76   \\
D & -0.47$\pm$1.57 & 0.07$\pm$0.11 & 0.15 & 0.56  \\
E & 1.11$\pm$0.46 & -0.04$\pm$0.03 & -0.28 & 0.31 \\
\hline
\end{tabular}
\label{table-col_mag}
\end{table}

\subsection{$\gamma$-ray spectral analysis}
\begin{figure*}
\begin{center}$
\begin{array}{rrr}
\includegraphics[width=60mm,height=50mm]{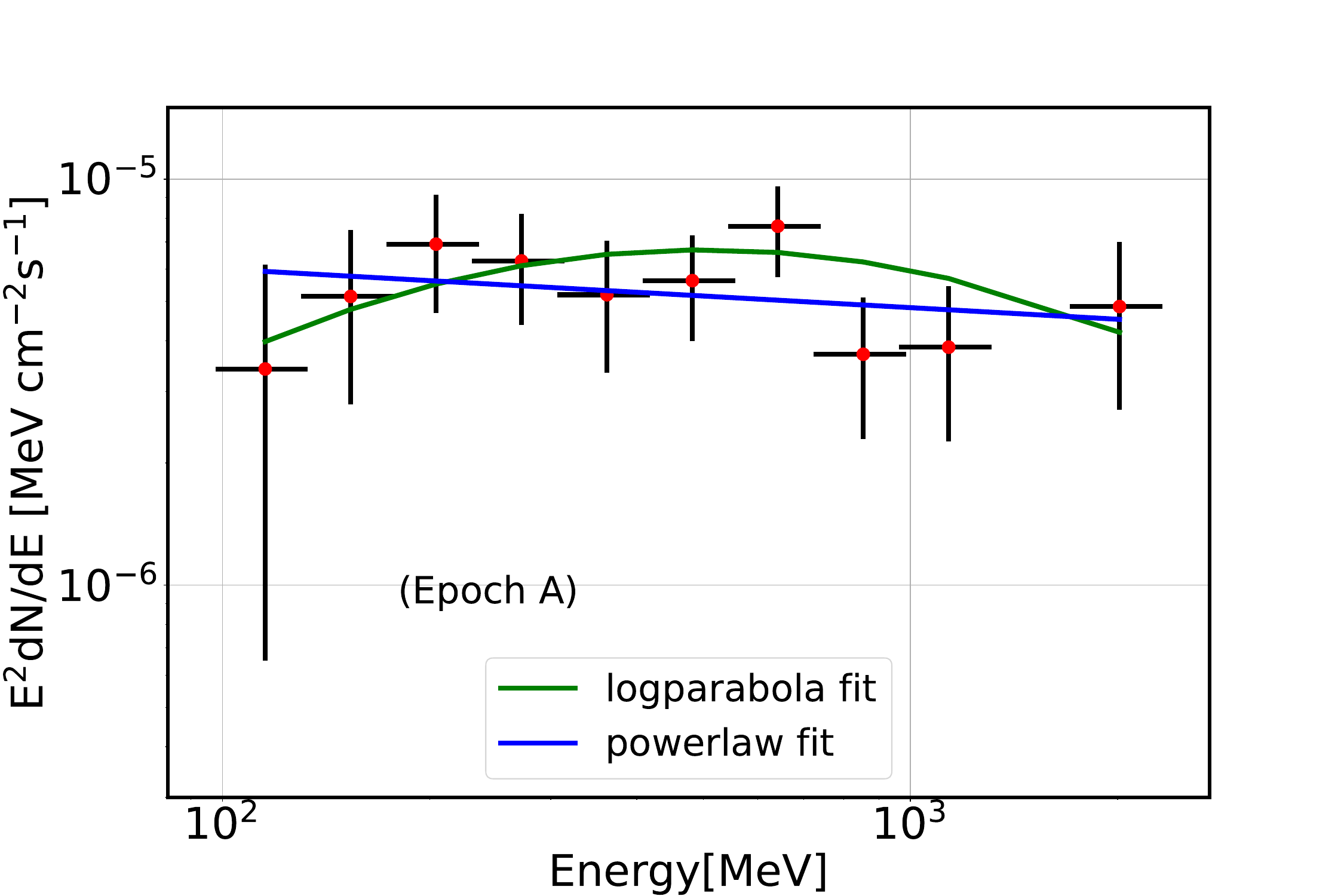}
\includegraphics[width=60mm,height=50mm]{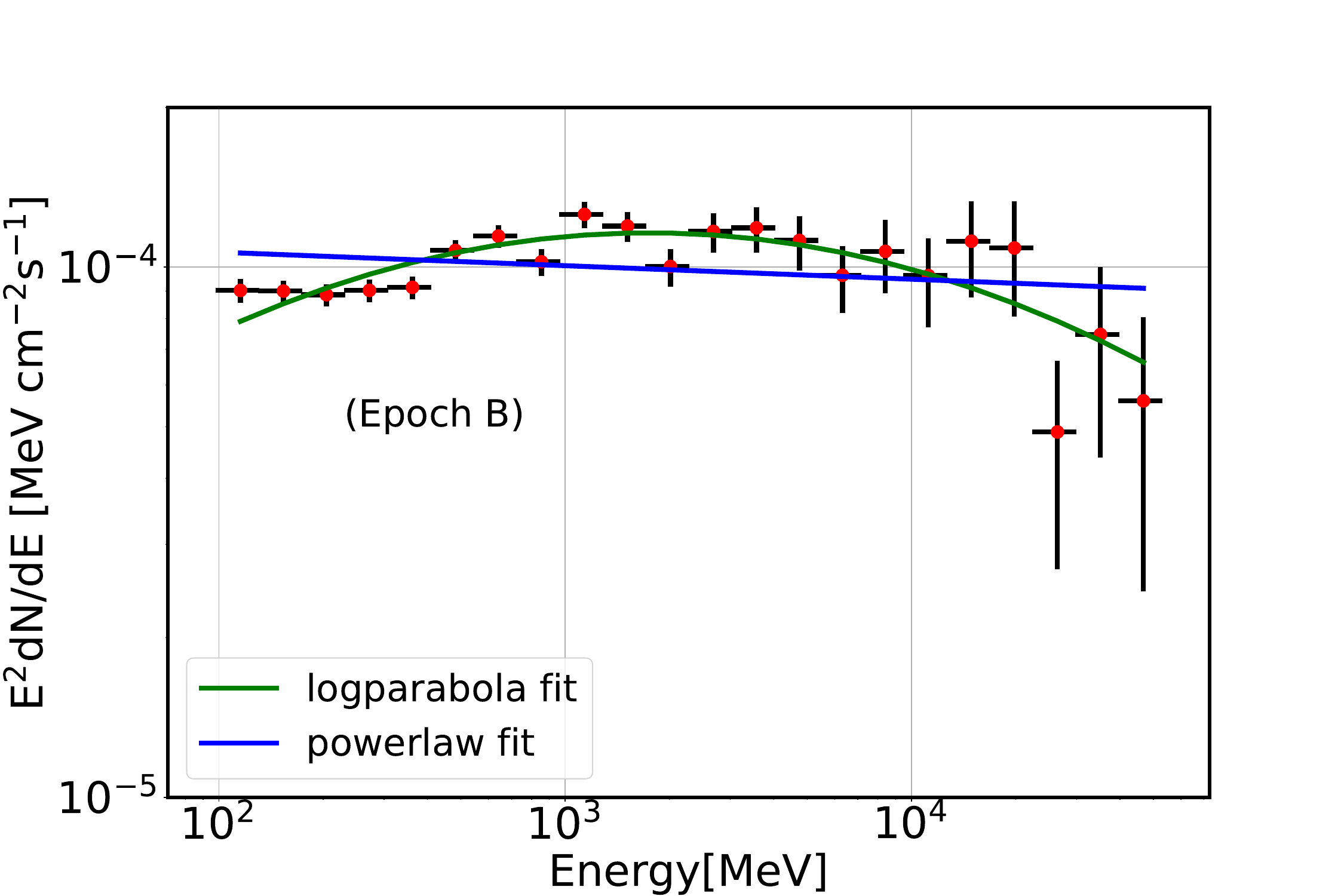}
\includegraphics[width=60mm,height=50mm]{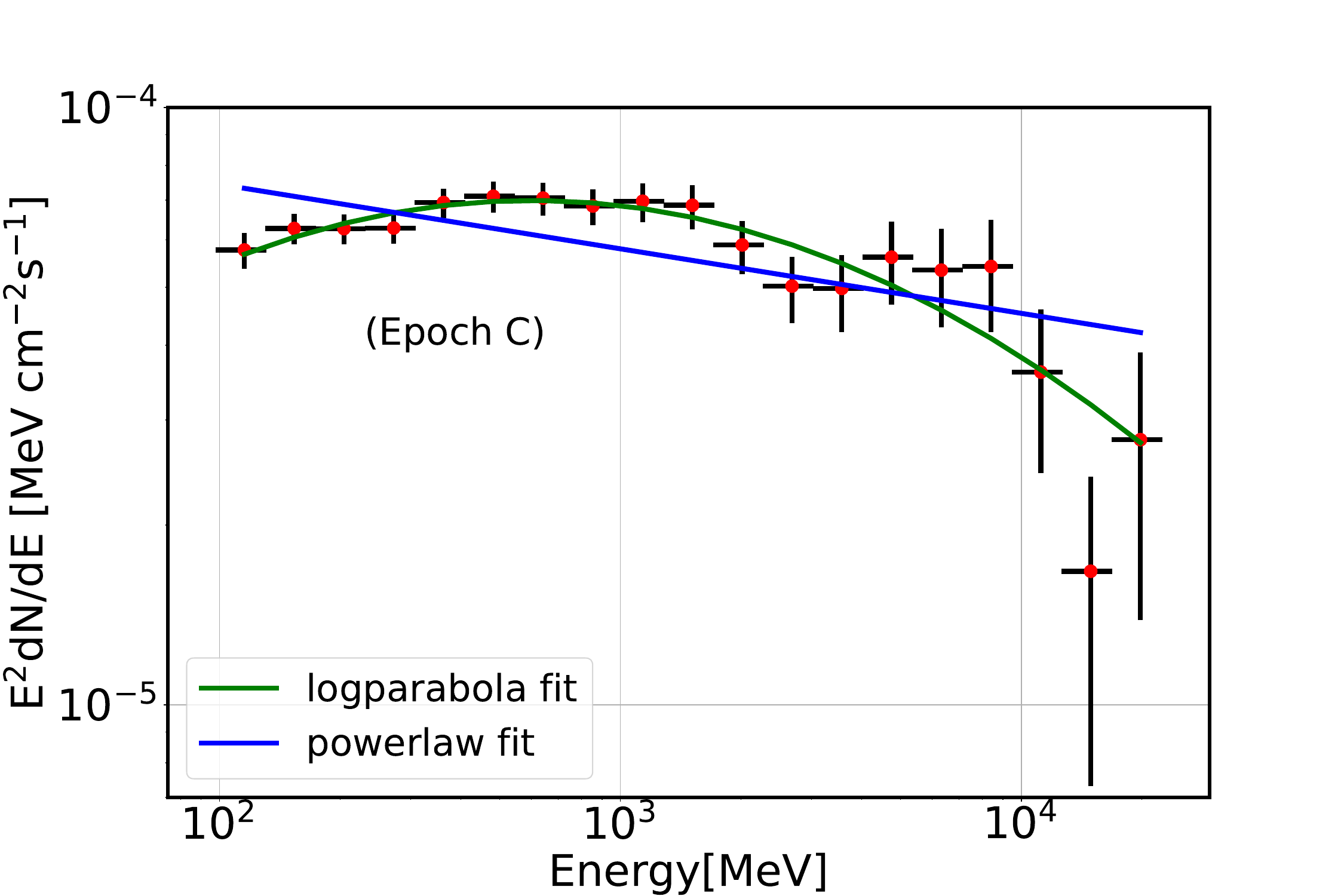}
\end{array}$
\end{center}
\begin{center}$
\begin{array}{rr}
\includegraphics[width=60mm,height=50mm]{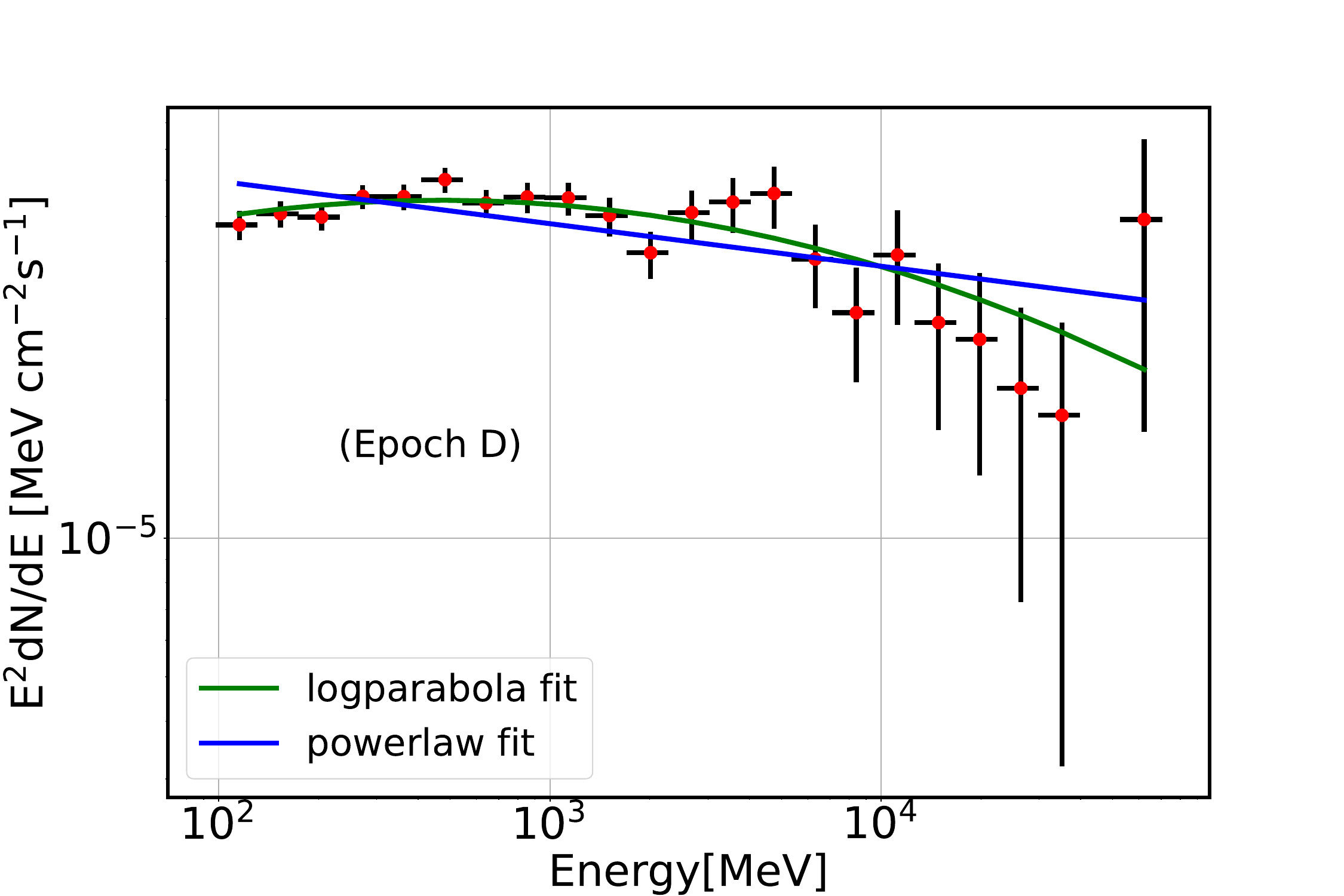}
\includegraphics[width=60mm,height=50mm]{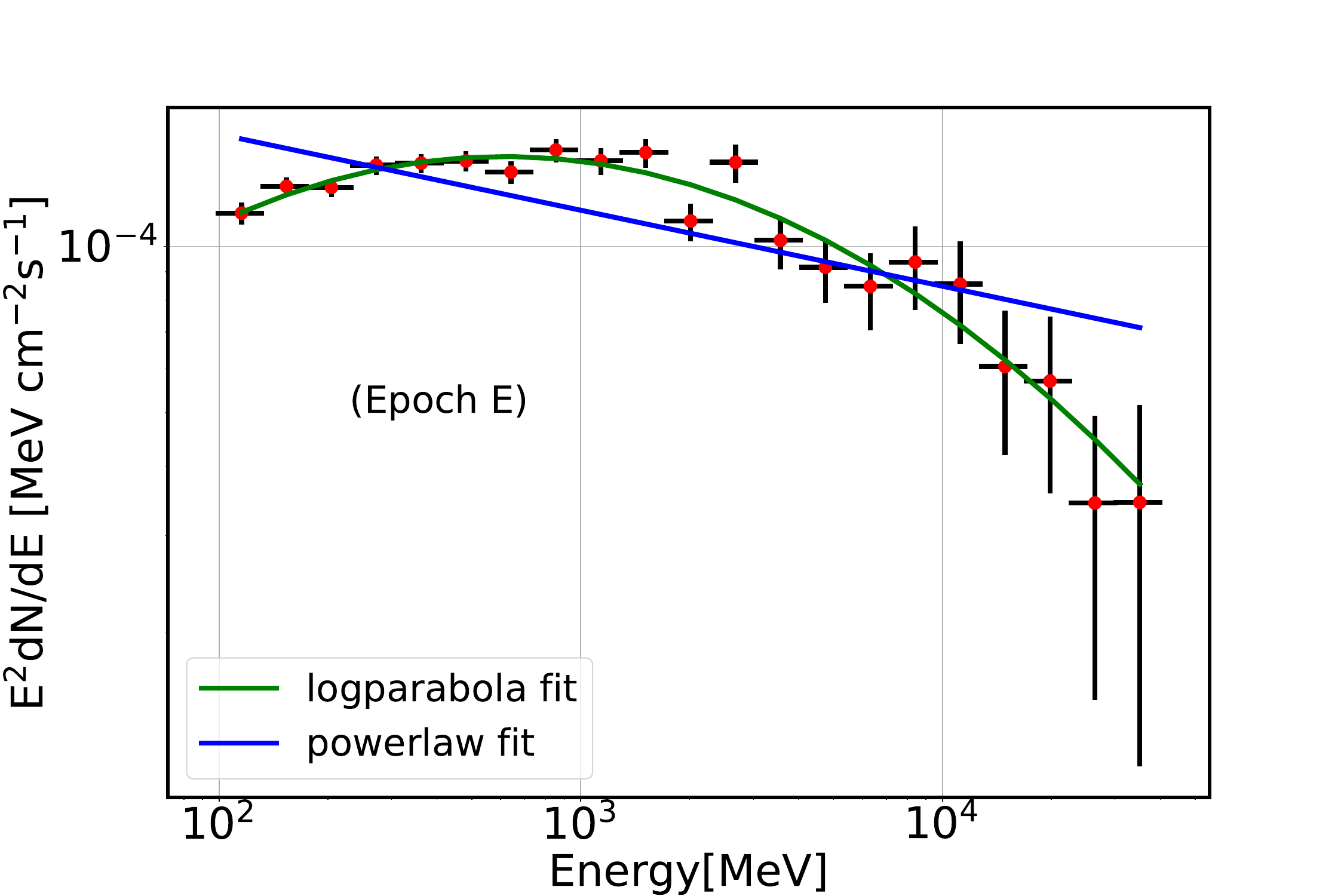}
\end{array}$
\end{center}
\caption{PL and LP fits to the $\gamma$-ray spectra for epochs A, B, C, D and E. Among these, the PL fit provides a better description of the $\gamma$-ray spectra in epoch A, while the LP fit better explains the $\gamma$-ray spectra in epochs B, C, D, and E. Refer to Table \ref{tab:res_spectrum} for detailed results.}
\label{figure-spectrum}
\end{figure*}

The inherent distribution of the emitting electrons in blazars may be reflected in the curvature of the $\gamma$-ray spectra, which also provides information on the potential acceleration and cooling processes in the jets. The $\gamma$-ray spectra of the blazars is either fit by power law (PL) or log parabola (LP) models. The PL model is defined as:
\begin{equation}
    dN(E)/dE = N_{\circ}(E/E_{\circ})^{{\Gamma_{\rm {P}}}}
\end{equation}
Here dN(E)/dE is the number of photons in {\rm cm$^{-2}$ s$^{-1}$ MeV$^{-1}$}, N$_{\circ}$ is the normalization factor of the energy spectrum, E$_{\circ}$ is the scaling factor and $\Gamma_{\rm {P}}$ is the photon index.\\
The LP model is defined as
\begin{equation}
    dN(E)/dE = N_{\circ}(E/E_{\circ})^{-\alpha-\beta ln(E/E_{\circ})}
\end{equation}
Here, $\alpha$ is the photon index at E and $\beta$ is the curvature index that determines the curvature around the peak.\\
To evaluate the model that more accurately characterizes the $\gamma$-ray spectra (PL against LP), we applied the maximum likelihood estimation technique outlined in \cite{1996ApJ...461..396M}, utilizing the analysis tool {\tt gtlike}. The {\tt gtlike} tool maximizes the likelihood by employing different optimizers, which determine the optimal spectral parameter values for the best fit. We also calculated the TS$_{\rm {curve}}$ value, as specified in \cite{2012ApJS..199...31N}, using the following equation:
\begin{equation}
TS_{\rm {curve}} = 2(log L_{\rm {LP}} - log L_{\rm {PL}})
\end{equation} 
In the above equation, L is the likelihood function. In order to detect significant curvature in the $\gamma$-ray spectra, we employed the threshold value TS$_{\rm {curve}}$ $>$ 16 (at 4$\sigma$ level; \citealt{1996ApJ...461..396M}). The results of the $\gamma$-ray spectral analysis are given in Table \ref{tab:res_spectrum} and the $\gamma$-ray spectra for the selected epochs are shown in Figure \ref{figure-spectrum}. The $\gamma$-ray spectra are well fit with the LP function for all the epochs, except the quiescent epoch A, where the value of TS$_{\rm {curve}}$ is less than 16.

\begin{table*}
\centering
\caption{\label{tab:res_spectrum} Spectral analysis results for the $\gamma$-ray observations. The $\gamma$-ray flux values are expressed in units of 10$^{-6}$ {\rm ph cm$^{-2}$ s$^{-1}$}. The PL and LP fits to the $\gamma$-ray spectra during these time periods are shown in Figure \ref{figure-spectrum}.}
\resizebox{\textwidth}{!}{
\begin{tabular}{@{\extracolsep{6pt}}lcccccccccc@{}}  \hline
&  \multicolumn{4}{c}{PL} & \multicolumn{5}{c}{LP} & \\ 
\cline{2-5}
\cline{6-10}
Epochs & $\Gamma_{\rm {P}}$ & Flux & TS & -{\rm Log} L$_{\rm {PL}}$ & $\alpha$ & $\beta$ & Flux & TS & -{\rm Log} L$_{\rm LP}$ & TS$_{\rm {curve}}$\\ \hline
A & 2.22$\pm$0.01 & 0.04$\pm$0.001 & 117.50 & 315166.67 & 2.16$\pm$0.16 & 0.42$\pm$0.78 & 0.04$\pm$0.008 & 117.82 & 315166.50 & 0.34\\
B & 1.95$\pm$0.01 & 0.92$\pm$0.01 & 19702.7 & 385467.61 & 1.88$\pm$0.002 & 0.49$\pm$0.01 & 0.88$\pm$0.003 & 19754.5 & 385446.44 & 42.34\\ 
C & 2.05$\pm$0.02 & 0.61$\pm$0.01 & 10498.9 & 406121.01 & 1.96$\pm$0.02 & 0.82$\pm$0.13 & 0.58$\pm$0.01 & 10552.7 & 406094.64 & 52.74\\
D & 2.04$\pm$0.001 & 0.49$\pm$0.001 & 8837.58 & 415533.94 & 1.97$\pm$0.02 & 0.56$\pm$0.12 & 0.47$\pm$0.01 & 8862.74 & 415521.70 & 24.48 \\
E & 2.05$\pm$0.001 & 1.27$\pm$0.007 & 23812.1 & 378575.73 & 1.97$\pm$0.02 & 0.83$\pm$0.09 & 1.20$\pm$0.02 & 23812.4 & 378520.98 & 109.50 \\
\hline
\end{tabular}}
\end{table*}

\subsection{Broadband SED modelling}
The source has shown varied flux variability behaviour over the last 15 years, during which we have detected instances of correlated optical and GeV variations as well as optical flare without $\gamma$-ray counterpart. To understand the diverse behaviour of the source during different epochs we constructed broadband spectral energy distributions and modelled them using one-zone leptonic emission model. To model the broadband SED, we used the publicly available code, JetSet \citep{2009A&A...501..879T, 2011ApJ...739...66T, 2020ascl.soft09001T}. JetSet (version 1.2.2) fits numerical models to the data in order to determine the optimal parameter values that best describe the observed data. 

The photometric observations taken in both UV and optical wavelengths during each epoch were averaged filter-wise to generate a single photometric data point for each epoch. On the other hand, the average spectra for X-rays and $\gamma$-rays were created using all of the data that was available for the 100-day span in each of the epochs. Since the EC process primarily accounts for the high-energy $\gamma$-ray emission in the FSRQs \citep{2015ApJ...803...15P, 2017MNRAS.470.3283S, 2019MNRAS.486.1781R, 2020MNRAS.498.5128R}, we used synchrotron, SSC, and EC processes to model our SEDs. In the one-zone leptonic emission model, the spherical blob of radius R, located at a distance of R$_{\rm {H}}$ form the central black hole. It is postulated that this spherical region is injected with non-thermally accelerated electrons that follow a distribution characterized by a broken power law. The broken power law distribution of electrons is defined as follows:
\begin{equation}
      N(\gamma)d\gamma =
      \begin{cases}
        K \gamma^{\rm {-p}} d\gamma, & \gamma_{\rm {min}} < \gamma < \gamma_{\rm {break}}\\
        K \gamma_{\rm {break}}^{\rm {p_{1}-p}} \gamma^{\rm {-p_{1}}} d\gamma, & \gamma_{\rm {break}} < \gamma < \gamma_{\rm {max}}
    \end{cases}       
\end{equation}  
Where, $\gamma$ is the electron Lorentz factor, N is the electron density in units of cm$^{\rm {-3}}$ and K is the normalization constant. p and p$_{1}$ are the low and high energy spectral slopes, and $\gamma_{\rm {min}}$, $\gamma_{\rm {max}}$ and $\gamma_{\rm {break}}$ are the Lorentz factors corresponding to the low-energy cut off, high-energy cut off and turn over energy, respectively. The emission region (spherical blob), which is moving down the jet with the bulk Lorentz factor $\Gamma$, is permeated with the magnetic field B. The relativistic electrons interact with the magnetic field and produce synchrotron radiation. The SSC emission is produced by the same synchrotron photons generated within the jet \citep{2009A&A...501..879T}. The seed photons for the EC process are external to the jet, which are produced (1) from the accretion disk \citep{2002ApJ...575..667D, 2009MNRAS.399.2041G}, (2) from broad line region \citep{2009ApJ...692...32D}, and (3) from dust torus \citep{2008ApJ...675...71S}. The external photons that directly come from the accretion disk and participate in the EC process have luminosity L$_{\rm {disk}}$ and temperature T$_{\rm {disk}}$. The reprocessed UV photons from the BLR, which is located at a distance of R$_{\rm {BLR}}$ from the black hole, and IR photons from the dust torus, which is located at a distance of R$_{\rm {DT}}$ with a temperature of T$_{\rm {DT}}$, are involved in the EC process. The fraction of L$_{\rm {disk}}$ reprocessed by the BLR and dust torus are given as $\tau_{\rm {BLR}}$ and $\tau_{\rm {DT}}$.

We have determined the size of the emission region R, which is defined as R $\leq$ cT$_{\rm {min}}$$\delta$/(1+z), where T$_{\rm {min}}$ is the minimum variability timescale and $\delta$ denotes the Doppler factor. We have calculated the minimum variability timescale for the $\gamma$-ray light curve using the methodology that models flares with an exponential profile, as outlined in \cite{1974ApJ...193...43B}:
\begin{equation}
    T_{\rm {min}} = \Delta t/ln(F_{2}/F_{1})
\end{equation}
Here, $\Delta$t = $\lvert$ t$_{2}$-t$_{1}$$\rvert$ and F$_{2}$ and F$_{1}$ are the flux values at a time t$_{2}$ and t$_{1}$, respectively.
The minimum variability timescale is calculated to be $\sim$ 7.23 hours. We have adopted the Doppler factor value for the source as $\delta$ = 18.2, which has been determined using the one-zone leptonic emission model by \cite{2017ApJ...851...33P}. Using the values of T$_{\rm {min}}$ and $\delta$, we have approximated the size of emission region to be 8.24 $\times$ 10$^{\rm {15}}$ cm $\sim$ 8 $\times$ 10$^{15}$ cm. Furthermore, assuming that this blob covers the whole cross-section of the jets, we have estimated the distance of this blob from the central black hole using the relation R$_{\rm {H}}$ = R/$\phi$, where $\phi$ represents the opening angle. The intrinsic opening angle for the Ton 599 is estimated as 0.58$^{\circ}$ ($\sim$ 0.01 radian) based on observations conducted with the 15.4 GHz Very Long Baseline Array (VLBA) from the 2 cm VLBA MOJAVE program \citep{2009A&A...507L..33P}. Utilizing the values of R and $\phi$, we estimate that the location of the emission region is located at roughly 8 $\times$ 10$^{17}$ cm away from the central black hole.

We have considered a value for L$_{\rm {disk}}$ equal to 4.5 $\times$ 10$^{45}$ erg/sec, which was determined through the application of a one-zone leptonic emission model in studies by \cite{2010MNRAS.402..497G, 2017ApJ...851...33P}. The inner radius of the spherical shell like BLR (R$_{\rm {BLR_{in}}}$) is considered to be 0.01 pc $\sim$ 3.1 $\times$ 10$^{16}$ cm \citep{2003APh....18..377D}. The outer radius of the BLR (R$_{\rm {BLR_{out}}}$) is estimated to be around 2.1 $\times$ 10$^{17}$ cm using the relationship R$_{\rm {BLR}}$ = 10$^{17}$ L$^{1/2}_{\rm {disk,45}}$ cm, where L$_{\rm {disk,45}}$ represents the luminosity of the accretion disk in units of 10$^{45}$ erg/sec. This estimation is based on the assumption that the BLR radius scales in relation to the square root of the luminosity of the accretion disk \citep{2008MNRAS.387.1669G}. Moreover, it is posited that the spherical shell like torus is situated at a distance R$_{\rm {DT}}$ = 2.5 $\times$ 10$^{18}$ L$^ {\rm {1/2}}_{\rm {disk,45}}$ cm \citep{2008MNRAS.387.1669G}, which is approximately equivalent to 5 $\times$ 10$^{18}$ cm. We have assumed that the fraction of the L$_{\rm {disk}}$ is reprocessed by the BLR ($\tau_{\rm {BLR}}$) is 0.1 and by the dust torus ($\tau_{\rm {DT}}$)is 0.5 \citep{2010MNRAS.402..497G}. Furthermore, the disk temperature T$_{\rm {disk}}$ and dust torus temperature T$_{\rm {DT}}$ are assumed to be at their typical values of 1.0 $\times$ 10$^5$ K \citep{2020MNRAS.492...72P} and 1.0 $\times$ 10$^3$ K \citep{2003APh....18..377D}, respectively, in our study. And the ratio of cold protons to relativistic electrons is taken as 0.1 in our SED analysis \citep{2012MNRAS.424L..26G}.

\begin{table}
\hspace{-1cm}
\caption{\label{tab:frozen_pars} Values of the frozen parameters for the one-zone leptonic model fits to the observed SEDs during the five epochs.}
\resizebox{\columnwidth}{!}{
\begin{tabular} {lc} \hline
name of the parameters & values\\ \hline
$\gamma_{\rm {min}}$ & 40\\
$\gamma_{\rm {break}}$ & 1.1 $\times$ 10$^{3}$\\
$\gamma_{\rm {max}}$ & 1.0 $\times$ 10$^{4}$\\
T$_{\rm {DT}}$ & 1000 K\\
R$_{\rm {DT}}$ & 5 $\times$ 10$^{18}$ cm\\
$\tau_{\rm {DT}}$ & 0.5\\
$\tau_{\rm {BLR}}$ & 0.1\\
R$_{\rm {BLR_{in}}}$ & 3.1 $\times$ 10$^{16}$ cm\\
R$_{\rm {BLR_{out}}}$ & 2.1 $\times$ 10$^{17}$ cm\\
L$_{\rm {disk}}$ & 4.5 $\times$ 10$^{45}$ erg/sec\\
T$_{\rm {disk}}$ & 1.0 $\times$ 10$^{5}$ K\\
R & 8.0 $\times$ 10$^{15}$ cm\\
R$_{\rm {H}}$ & 8.0 $\times$ 10$^{17}$ cm\\
cold proton to relativistic electron ratio & 0.1\\
\hline
\end{tabular}}
\end{table}

\begin{figure*}
\vspace{-1cm}
\hspace{2cm}$
\begin{array}{r}
\includegraphics[width=120mm,height=75mm]{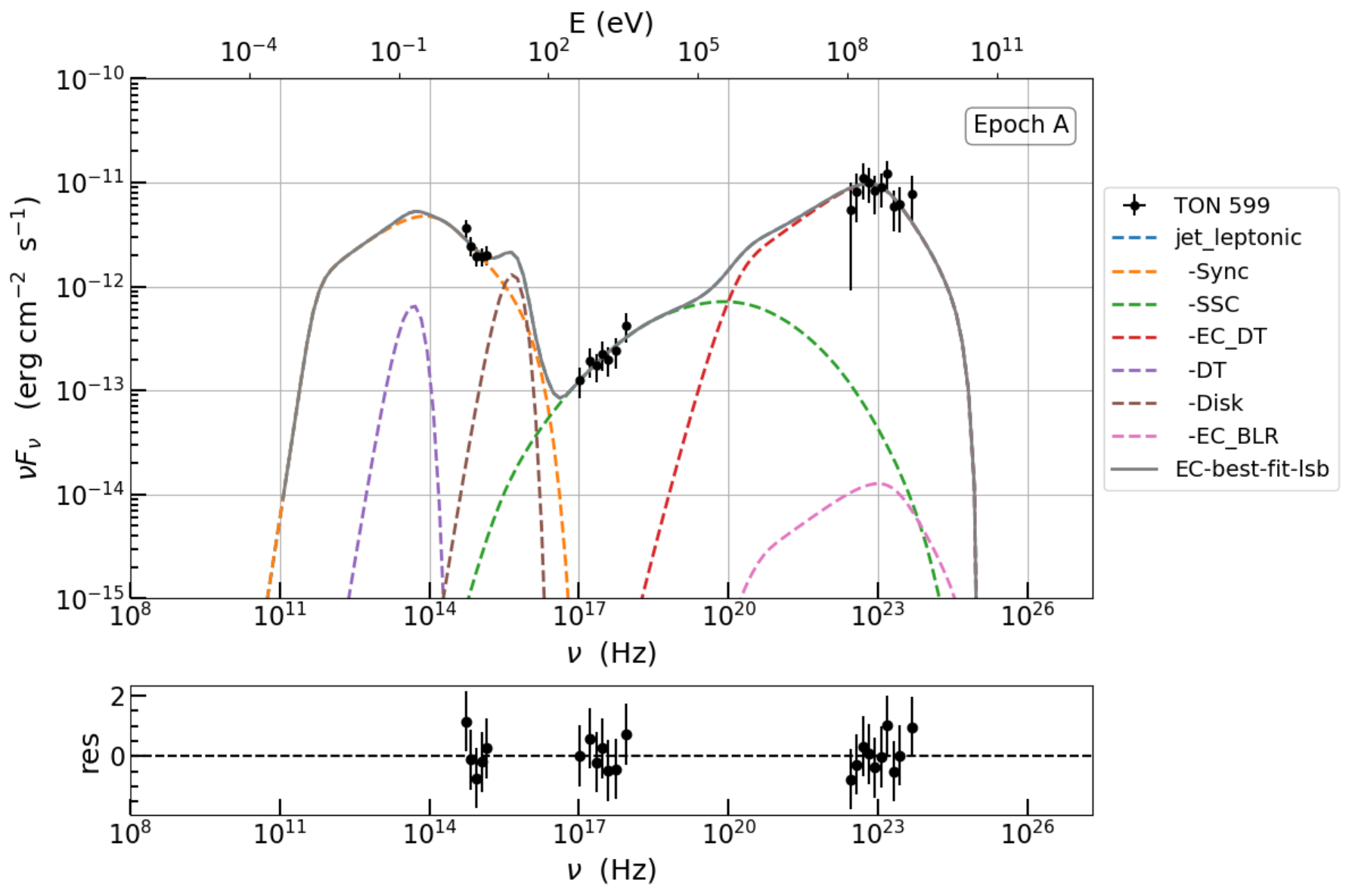}
\end{array}$

\begin{center}$
\begin{array}{rr}
\includegraphics[width=90mm,height=75mm]{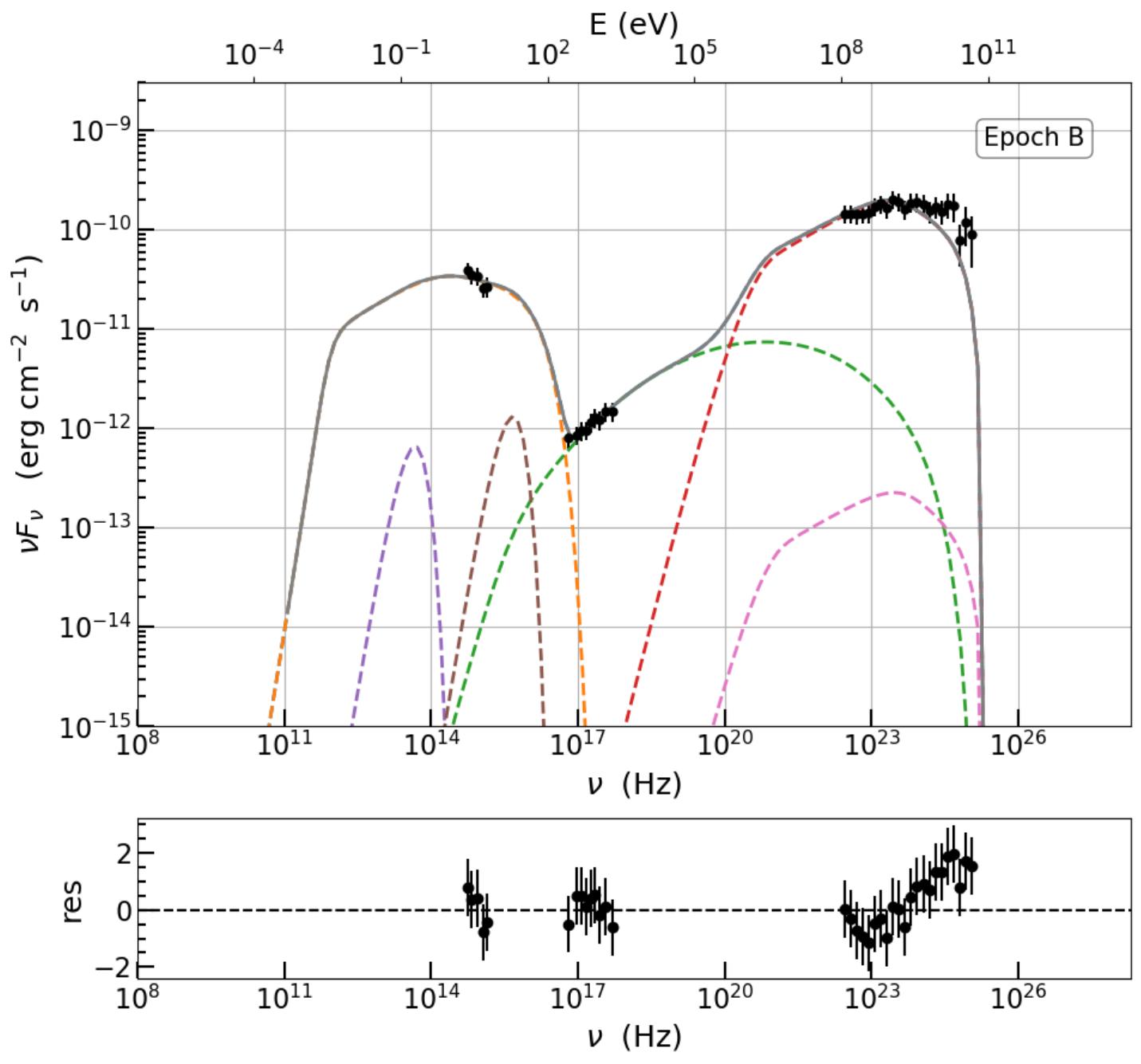}
\includegraphics[width=90mm,height=75mm]{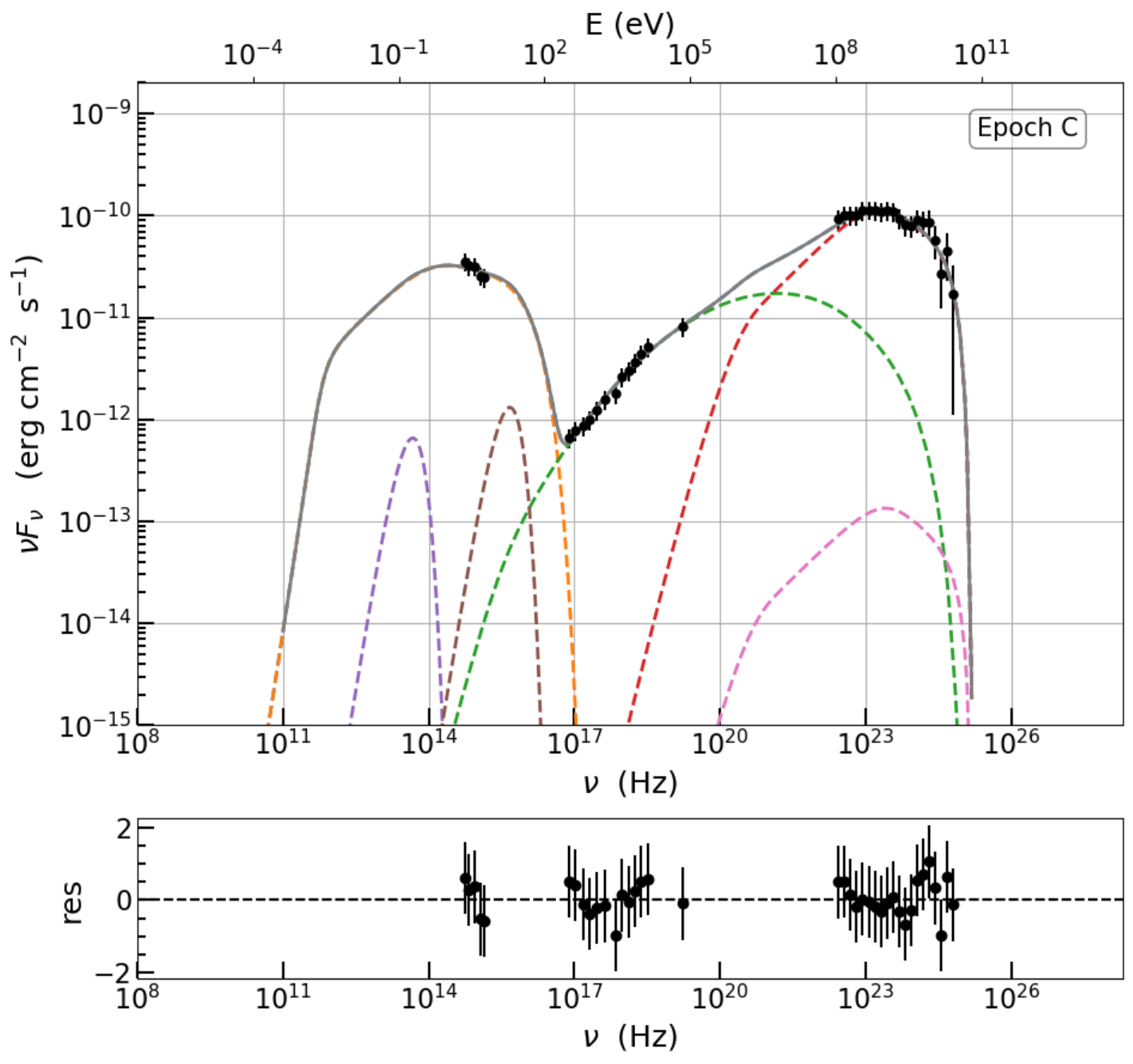}
\end{array}$
\end{center}
\begin{center}$
\begin{array}{rr}
\includegraphics[width=90mm,height=75mm]{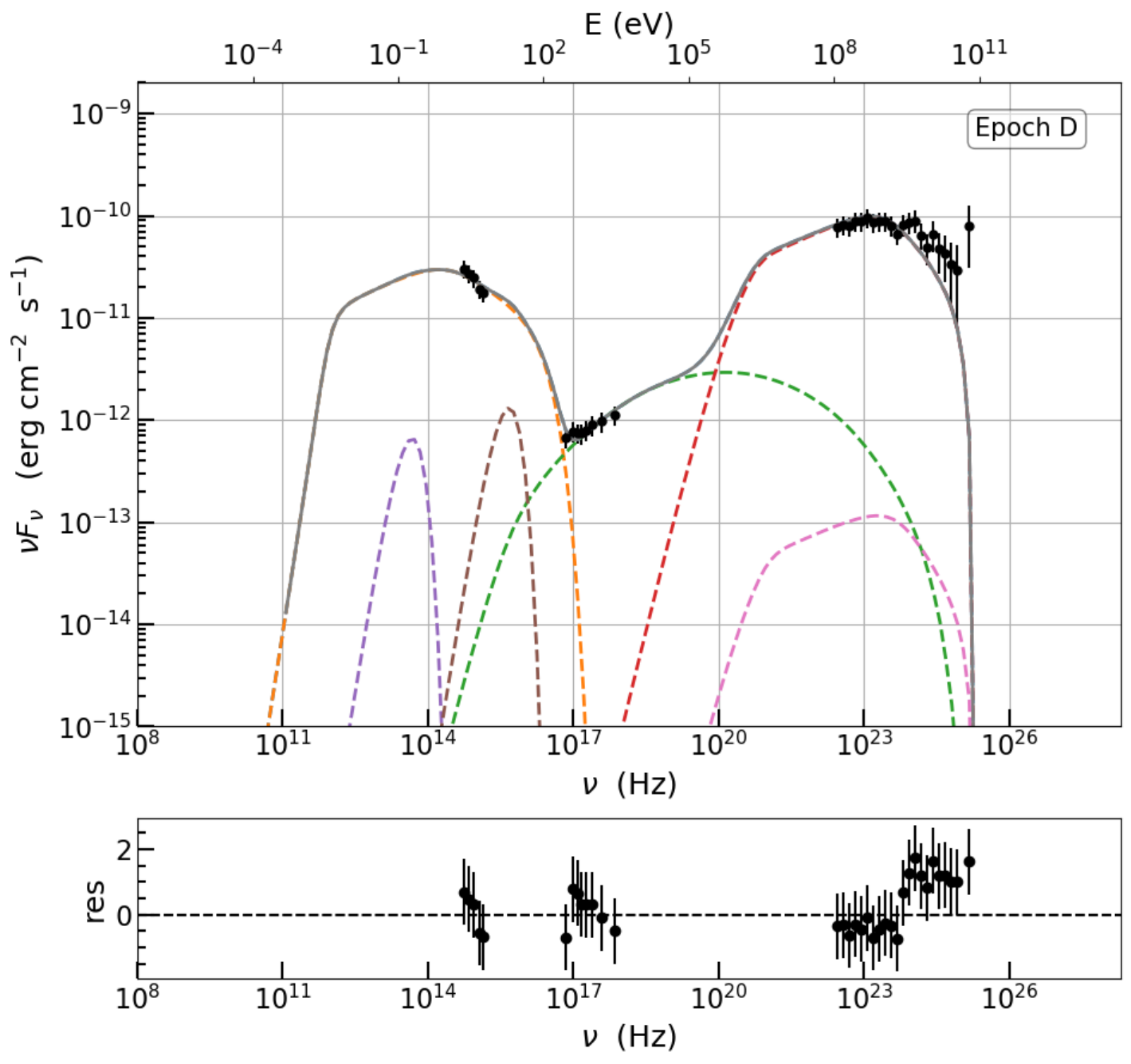}
\includegraphics[width=90mm,height=75mm]{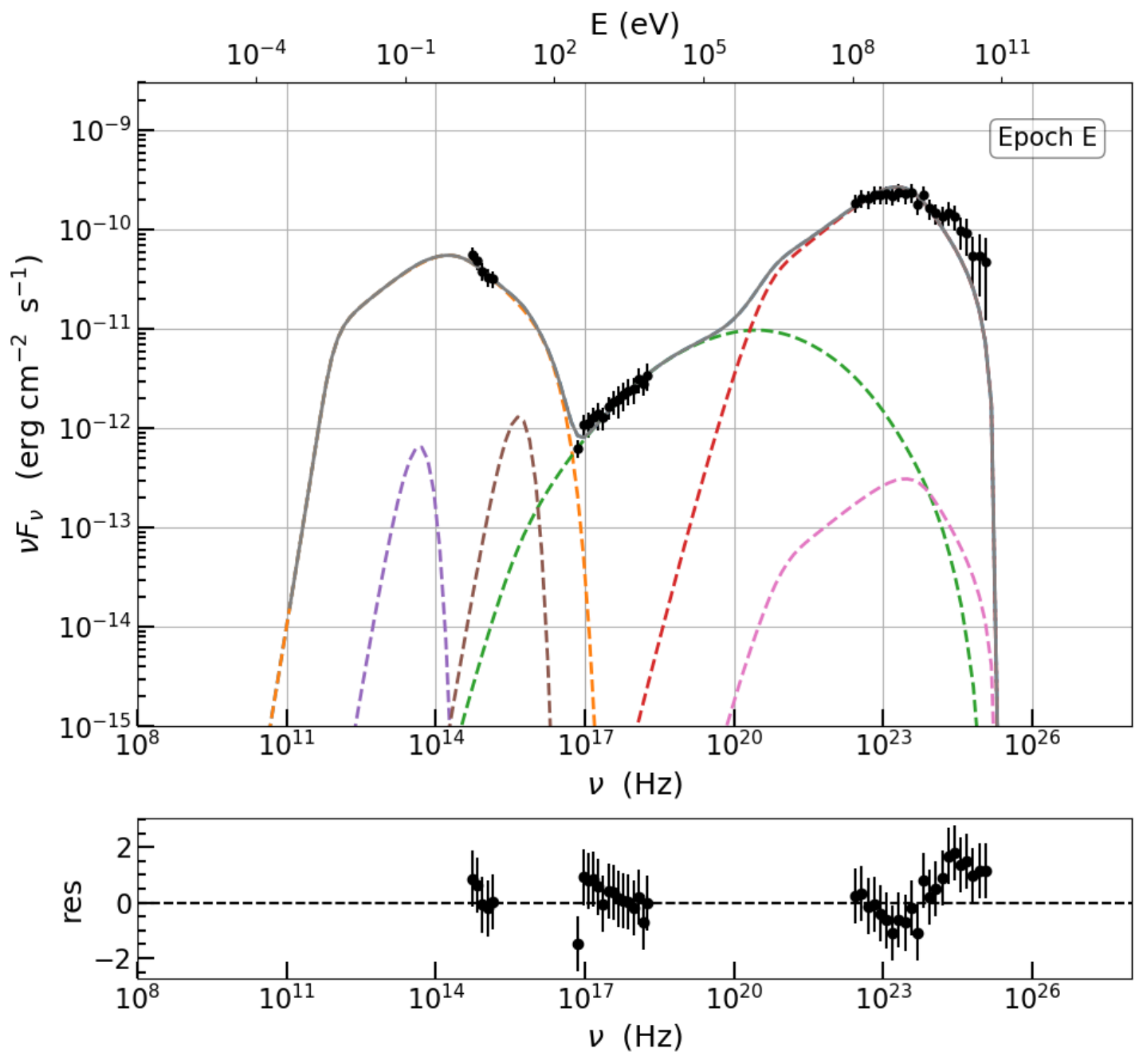}
\end{array}$
\end{center}
\caption{\label{fig:epoch_a_sed}Observed broad band SED along with the one zone leptonic emission model fit during epochs A, B, C, D and E (top panel) and the residuals (bottom panel). Here the solid line is the best fit to the SED, orange dashed represents the emission from synchrotron process, green dashed line represents the emission from SSC process, red dashed line represents the EC emission from dusty torus and pink dashed line represents the EC emission from BLR. Purple and brown dashed lines represent the thermal emissions from the dusty torus and accretion disk, respectively. The fit to the quiescent state is applied for determining the values of the parameters $\gamma_{\rm {min}}$, $\gamma_{\rm {max}}$, and $\gamma_{\rm {break}}$ in the subsequent epochs modelled in this study.}
\end{figure*}

In one-zone leptonic model, the characteristics of the observed SED are primarily determined by nineteen parameters. Five parameters, namely, p, p$_{1}$, N, B, and $\Gamma$, are chosen to be free, while the remaining fourteen parameters are chosen to be frozen. The purpose of parameter freezing is to compare the changes in certain parameters, such as the magnetic field, bulk Lorentz factor, and particle density, etc., between quiescent and flaring epochs. Details of frozen parameters are given in Table \ref{tab:frozen_pars}. 
In order to investigate the distinct correlation behavior between optical and $\gamma$-ray flux variations during different flaring epochs, we initially fitted the quiescent epoch to acquire the values for the parameters $\gamma_{\rm {min}}$, $\gamma_{\rm {max}}$ and $\gamma_{\rm {break}}$. The fitting process using the same frozen parameters and the previously acquired values of $\gamma_{\rm {min}}$, $\gamma_{\rm {max}}$, and $\gamma_{\rm {break}}$ from the fitting during the quiescent epoch, was repeated for the other epochs B, C, D, and E with the selection of free parameters. Our SED analysis shows that the leptonic scenario provides a good fit to the observed broadband SED across all the epochs. The resultant best fit parameters for the selected epochs are given in Table \ref{tab:res_sed} and the fitted SEDs are shown in Figures \ref{fig:epoch_a_sed}.

\begin{table*}
\caption{\label{tab:res_sed} Results of the broadband SED fitting during the epochs A, B, C, D, and E. }
\begin{center}
\begin{tabular} {lccccc} \hline
Parameters & Epoch A & Epoch B & Epoch C & Epoch D & Epoch E \\ \hline
p & 2.29$\pm$0.31 & 2.40$\pm$0.11 & 2.04$\pm$0.06 & 2.55$\pm$0.11 & 2.17$\pm$0.13 \\
p$_{1}$ & 4.03$\pm$0.37 & 3.20$\pm$0.15 & 3.18$\pm$0.08 & 3.60$\pm$0.19 & 3.85$\pm$0.21\\ 
N ($\times$ 10$^{3}$ cm$^{-3})$ & 0.98$\pm$0.20 & 1.19$\pm$0.15 & 1.31$\pm$0.09 & 0.89$\pm$0.11 & 0.84$\pm$0.11\\
B (Gauss) & 1.63$\pm$0.29 & 1.51$\pm$0.10 & 1.51$\pm$0.07 & 2.05$\pm$0.15 & 1.73$\pm$0.14\\
Bulk Lorentz factor & 22.48$\pm$2.78 & 37.31$\pm$1.75 & 28.64$\pm$0.83 & 37.37$\pm$1.87 & 38.54$\pm$2.01\\
\hline
\end{tabular}
\end{center}
\end{table*}

\section{Discussion}\label{sec:discussion}
\subsection{Connection between optical and $\gamma$-ray flux variations}
\begin{figure}
\hspace{-1.5cm}
\includegraphics[width=110mm,height=100mm]{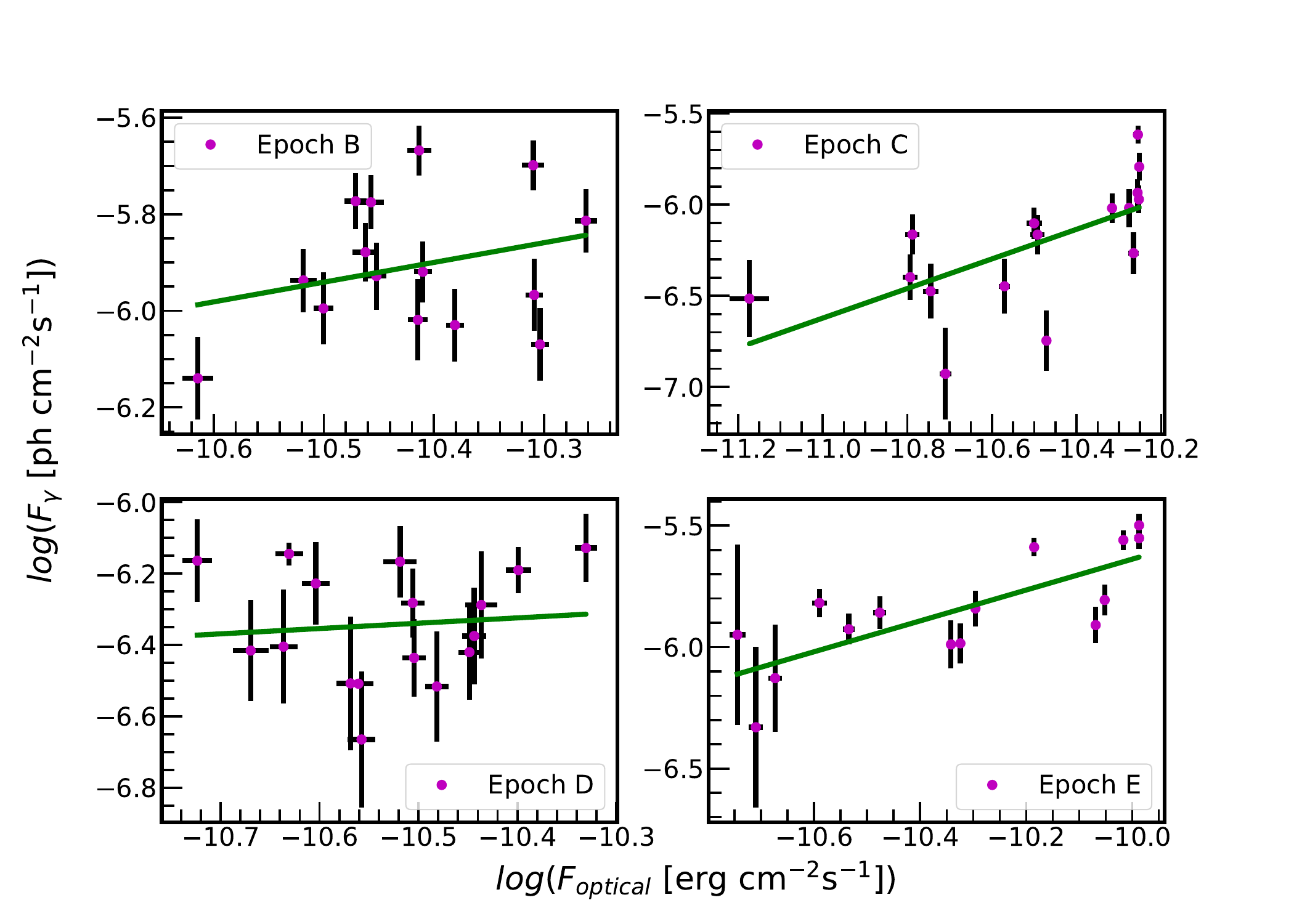}

\caption{$\gamma$-ray v/s Optical V-band fluxes during epochs B, C, D, and E.} 
\label{figure-correlation}
\end{figure}

Over a span of around 15 years of study, Ton 599 has shown varied correlated behavior between optical V-band flux and $\gamma$-ray flux variations. We found two cases, where, (1) both the optical flux and the $\gamma$-ray flux showed flaring behavior, (2) an increase in optical flux without a corresponding increase in $\gamma$-ray flux. We chose five epochs A, B, C, D, and E based on the distinct behaviour between optical V-band and $\gamma$-ray flux variations. During epoch A, the source was in its quiescent state in each waveband. During epochs B, C and E, there was a structural correlation between the variations in optical and $\gamma$-ray flux. However, during epoch D, the source exhibited an optical flare without a corresponding $\gamma$-ray response. Our investigation involving linear regression analysis on the logarithmic values of $\gamma$-ray flux and optical flux suggests a similar trend. The linear least square fits are shown in Figure \ref{figure-correlation}, depicting the logarithm of $\gamma$-ray flux plotted against the logarithm of optical flux. The outcomes of the linear regression analysis are given in Table \ref{table-correlation}. Moreover, we employed the sophisticated Discrete Cross Correlation Function to cross-correlate the optical V-band and $\gamma$-ray light curves, aiming to establish whether there is a correlation between variations in the optical flux and $\gamma$-ray flux. We found that there are weakly significant correlations between the optical and $\gamma$-ray flux variations. Given the observed structural similarities between these two energy bands, this low significance may result from the relatively low amount of optical data used. 

A time lag of a few days within the period of October-December 2017, falling within the timeframe of epoch B considered here, was previously identified by \cite{2019ApJ...871..101P}. In the cross-correlation study of bright blazars, \cite{2018MNRAS.480.5517L, 2019ApJ...880...32L} found  that the time lag between optical and $\gamma$-ray bands are generally small. Moreover, \cite{2023MNRAS.519.6349D} found that the time lag distribution between optical and $\gamma$-ray bands is consistent with zero lags. These results indicate that the optical and $\gamma$-ray flux emission regions are co-spatial and favour leptonic emission model of blazars, where the same relativistic electron population is accountable for both the optical and $\gamma$-ray emission. Given the scenario that both optical and $\gamma$-ray emissions are proportionate to the number of emitting electrons, it is plausible that variations in the electron population of the jet lead to the correlation between these two types of flux variations, as proposed by \cite{2011A&A...529A.145D}.

\begin{table}
\caption{The results obtained by applying linear regression on the logarithm of $\gamma$-ray flux and optical flux during epochs B, C, D, and E.}
\resizebox{\columnwidth}{!}{
\begin{tabular} {lcccr} \hline
Epoch & Intercept & Slope & Correlation coefficient & p-value \\ \hline
B & -1.65$\pm$4.05 & 0.41$\pm$0.39 & 0.28 & 0.31  \\
C & 2.31$\pm$2.76 & 0.81$\pm$0.26 & 0.64 & 0.008   \\
D & -4.77$\pm$4.17 & 0.15$\pm$0.40 & 0.10 & 0.71  \\
E & 0.71$\pm$1.53 & 0.64$\pm$0.15 & 0.77 & 0.001 \\
\hline
\end{tabular}}
\label{table-correlation}
\end{table}

\subsection{$\gamma$-ray spectra}
 The $\gamma$-ray spectral shape of blazars can yield valuable information about the inherent characteristics of the jets such as the distribution of emitting particles. In particular, these $\gamma$-ray spectra can help us determine if there is any absorption of $\gamma$-rays within the emission region and can also reveal whether $\gamma$-ray emissions are attributed to a single component of the jet or multiple components. The high-energy $\gamma$-ray spectra of FSRQ sources are better explained by the broken power law or log parabola model \citep{2010ApJ...710.1271A, 2015ApJ...803...15P, 2019MNRAS.486.1781R, 2020A&A...635A..25S, 2020MNRAS.498.5128R, 2021Galax...9..118R}. The curvature observed in the $\gamma$-ray spectra of FSRQs may be attributed to external factors, such as the absorption of $\gamma$-rays through photon-photon pair production within the BLR \citep{2010ApJ...717L.118P}, and the impact of the Klein-Nishina effect on the inverse Compton scattering of BLR photons by relativistic jet electrons with a curved distribution \citep{2013ApJ...771L...4C}. However, both the effects account for the IC scattering of the BLR photons, whereas in the large number of FSRQs the $\gamma$-ray emission site lies outside the BLR and $\gamma$-ray emission is not produced by the IC scattering of the BLR photons \citep{2018MNRAS.477.4749C}. Our SED analysis also indicates that the $\gamma$-ray emission is produced by the IC scattering of the IR photons from the dusty torus.

\begin{figure*}
\begin{center}$
\begin{array}{rrr}
\includegraphics[width=80mm,height=75mm]{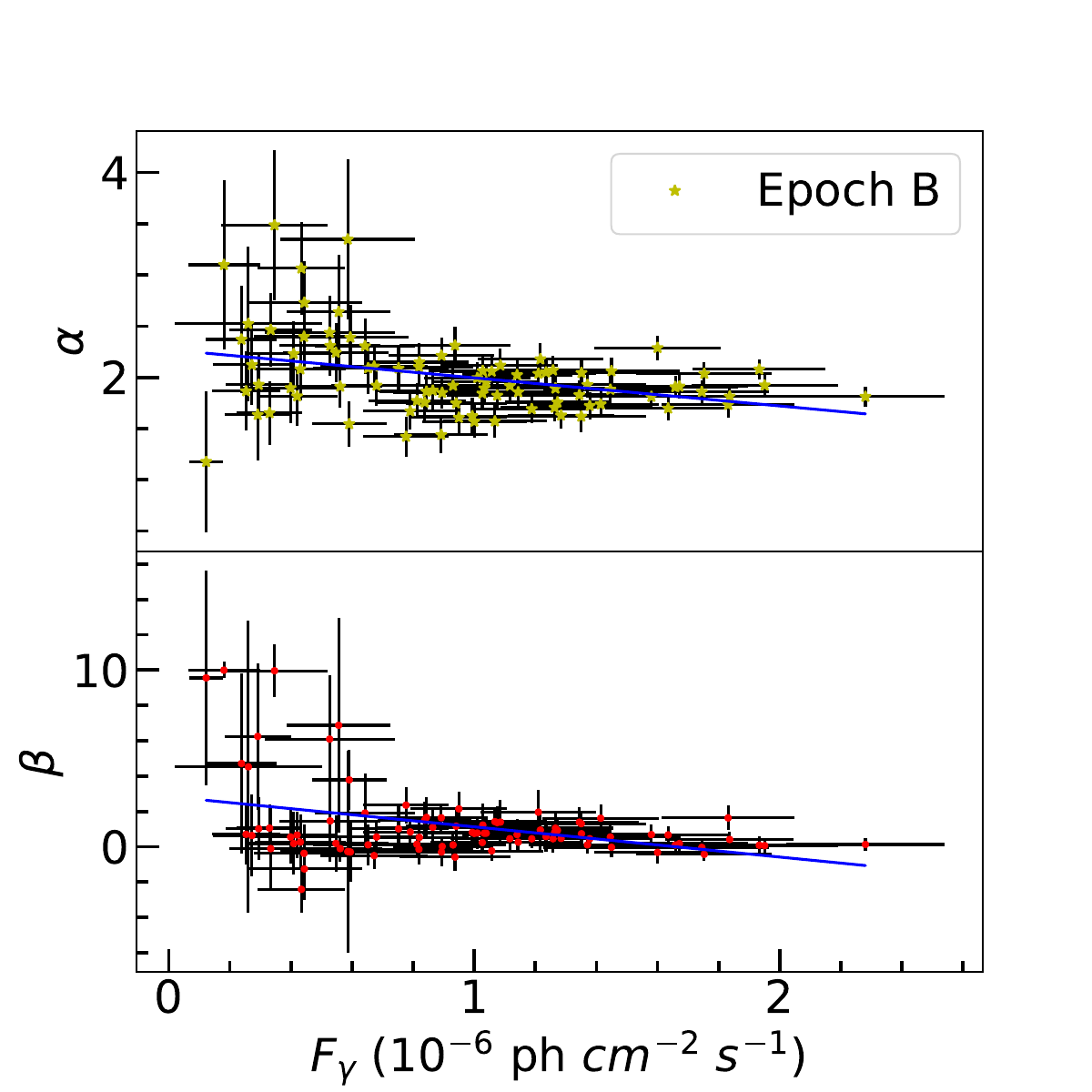}
\includegraphics[width=80mm,height=75mm]{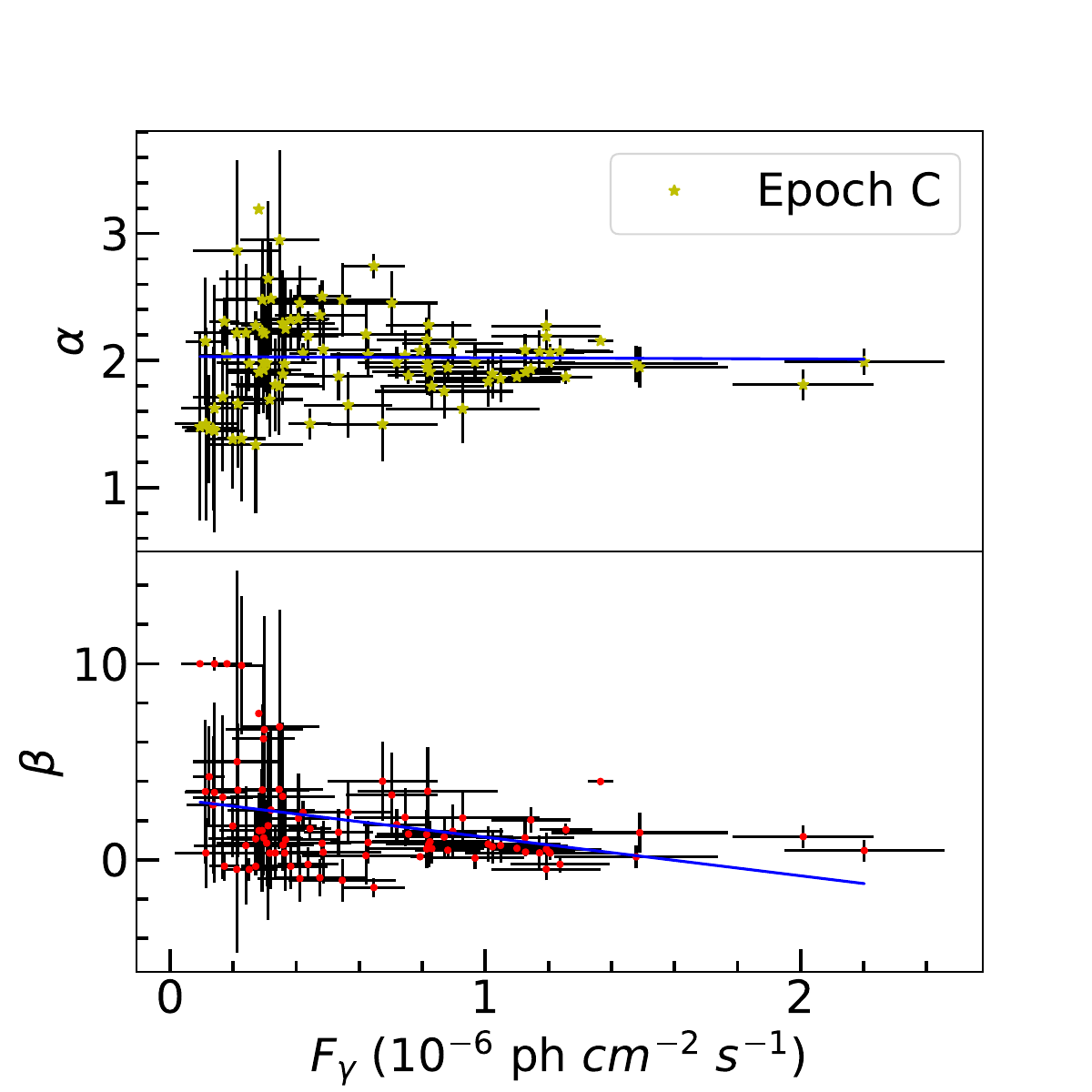}
\end{array}$
\end{center}
\begin{center}$
\begin{array}{rr}
\includegraphics[width=80mm,height=75mm]{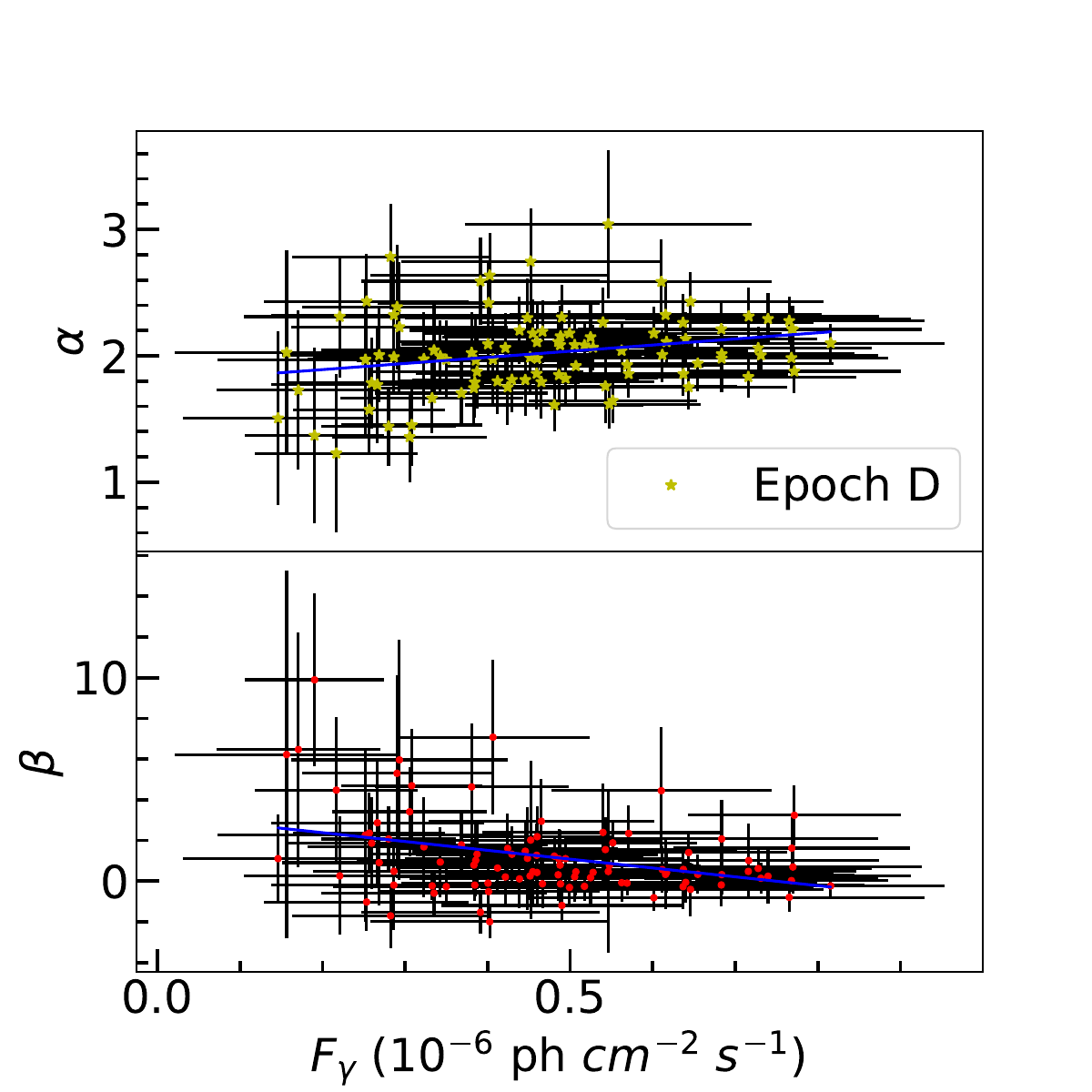}
\includegraphics[width=80mm,height=75mm]{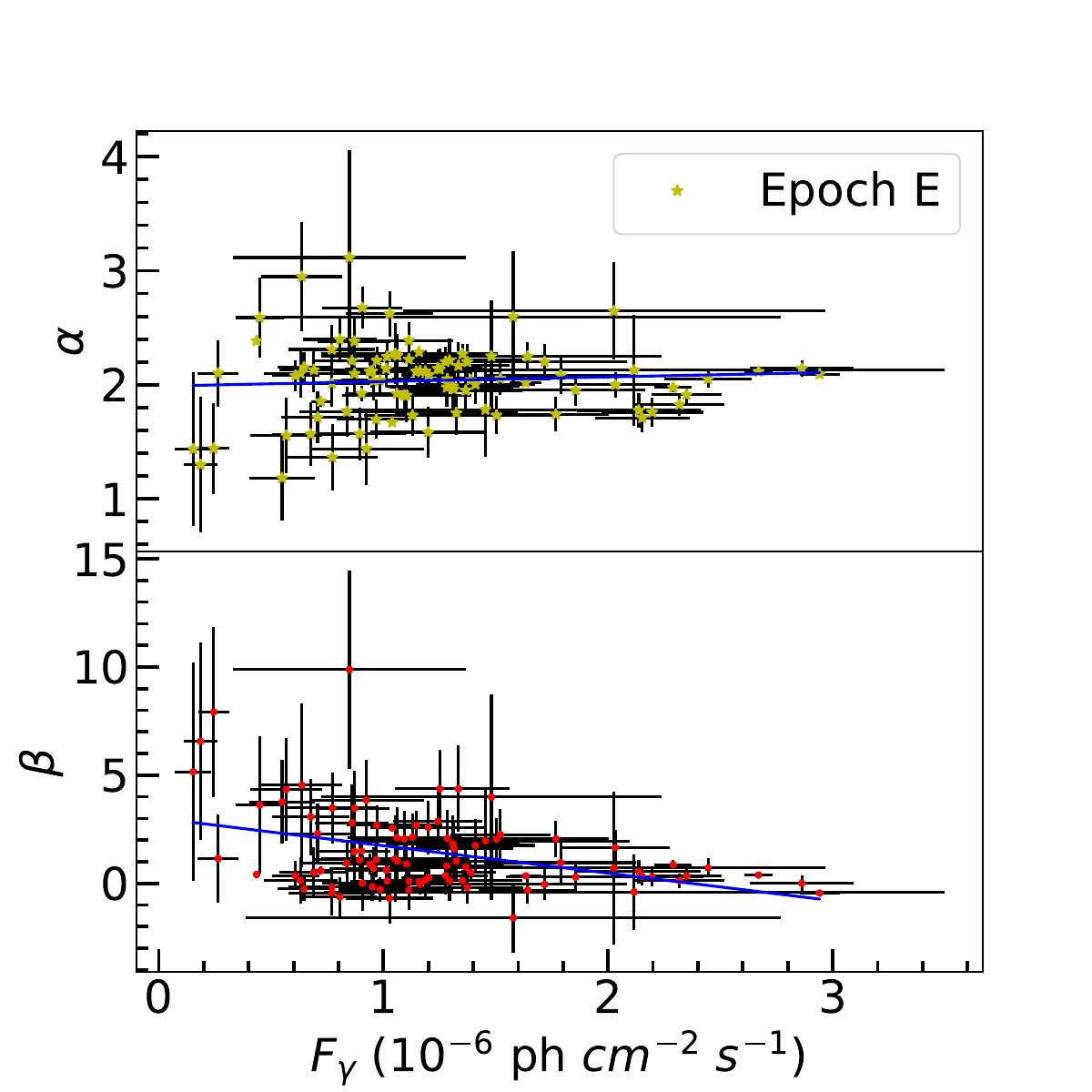}
\end{array}$
\end{center}
\caption{Correlation between the parameters $\alpha$ and $\beta$ and the $\gamma$-ray flux. The blue line represents the least-square fit applied to the data.}
\label{figure-alpha}
\end{figure*}

In our work, the LP model fits the $\gamma$-ray spectra in the flaring epochs (and orphan optical flare epoch) adequately, whereas the PL model fits the $\gamma$-ray spectra in the quiescent epoch well. The curvature observed in the $\gamma$-ray spectra is probably caused by the inherent characteristics of the electron distribution that produces emission. These electrons follow either a cutoff energy distribution or a log parabola energy distribution. However, during quiescent epoch the low quality of photons data available for analysis makes it challenging for us to identify this curvature. In the LP model fits to $\gamma$-ray spectra, the parameters $\alpha$ and $\beta$ convey crucial information about the properties of the $\gamma$-ray spectra. Changes in the $\alpha$ and $\beta$ values are reflected in the shape of $\gamma$-ray spectra. We looked for the variations in the $\alpha$ and $\beta$ parameters against $\gamma$-ray flux during epochs B, C, D and E, which are shown in the Figure \ref{figure-alpha}. We conducted an analysis to find a correlation between the $\alpha$ and $\beta$ parameters and the $\gamma$-ray flux, employing the Spearman rank test. We considered the correlation is significant if the Spearman rank correlation coefficient $>$ 0.5 or $<$ $-$0.5, along with a p-value $<$ 0.05. We observed that the parameter $\alpha$ showed no significant pattern, and the curvature parameter $\beta$ exhibited a mild negative correlation with $\gamma$-ray flux throughout all epochs. The results of Spearman rank test are given in Table \ref{table-curvature_parameters}. Decreasing trend of $\beta$ with increasing flux was also noticed in the study of 3LAC (The Third Catalog of Active Galactic Nuclei Detected by the {\it Fermi}) sources by \cite{2015ApJ...810...14A}. According to \cite{2016MNRAS.458..354C}, the variations in the location of the $\gamma$-ray emission area during various activity levels of the sources may be responsible for these distinct spectral behaviors of the $\gamma$-ray spectra. Such varied spectral behavior could also be caused by a change in the IC peak frequency, where the external photon field is responsible for producing the $\gamma$-ray emission in FSRQs \citep{2020MNRAS.498.5128R}. 

\begin{table}
\caption{Results of the Spearman rank test for the correlation of $\alpha$ and $\beta$ parameters with $\gamma$-ray flux. The plots of $\alpha$ and $\beta$ parameters v/s $\gamma$-ray flux are shown in Figure \ref{figure-alpha}.}
\begin{tabular} {lcccr} \hline
 & \multicolumn{2}{c}{$\alpha$}  & \multicolumn{2}{c}{$\beta$} \\
Epoch & R & p-value & R & p-value \\ \hline
B & -0.33 & 0.002 & -0.21 & 0.04  \\
C & 0.06 & 0.58 & -0.34 & 0.001   \\
D & 0.26 & 0.01 & -0.27 & 0.007  \\
E & 0.02 & 0.84 & -0.30 & 0.003 \\
\hline
\end{tabular}
\label{table-curvature_parameters}
\end{table}

\subsection{Broadband spectral energy distribution}
In our study, we found instances of both optical and $\gamma$-ray flares as well as optical flare without its corresponding $\gamma$-ray counterpart. \cite{2019ApJ...880...32L} investigated for a correlation between optical and GeV flux changes in 178 blazars and found that in nearly 50\% of the blazars, optical flares occurred without their $\gamma$-ray counterparts. A Few blazars with complex behavior between changes in optical and $\gamma$-ray flux were also studied by \cite{2013ApJ...763L..11C, 2014ApJ...797..137C, 2019MNRAS.486.1781R, 2020MNRAS.498.5128R}. Hadronic processes could provide a possible explanation for optical flux variations without $\gamma$-ray counterparts or vice versa \citep{2001APh....15..121M}. Furthermore, the leptonic scenario of blazars favours a close correlation between changes in the optical and GeV flux. However, results from \cite{2019MNRAS.486.1781R, 2020MNRAS.498.5128R} indicate that different correlation behaviour between optical and $\gamma$-ray flux changes are well explained by one-zone leptonic model scenario.

In this work, we performed broadband SED modeling using a one-zone leptonic emission model to model the five epochs considered here. The leptonic model provides a good fit during all the epochs with the seed photons for the EC process from the torus. The source was in a quiescent condition during epoch A, therefore we first fitted the broadband SED for epoch A and then looked for changes in parameters in other epochs relative to the quiescent epoch. The accretion disk component is clearly evident in the broadband SED during the quiescent phase, but synchrotron radiation dominates the optical-UV emission during epochs B, C, D, and E. The source showed both optical and $\gamma$-ray flares during epochs B, C and E. During epochs B and E, the bulk Lorentz factor was roughly 1.5 times higher than the quiescent epoch A. Furthermore, there was an increase in electron density and a decrease in the magnetic field during period B. In contrast, during period E, there was a decrease in electron density and an increase in the magnetic field. The enhanced optical and $\gamma$-ray fluxes in these two time intervals may be attributed to the increase bulk Lorentz factor and changes in electron density and magnetic field. In comparison to the quiescent period, the bulk Lorentz factor changed slightly during epoch C, and there was also decrease in the magnetic field. However, the electron density increased during epoch C, which might be accountable for the optical and $\gamma$-ray flares during this epoch.

During Epoch D, when the source showed an orphan optical flare, the bulk Lorentz factor increased almost 1.5 times, while the electron density decreased when compared to Epoch A. Additionally, the magnetic field was at its peak during this epoch. Consequently, the orphan optical flare can be the result of the increased magnetic field and bulk Lorentz factor during epoch D. According to \cite{1985ApJ...298..114M} the change in magnetic field can affect the synchrotron emission in the jet, however the $\gamma$-ray emission will be unperturbed by the change in the magnetic field. The presence of increased optical polarization measurements and X-ray flare can support this argument. A significant X-ray flare is observed during this time (see the Figure \ref{figure-lightcurve}), which is dominated by SSC process (the same photons emitted by the synchrotron process are involved in the EC process), however, the optical polarization measurements are not available at this time. From SED analysis, we draw the conclusion that the leptonic model explains the optical-GeV variations during the various epochs in Ton 599.

\section{Summary}\label{sec:summary}
We investigated the correlation between the changes in optical V-band and GeV flux in the FSRQ Ton 599 for the time period MJD 54686–60008. Our study includes 1) the identification of the epochs exhibiting optical or $\gamma$-ray flare, 2) cross-correlation analysis between optical flux and $\gamma$-ray flux variations, 3) analysis of optical spectral variations, 4) analysis of $\gamma$-ray spectra 5) broadband SED modeling with a one-zone leptonic emission model for the chosen epochs. We summarize the results below
\begin{enumerate}
\item In the period of around 15 years, we selected five epochs A, B, C, D, and E. The source was in its quiescent state during epoch A. Our visual inspections indicated that, in epochs B, C, and E, the source showed variations in both optical and $\gamma$-ray emissions, showing structural similarities. However, during epoch, D the source showed an optical flare, without a clear $\gamma$-ray counterpart. The source was in its highest flaring state during epoch E.
\item Through cross-correlation analysis, we identified a weak linear correlation between optical and $\gamma$-ray light curves. 
\item We found no optical spectral variations during the epochs.
\item During the quiescent period, the $\gamma$-ray spectra of the source was well described by the PL model, while during the flaring periods, the $\gamma$-ray spectra was best described by the LP model.
\item During all the epochs, from SED modelling, we found the $\gamma$-ray emission to be produced by EC process, where the source of seed photons is the torus. Our SED analysis showed that during the quiescent period, the optical-UV emission was caused by the accretion disk component, but during the flaring periods, the synchrotron emission was the cause for the optical-UV emission. The varied behaviour observed in our work between the changes in optical and GeV flux, could be explained by changes in the magnetic field, the bulk Lorentz factor, and the electron density.
\end{enumerate}

\section*{Acknowledgements}
We are thankful to the anonymous referee for providing valuable feedback, which helped to improve the manuscript. B.R. gratefully acknowledges support by ANID BASAL project FB210003 and FONDECYT Postdoctorado 3230631. W.M. gratefully acknowledges support by the ANID BASAL project FB210003. Part of this work was supported by the Polish Funding Agency National Science Centre, project 2017/26/A/ST9/00756 (MAESTRO 9). This research has made use of data from {\it Fermi} Science Support Center, which was made available by NASA's Goddard Space Flight Center (GSFC). The High Energy Astrophysics Science Archive Research Center (HEASARC), a service of GSFC, is used to obtain the data, software, and online tools. This study used {\it NuSTAR} data as well as {\it NuSTAR} Data Analysis Software ({\tt NuSTARDAS}), which was collaboratively developed by the California Institute of Technology (Caltech, USA) and the ASI Science Data Center (ASDC, Italy). We are thankful for photometry program from Paul Smiths monitoring program at the Steward Observatory, which is supported by {\it Fermi} Guest Investigator grants NNX08AW56G, NNX09AU10G, NNX12AO93G, and NNX15AU81G. This study utilized ASAS-SN data, which was partially funded by the Alfred P. Sloan Foundation grant G-2021-14192 and Gordon and Betty Moore Foundation grants GBMF5490 and GBMF10501 to the Ohio State University. This research utilized the publicly accessible numerical code JetSet\footnote{https://github.com/andreatramacere/jetset}, developed by Tramacera A.

\section*{Data Availability}
 The multi-wavelength data utilized in this study are publicly accessible from the {\it Fermi-LAT}\footnote{https://fermi.gsfc.nasa.gov/ssc/data/access/lat/}, {\it Swift}-XRT\footnote{https://heasarc.gsfc.nasa.gov/docs/archive.html}, {\it NuSTAR}\footnote{http://heasarc.gsfc.nasa.gov/docs/nustar/analysis/}, {\it Swift}-UVOT\footnote{https://www.ssdc.asi.it/cgi-bin/swiftuvarchint}, {\it Steward observatory}\footnote{http://james.as.arizona.edu/$\sim$psmith/Fermi/} and {\it ASAS-SN}\footnote{https://asas-sn.osu.edu/}.



\bibliographystyle{mnras}
\bibliography{example} 

\begin{thebibliography}{}
\makeatletter
\relax
\def\mn@urlcharsother{\let\do\@makeother \do\$\do\&\do\#\do\^\do\_\do\%\do\~}
\def\mn@doi{\begingroup\mn@urlcharsother \@ifnextchar [ {\mn@doi@}
  {\mn@doi@[]}}
\def\mn@doi@[#1]#2{\def\@tempa{#1}\ifx\@tempa\@empty \href
  {http://dx.doi.org/#2} {doi:#2}\else \href {http://dx.doi.org/#2} {#1}\fi
  \endgroup}
\def\mn@eprint#1#2{\mn@eprint@#1:#2::\@nil}
\def\mn@eprint@arXiv#1{\href {http://arxiv.org/abs/#1} {{\tt arXiv:#1}}}
\def\mn@eprint@dblp#1{\href {http://dblp.uni-trier.de/rec/bibtex/#1.xml}
  {dblp:#1}}
\def\mn@eprint@#1:#2:#3:#4\@nil{\def\@tempa {#1}\def\@tempb {#2}\def\@tempc
  {#3}\ifx \@tempc \@empty \let \@tempc \@tempb \let \@tempb \@tempa \fi \ifx
  \@tempb \@empty \def\@tempb {arXiv}\fi \@ifundefined
  {mn@eprint@\@tempb}{\@tempb:\@tempc}{\expandafter \expandafter \csname
  mn@eprint@\@tempb\endcsname \expandafter{\@tempc}}}

\bibitem[\protect\citeauthoryear{{Abdo} et~al.,}{{Abdo}
  et~al.}{2010a}]{2010ApJ...710.1271A}
{Abdo} A.~A.,  et~al., 2010a, \mn@doi [\apj] {10.1088/0004-637X/710/2/1271},
  \href {https://ui.adsabs.harvard.edu/abs/2010ApJ...710.1271A} {710, 1271}

\bibitem[\protect\citeauthoryear{{Abdo} et~al.,}{{Abdo}
  et~al.}{2010b}]{2010ApJ...715..429A}
{Abdo} A.~A.,  et~al., 2010b, \mn@doi [\apj] {10.1088/0004-637X/715/1/429},
  \href {https://ui.adsabs.harvard.edu/abs/2010ApJ...715..429A} {715, 429}

\bibitem[\protect\citeauthoryear{{Abdo} et~al.,}{{Abdo}
  et~al.}{2010c}]{2010ApJ...716...30A}
{Abdo} A.~A.,  et~al., 2010c, \mn@doi [\apj] {10.1088/0004-637X/716/1/30},
  \href {http://adsabs.harvard.edu/abs/2010ApJ...716...30A} {716, 30}

\bibitem[\protect\citeauthoryear{{Abdollahi} et~al.,}{{Abdollahi}
  et~al.}{2020}]{2020ApJS..247...33A}
{Abdollahi} S.,  et~al., 2020, \mn@doi [\apjs] {10.3847/1538-4365/ab6bcb},
  \href {https://ui.adsabs.harvard.edu/abs/2020ApJS..247...33A} {247, 33}

\bibitem[\protect\citeauthoryear{{Abdollahi} et~al.,}{{Abdollahi}
  et~al.}{2022}]{2022ApJS..260...53A}
{Abdollahi} S.,  et~al., 2022, \mn@doi [\apjs] {10.3847/1538-4365/ac6751},
  \href {https://ui.adsabs.harvard.edu/abs/2022ApJS..260...53A} {260, 53}

\bibitem[\protect\citeauthoryear{{Ackermann} et~al.,}{{Ackermann}
  et~al.}{2015}]{2015ApJ...810...14A}
{Ackermann} M.,  et~al., 2015, \mn@doi [\apj] {10.1088/0004-637X/810/1/14},
  \href {https://ui.adsabs.harvard.edu/abs/2015ApJ...810...14A} {810, 14}

\bibitem[\protect\citeauthoryear{{Acosta Pulido}, {Redondo}, {Carnerero},
  {Raiteri}, {Villata}, {Larionov}, {Kopatskaya}  \& {GASP-WEBT
  Collaboration}}{{Acosta Pulido} et~al.}{2017}]{2017ATel10949....1A}
{Acosta Pulido} J.~A.,  {Redondo} J.,  {Carnerero} M.~I.,  {Raiteri} C.~M.,
  {Villata} M.,  {Larionov} V.~M.,  {Kopatskaya} E.~N.,   {GASP-WEBT
  Collaboration} 2017, The Astronomer's Telegram, \href
  {https://ui.adsabs.harvard.edu/abs/2017ATel10949....1A} {10949, 1}

\bibitem[\protect\citeauthoryear{{Aharonian}}{{Aharonian}}{2000}]{2000NewA....5..377A}
{Aharonian} F.~A.,  2000, \mn@doi [\na] {10.1016/S1384-1076(00)00039-7}, \href
  {https://ui.adsabs.harvard.edu/abs/2000NewA....5..377A} {5, 377}

\bibitem[\protect\citeauthoryear{{Antonucci}}{{Antonucci}}{1993}]{1993ARA&A..31..473A}
{Antonucci} R.,  1993, \mn@doi [\araa] {10.1146/annurev.aa.31.090193.002353},
  \href {https://ui.adsabs.harvard.edu/abs/1993ARA&A..31..473A} {31, 473}

\bibitem[\protect\citeauthoryear{{Arnaud}}{{Arnaud}}{1996}]{1996ASPC..101...17A}
{Arnaud} K.~A.,  1996, in {Jacoby} G.~H.,  {Barnes} J.,  eds,  Astronomical
  Society of the Pacific Conference Series Vol. 101, Astronomical Data Analysis
  Software and Systems V. p.~17

\bibitem[\protect\citeauthoryear{{Atwood} et~al.,}{{Atwood}
  et~al.}{2009}]{2009ApJ...697.1071A}
{Atwood} W.~B.,  et~al., 2009, \mn@doi [\apj] {10.1088/0004-637X/697/2/1071},
  \href {https://ui.adsabs.harvard.edu/abs/2009ApJ...697.1071A} {697, 1071}

\bibitem[\protect\citeauthoryear{{Bhatta} et~al.,}{{Bhatta}
  et~al.}{2023}]{2023MNRAS.tmp..320B}
{Bhatta} G.,  et~al., 2023, \mn@doi [\mnras] {10.1093/mnras/stad280}, \href
  {https://ui.adsabs.harvard.edu/abs/2023MNRAS.tmp..320B} {}

\bibitem[\protect\citeauthoryear{{Blandford} \& {Rees}}{{Blandford} \&
  {Rees}}{1978}]{1978bllo.conf..328B}
{Blandford} R.~D.,  {Rees} M.~J.,  1978, in {Wolfe} A.~M.,  ed., BL Lac
  Objects. pp 328--341

\bibitem[\protect\citeauthoryear{{B{\l}a{\.z}ejowski}, {Sikora}, {Moderski}  \&
  {Madejski}}{{B{\l}a{\.z}ejowski} et~al.}{2000}]{2000ApJ...545..107B}
{B{\l}a{\.z}ejowski} M.,  {Sikora} M.,  {Moderski} R.,   {Madejski} G.~M.,
  2000, \mn@doi [\apj] {10.1086/317791}, \href
  {https://ui.adsabs.harvard.edu/abs/2000ApJ...545..107B} {545, 107}

\bibitem[\protect\citeauthoryear{{Boettcher}, {Mause}  \&
  {Schlickeiser}}{{Boettcher} et~al.}{1997}]{1997A&A...324..395B}
{Boettcher} M.,  {Mause} H.,   {Schlickeiser} R.,  1997, \mn@doi [\aap]
  {10.48550/arXiv.astro-ph/9604003}, \href
  {https://ui.adsabs.harvard.edu/abs/1997A&A...324..395B} {324, 395}

\bibitem[\protect\citeauthoryear{{Bonning} et~al.,}{{Bonning}
  et~al.}{2009}]{2009ApJ...697L..81B}
{Bonning} E.~W.,  et~al., 2009, \mn@doi [\apjl] {10.1088/0004-637X/697/2/L81},
  \href {https://ui.adsabs.harvard.edu/abs/2009ApJ...697L..81B} {697, L81}

\bibitem[\protect\citeauthoryear{{Bonning} et~al.,}{{Bonning}
  et~al.}{2012}]{2012ApJ...756...13B}
{Bonning} E.,  et~al., 2012, \mn@doi [\apj] {10.1088/0004-637X/756/1/13}, \href
  {https://ui.adsabs.harvard.edu/abs/2012ApJ...756...13B} {756, 13}

\bibitem[\protect\citeauthoryear{{B{\"o}ttcher}}{{B{\"o}ttcher}}{2007}]{2007Ap&SS.309...95B}
{B{\"o}ttcher} M.,  2007, \mn@doi [\apss] {10.1007/s10509-007-9404-0}, \href
  {https://ui.adsabs.harvard.edu/abs/2007Ap&SS.309...95B} {309, 95}

\bibitem[\protect\citeauthoryear{{B{\"o}ttcher}, {Reimer}, {Sweeney}  \&
  {Prakash}}{{B{\"o}ttcher} et~al.}{2013}]{2013ApJ...768...54B}
{B{\"o}ttcher} M.,  {Reimer} A.,  {Sweeney} K.,   {Prakash} A.,  2013, \mn@doi
  [\apj] {10.1088/0004-637X/768/1/54}, \href
  {https://ui.adsabs.harvard.edu/abs/2013ApJ...768...54B} {768, 54}

\bibitem[\protect\citeauthoryear{{Breeveld}, {Landsman}, {Holland}, {Roming},
  {Kuin}  \& {Page}}{{Breeveld} et~al.}{2011}]{2011AIPC.1358..373B}
{Breeveld} A.~A.,  {Landsman} W.,  {Holland} S.~T.,  {Roming} P.,  {Kuin}
  N.~P.~M.,   {Page} M.~J.,  2011, in {McEnery} J.~E.,  {Racusin} J.~L.,
  {Gehrels} N.,  eds,  American Institute of Physics Conference Series Vol.
  1358, Gamma Ray Bursts 2010. pp 373--376 (\mn@eprint {arXiv} {1102.4717}),
  \mn@doi{10.1063/1.3621807}

\bibitem[\protect\citeauthoryear{{Burbidge}, {Jones}  \& {O'Dell}}{{Burbidge}
  et~al.}{1974}]{1974ApJ...193...43B}
{Burbidge} G.~R.,  {Jones} T.~W.,   {O'Dell} S.~L.,  1974, \mn@doi [\apj]
  {10.1086/153125}, \href
  {https://ui.adsabs.harvard.edu/abs/1974ApJ...193...43B} {193, 43}

\bibitem[\protect\citeauthoryear{{Burrows} et~al.,}{{Burrows}
  et~al.}{2005}]{2005SSRv..120..165B}
{Burrows} D.~N.,  et~al., 2005, \mn@doi [\ssr] {10.1007/s11214-005-5097-2},
  \href {https://ui.adsabs.harvard.edu/abs/2005SSRv..120..165B} {120, 165}

\bibitem[\protect\citeauthoryear{{Carrasco}, {Escobedo}, {Recillas}, {Porras},
  {Chavushyan}  \& {Mayya}}{{Carrasco} et~al.}{2017}]{2017ATel10932....1C}
{Carrasco} L.,  {Escobedo} G.,  {Recillas} E.,  {Porras} A.,  {Chavushyan} V.,
   {Mayya} D.~Y.,  2017, The Astronomer's Telegram, \href
  {https://ui.adsabs.harvard.edu/abs/2017ATel10932....1C} {10932, 1}

\bibitem[\protect\citeauthoryear{{Cerruti}, {Dermer}, {Lott}, {Boisson}  \&
  {Zech}}{{Cerruti} et~al.}{2013}]{2013ApJ...771L...4C}
{Cerruti} M.,  {Dermer} C.~D.,  {Lott} B.,  {Boisson} C.,   {Zech} A.,  2013,
  \mn@doi [\apjl] {10.1088/2041-8205/771/1/L4}, \href
  {https://ui.adsabs.harvard.edu/abs/2013ApJ...771L...4C} {771, L4}

\bibitem[\protect\citeauthoryear{{Cerruti}, {Zech}, {Boisson}, {Emery}, {Inoue}
   \& {Lenain}}{{Cerruti} et~al.}{2019}]{2019MNRAS.483L..12C}
{Cerruti} M.,  {Zech} A.,  {Boisson} C.,  {Emery} G.,  {Inoue} S.,   {Lenain}
  J.~P.,  2019, \mn@doi [\mnras] {10.1093/mnrasl/sly210}, \href
  {https://ui.adsabs.harvard.edu/abs/2019MNRAS.483L..12C} {483, L12}

\bibitem[\protect\citeauthoryear{{Chatterjee} et~al.,}{{Chatterjee}
  et~al.}{2012}]{2012ApJ...749..191C}
{Chatterjee} R.,  et~al., 2012, \mn@doi [\apj] {10.1088/0004-637X/749/2/191},
  \href {https://ui.adsabs.harvard.edu/abs/2012ApJ...749..191C} {749, 191}

\bibitem[\protect\citeauthoryear{{Chatterjee} et~al.,}{{Chatterjee}
  et~al.}{2013}]{2013ApJ...763L..11C}
{Chatterjee} R.,  et~al., 2013, \mn@doi [\apjl] {10.1088/2041-8205/763/1/L11},
  \href {https://ui.adsabs.harvard.edu/abs/2013ApJ...763L..11C} {763, L11}

\bibitem[\protect\citeauthoryear{{Cheung}, {Gasparrini}  \& {Buson}}{{Cheung}
  et~al.}{2017}]{2017ATel10931....1C}
{Cheung} C.~C.,  {Gasparrini} D.,   {Buson} S.,  2017, The Astronomer's
  Telegram, \href {https://ui.adsabs.harvard.edu/abs/2017ATel10931....1C}
  {10931, 1}

\bibitem[\protect\citeauthoryear{{Cohen}, {Romani}, {Filippenko}, {Cenko},
  {Lott}, {Zheng}  \& {Li}}{{Cohen} et~al.}{2014}]{2014ApJ...797..137C}
{Cohen} D.~P.,  {Romani} R.~W.,  {Filippenko} A.~V.,  {Cenko} S.~B.,  {Lott}
  B.,  {Zheng} W.,   {Li} W.,  2014, \mn@doi [\apj]
  {10.1088/0004-637X/797/2/137}, \href
  {https://ui.adsabs.harvard.edu/abs/2014ApJ...797..137C} {797, 137}

\bibitem[\protect\citeauthoryear{{Coogan}, {Brown}  \& {Chadwick}}{{Coogan}
  et~al.}{2016}]{2016MNRAS.458..354C}
{Coogan} R.~T.,  {Brown} A.~M.,   {Chadwick} P.~M.,  2016, \mn@doi [\mnras]
  {10.1093/mnras/stw199}, \href
  {https://ui.adsabs.harvard.edu/abs/2016MNRAS.458..354C} {458, 354}

\bibitem[\protect\citeauthoryear{{Costamante}, {Cutini}, {Tosti}, {Antolini}
  \& {Tramacere}}{{Costamante} et~al.}{2018}]{2018MNRAS.477.4749C}
{Costamante} L.,  {Cutini} S.,  {Tosti} G.,  {Antolini} E.,   {Tramacere} A.,
  2018, \mn@doi [\mnras] {10.1093/mnras/sty887}, \href
  {https://ui.adsabs.harvard.edu/abs/2018MNRAS.477.4749C} {477, 4749}

\bibitem[\protect\citeauthoryear{{D'Ammando} et~al.,}{{D'Ammando}
  et~al.}{2011}]{2011A&A...529A.145D}
{D'Ammando} F.,  et~al., 2011, \mn@doi [\aap] {10.1051/0004-6361/201016128},
  \href {https://ui.adsabs.harvard.edu/abs/2011A&A...529A.145D} {529, A145}

\bibitem[\protect\citeauthoryear{{Dermer} \& {Schlickeiser}}{{Dermer} \&
  {Schlickeiser}}{2002}]{2002ApJ...575..667D}
{Dermer} C.~D.,  {Schlickeiser} R.,  2002, \mn@doi [\apj] {10.1086/341431},
  \href {https://ui.adsabs.harvard.edu/abs/2002ApJ...575..667D} {575, 667}

\bibitem[\protect\citeauthoryear{{Dermer}, {Finke}, {Krug}  \&
  {B{\"o}ttcher}}{{Dermer} et~al.}{2009}]{2009ApJ...692...32D}
{Dermer} C.~D.,  {Finke} J.~D.,  {Krug} H.,   {B{\"o}ttcher} M.,  2009, \mn@doi
  [\apj] {10.1088/0004-637X/692/1/32}, \href
  {https://ui.adsabs.harvard.edu/abs/2009ApJ...692...32D} {692, 32}

\bibitem[\protect\citeauthoryear{{Dickey} \& {Lockman}}{{Dickey} \&
  {Lockman}}{1990}]{1990ARA&A..28..215D}
{Dickey} J.~M.,  {Lockman} F.~J.,  1990, \mn@doi [\araa]
  {10.1146/annurev.aa.28.090190.001243}, \href
  {https://ui.adsabs.harvard.edu/abs/1990ARA&A..28..215D} {28, 215}

\bibitem[\protect\citeauthoryear{{Diltz} \& {B{\"o}ttcher}}{{Diltz} \&
  {B{\"o}ttcher}}{2016}]{2016ApJ...826...54D}
{Diltz} C.,  {B{\"o}ttcher} M.,  2016, \mn@doi [\apj]
  {10.3847/0004-637X/826/1/54}, \href
  {https://ui.adsabs.harvard.edu/abs/2016ApJ...826...54D} {826, 54}

\bibitem[\protect\citeauthoryear{{Donea} \& {Protheroe}}{{Donea} \&
  {Protheroe}}{2003}]{2003APh....18..377D}
{Donea} A.-C.,  {Protheroe} R.~J.,  2003, \mn@doi [Astroparticle Physics]
  {10.1016/S0927-6505(02)00155-X}, \href
  {https://ui.adsabs.harvard.edu/abs/2003APh....18..377D} {18, 377}

\bibitem[\protect\citeauthoryear{{Dutka} et~al.,}{{Dutka}
  et~al.}{2013}]{2013ApJ...779..174D}
{Dutka} M.~S.,  et~al., 2013, \mn@doi [\apj] {10.1088/0004-637X/779/2/174},
  \href {https://ui.adsabs.harvard.edu/abs/2013ApJ...779..174D} {779, 174}

\bibitem[\protect\citeauthoryear{{Edelson} \& {Krolik}}{{Edelson} \&
  {Krolik}}{1988}]{1988ApJ...333..646E}
{Edelson} R.~A.,  {Krolik} J.~H.,  1988, \mn@doi [\apj] {10.1086/166773}, \href
  {https://ui.adsabs.harvard.edu/abs/1988ApJ...333..646E} {333, 646}

\bibitem[\protect\citeauthoryear{{Fan} et~al.,}{{Fan}
  et~al.}{2006}]{2006PASJ...58..797F}
{Fan} J.~H.,  et~al., 2006, \mn@doi [\pasj] {10.1093/pasj/58.5.797}, \href
  {https://ui.adsabs.harvard.edu/abs/2006PASJ...58..797F} {58, 797}

\bibitem[\protect\citeauthoryear{{Foffano}, {Prandini}, {Franceschini}  \&
  {Paiano}}{{Foffano} et~al.}{2019}]{2019MNRAS.486.1741F}
{Foffano} L.,  {Prandini} E.,  {Franceschini} A.,   {Paiano} S.,  2019, \mn@doi
  [\mnras] {10.1093/mnras/stz812}, \href
  {https://ui.adsabs.harvard.edu/abs/2019MNRAS.486.1741F} {486, 1741}

\bibitem[\protect\citeauthoryear{{Fossati}, {Maraschi}, {Celotti}, {Comastri}
  \& {Ghisellini}}{{Fossati} et~al.}{1998}]{1998MNRAS.299..433F}
{Fossati} G.,  {Maraschi} L.,  {Celotti} A.,  {Comastri} A.,   {Ghisellini} G.,
   1998, \mn@doi [\mnras] {10.1046/j.1365-8711.1998.01828.x}, \href
  {https://ui.adsabs.harvard.edu/abs/1998MNRAS.299..433F} {299, 433}

\bibitem[\protect\citeauthoryear{{Garrappa} \& {Valverd}}{{Garrappa} \&
  {Valverd}}{2023}]{2023ATel15859....1G}
{Garrappa} S.,  {Valverd} J.,  2023, The Astronomer's Telegram, \href
  {https://ui.adsabs.harvard.edu/abs/2023ATel15859....1G} {15859, 1}

\bibitem[\protect\citeauthoryear{{Gehrels} et~al.,}{{Gehrels}
  et~al.}{2004}]{2004ApJ...611.1005G}
{Gehrels} N.,  et~al., 2004, \mn@doi [\apj] {10.1086/422091}, \href
  {https://ui.adsabs.harvard.edu/abs/2004ApJ...611.1005G} {611, 1005}

\bibitem[\protect\citeauthoryear{{Ghisellini}}{{Ghisellini}}{2012}]{2012MNRAS.424L..26G}
{Ghisellini} G.,  2012, \mn@doi [\mnras] {10.1111/j.1745-3933.2012.01280.x},
  \href {https://ui.adsabs.harvard.edu/abs/2012MNRAS.424L..26G} {424, L26}

\bibitem[\protect\citeauthoryear{{Ghisellini} \& {Madau}}{{Ghisellini} \&
  {Madau}}{1996}]{1996MNRAS.280...67G}
{Ghisellini} G.,  {Madau} P.,  1996, \mn@doi [\mnras] {10.1093/mnras/280.1.67},
  \href {https://ui.adsabs.harvard.edu/abs/1996MNRAS.280...67G} {280, 67}

\bibitem[\protect\citeauthoryear{{Ghisellini} \& {Maraschi}}{{Ghisellini} \&
  {Maraschi}}{1989}]{1989ApJ...340..181G}
{Ghisellini} G.,  {Maraschi} L.,  1989, \mn@doi [\apj] {10.1086/167383}, \href
  {https://ui.adsabs.harvard.edu/abs/1989ApJ...340..181G} {340, 181}

\bibitem[\protect\citeauthoryear{{Ghisellini} \& {Tavecchio}}{{Ghisellini} \&
  {Tavecchio}}{2008}]{2008MNRAS.387.1669G}
{Ghisellini} G.,  {Tavecchio} F.,  2008, \mn@doi [\mnras]
  {10.1111/j.1365-2966.2008.13360.x}, \href
  {https://ui.adsabs.harvard.edu/abs/2008MNRAS.387.1669G} {387, 1669}

\bibitem[\protect\citeauthoryear{{Ghisellini}, {Tavecchio}  \&
  {Ghirlanda}}{{Ghisellini} et~al.}{2009}]{2009MNRAS.399.2041G}
{Ghisellini} G.,  {Tavecchio} F.,   {Ghirlanda} G.,  2009, \mn@doi [\mnras]
  {10.1111/j.1365-2966.2009.15397.x}, \href
  {https://ui.adsabs.harvard.edu/abs/2009MNRAS.399.2041G} {399, 2041}

\bibitem[\protect\citeauthoryear{{Ghisellini}, {Tavecchio}, {Foschini},
  {Ghirlanda}, {Maraschi}  \& {Celotti}}{{Ghisellini}
  et~al.}{2010}]{2010MNRAS.402..497G}
{Ghisellini} G.,  {Tavecchio} F.,  {Foschini} L.,  {Ghirlanda} G.,  {Maraschi}
  L.,   {Celotti} A.,  2010, \mn@doi [\mnras]
  {10.1111/j.1365-2966.2009.15898.x}, \href
  {https://ui.adsabs.harvard.edu/abs/2010MNRAS.402..497G} {402, 497}

\bibitem[\protect\citeauthoryear{{Ghisellini}, {Tavecchio}, {Foschini}  \&
  {Ghirland a}}{{Ghisellini} et~al.}{2011}]{2011MNRAS.414.2674G}
{Ghisellini} G.,  {Tavecchio} F.,  {Foschini} L.,   {Ghirland a} G.,  2011,
  \mn@doi [\mnras] {10.1111/j.1365-2966.2011.18578.x}, \href
  {https://ui.adsabs.harvard.edu/abs/2011MNRAS.414.2674G} {414, 2674}

\bibitem[\protect\citeauthoryear{{Goyal}}{{Goyal}}{2021}]{2021ApJ...909...39G}
{Goyal} A.,  2021, \mn@doi [\apj] {10.3847/1538-4357/abd7fb}, \href
  {https://ui.adsabs.harvard.edu/abs/2021ApJ...909...39G} {909, 39}

\bibitem[\protect\citeauthoryear{{Gu}, {Lee}, {Pak}, {Yim}  \& {Fletcher}}{{Gu}
  et~al.}{2006}]{2006A&A...450...39G}
{Gu} M.~F.,  {Lee} C.~U.,  {Pak} S.,  {Yim} H.~S.,   {Fletcher} A.~B.,  2006,
  \mn@doi [\aap] {10.1051/0004-6361:20054271}, \href
  {https://ui.adsabs.harvard.edu/abs/2006A&A...450...39G} {450, 39}

\bibitem[\protect\citeauthoryear{{Harrison} et~al.,}{{Harrison}
  et~al.}{2013}]{2013ApJ...770..103H}
{Harrison} F.~A.,  et~al., 2013, \mn@doi [\apj] {10.1088/0004-637X/770/2/103},
  \href {https://ui.adsabs.harvard.edu/abs/2013ApJ...770..103H} {770, 103}

\bibitem[\protect\citeauthoryear{{Hewett} \& {Wild}}{{Hewett} \&
  {Wild}}{2010}]{2010MNRAS.405.2302H}
{Hewett} P.~C.,  {Wild} V.,  2010, \mn@doi [\mnras]
  {10.1111/j.1365-2966.2010.16648.x}, \href
  {https://ui.adsabs.harvard.edu/abs/2010MNRAS.405.2302H} {405, 2302}

\bibitem[\protect\citeauthoryear{{Impey} \& {Neugebauer}}{{Impey} \&
  {Neugebauer}}{1988}]{1988AJ.....95..307I}
{Impey} C.~D.,  {Neugebauer} G.,  1988, \mn@doi [\aj] {10.1086/114638}, \href
  {https://ui.adsabs.harvard.edu/abs/1988AJ.....95..307I} {95, 307}

\bibitem[\protect\citeauthoryear{{Kalberla}, {Burton}, {Hartmann}, {Arnal},
  {Bajaja}, {Morras}  \& {P{\"o}ppel}}{{Kalberla}
  et~al.}{2005}]{2005A&A...440..775K}
{Kalberla} P.~M.~W.,  {Burton} W.~B.,  {Hartmann} D.,  {Arnal} E.~M.,  {Bajaja}
  E.,  {Morras} R.,   {P{\"o}ppel} W.~G.~L.,  2005, \mn@doi [\aap]
  {10.1051/0004-6361:20041864}, \href
  {https://ui.adsabs.harvard.edu/abs/2005A&A...440..775K} {440, 775}

\bibitem[\protect\citeauthoryear{{Kochanek} et~al.,}{{Kochanek}
  et~al.}{2017}]{2017PASP..129j4502K}
{Kochanek} C.~S.,  et~al., 2017, \mn@doi [\pasp] {10.1088/1538-3873/aa80d9},
  \href {https://ui.adsabs.harvard.edu/abs/2017PASP..129j4502K} {129, 104502}

\bibitem[\protect\citeauthoryear{{Konigl}}{{Konigl}}{1981}]{1981ApJ...243..700K}
{Konigl} A.,  1981, \mn@doi [\apj] {10.1086/158638}, \href
  {https://ui.adsabs.harvard.edu/abs/1981ApJ...243..700K} {243, 700}

\bibitem[\protect\citeauthoryear{{Liao}, {Bai}, {Liu}, {Weng}, {Chen}  \&
  {Li}}{{Liao} et~al.}{2014}]{2014ApJ...783...83L}
{Liao} N.~H.,  {Bai} J.~M.,  {Liu} H.~T.,  {Weng} S.~S.,  {Chen} L.,   {Li} F.,
   2014, \mn@doi [\apj] {10.1088/0004-637X/783/2/83}, \href
  {https://ui.adsabs.harvard.edu/abs/2014ApJ...783...83L} {783, 83}

\bibitem[\protect\citeauthoryear{{Liodakis}, {Romani}, {Filippenko},
  {Kiehlmann}, {Max-Moerbeck}, {Readhead}  \& {Zheng}}{{Liodakis}
  et~al.}{2018}]{2018MNRAS.480.5517L}
{Liodakis} I.,  {Romani} R.~W.,  {Filippenko} A.~V.,  {Kiehlmann} S.,
  {Max-Moerbeck} W.,  {Readhead} A.~C.~S.,   {Zheng} W.,  2018, \mn@doi
  [\mnras] {10.1093/mnras/sty2264}, \href
  {https://ui.adsabs.harvard.edu/abs/2018MNRAS.480.5517L} {480, 5517}

\bibitem[\protect\citeauthoryear{{Liodakis}, {Romani}, {Filippenko}, {Kocevski}
   \& {Zheng}}{{Liodakis} et~al.}{2019}]{2019ApJ...880...32L}
{Liodakis} I.,  {Romani} R.~W.,  {Filippenko} A.~V.,  {Kocevski} D.,   {Zheng}
  W.,  2019, \mn@doi [\apj] {10.3847/1538-4357/ab26b7}, \href
  {https://ui.adsabs.harvard.edu/abs/2019ApJ...880...32L} {880, 32}

\bibitem[\protect\citeauthoryear{{MacDonald}, {Marscher}, {Jorstad}  \&
  {Joshi}}{{MacDonald} et~al.}{2015}]{2015ApJ...804..111M}
{MacDonald} N.~R.,  {Marscher} A.~P.,  {Jorstad} S.~G.,   {Joshi} M.,  2015,
  \mn@doi [\apj] {10.1088/0004-637X/804/2/111}, \href
  {https://ui.adsabs.harvard.edu/abs/2015ApJ...804..111M} {804, 111}

\bibitem[\protect\citeauthoryear{{Mannheim}}{{Mannheim}}{1993}]{1993A&A...269...67M}
{Mannheim} K.,  1993, \mn@doi [\aap] {10.48550/arXiv.astro-ph/9302006}, \href
  {https://ui.adsabs.harvard.edu/abs/1993A&A...269...67M} {269, 67}

\bibitem[\protect\citeauthoryear{{Mao}, {Urry}, {Massaro}, {Paggi},
  {Cauteruccio}  \& {K{\"u}nzel}}{{Mao} et~al.}{2016}]{2016ApJS..224...26M}
{Mao} P.,  {Urry} C.~M.,  {Massaro} F.,  {Paggi} A.,  {Cauteruccio} J.,
  {K{\"u}nzel} S.~R.,  2016, \mn@doi [\apjs] {10.3847/0067-0049/224/2/26},
  \href {https://ui.adsabs.harvard.edu/abs/2016ApJS..224...26M} {224, 26}

\bibitem[\protect\citeauthoryear{{Marscher} \& {Gear}}{{Marscher} \&
  {Gear}}{1985}]{1985ApJ...298..114M}
{Marscher} A.~P.,  {Gear} W.~K.,  1985, \mn@doi [\apj] {10.1086/163592}, \href
  {https://ui.adsabs.harvard.edu/abs/1985ApJ...298..114M} {298, 114}

\bibitem[\protect\citeauthoryear{{Mattox} et~al.,}{{Mattox}
  et~al.}{1996}]{1996ApJ...461..396M}
{Mattox} J.~R.,  et~al., 1996, \mn@doi [\apj] {10.1086/177068}, \href
  {https://ui.adsabs.harvard.edu/abs/1996ApJ...461..396M} {461, 396}

\bibitem[\protect\citeauthoryear{{Max-Moerbeck} et~al.,}{{Max-Moerbeck}
  et~al.}{2014a}]{2014MNRAS.445..428M}
{Max-Moerbeck} W.,  et~al., 2014a, \mn@doi [\mnras] {10.1093/mnras/stu1749},
  \href {https://ui.adsabs.harvard.edu/abs/2014MNRAS.445..428M} {445, 428}

\bibitem[\protect\citeauthoryear{{Max-Moerbeck}, {Richards}, {Hovatta},
  {Pavlidou}, {Pearson}  \& {Readhead}}{{Max-Moerbeck}
  et~al.}{2014b}]{2014MNRAS.445..437M}
{Max-Moerbeck} W.,  {Richards} J.~L.,  {Hovatta} T.,  {Pavlidou} V.,  {Pearson}
  T.~J.,   {Readhead} A.~C.~S.,  2014b, \mn@doi [\mnras]
  {10.1093/mnras/stu1707}, \href
  {https://ui.adsabs.harvard.edu/abs/2014MNRAS.445..437M} {445, 437}

\bibitem[\protect\citeauthoryear{{M{\"u}cke} \& {Protheroe}}{{M{\"u}cke} \&
  {Protheroe}}{2001}]{2001APh....15..121M}
{M{\"u}cke} A.,  {Protheroe} R.~J.,  2001, \mn@doi [Astroparticle Physics]
  {10.1016/S0927-6505(00)00141-9}, \href
  {https://ui.adsabs.harvard.edu/abs/2001APh....15..121M} {15, 121}

\bibitem[\protect\citeauthoryear{{M{\"u}cke}, {Protheroe}, {Engel}, {Rachen}
  \& {Stanev}}{{M{\"u}cke} et~al.}{2003}]{2003APh....18..593M}
{M{\"u}cke} A.,  {Protheroe} R.~J.,  {Engel} R.,  {Rachen} J.~P.,   {Stanev}
  T.,  2003, \mn@doi [Astroparticle Physics] {10.1016/S0927-6505(02)00185-8},
  \href {https://ui.adsabs.harvard.edu/abs/2003APh....18..593M} {18, 593}

\bibitem[\protect\citeauthoryear{{Mukherjee} \& {VERITAS
  Collaboration}}{{Mukherjee} \& {VERITAS
  Collaboration}}{2017}]{2017ATel11075....1M}
{Mukherjee} R.,  {VERITAS Collaboration} 2017, The Astronomer's Telegram, \href
  {https://ui.adsabs.harvard.edu/abs/2017ATel11075....1M} {11075, 1}

\bibitem[\protect\citeauthoryear{{Negi}, {Joshi}, {Chand}, {Chand}, {Wiita},
  {Ho}  \& {Singh}}{{Negi} et~al.}{2022}]{2022MNRAS.510.1791N}
{Negi} V.,  {Joshi} R.,  {Chand} K.,  {Chand} H.,  {Wiita} P.,  {Ho} L.~C.,
  {Singh} R.~S.,  2022, \mn@doi [\mnras] {10.1093/mnras/stab3591}, \href
  {https://ui.adsabs.harvard.edu/abs/2022MNRAS.510.1791N} {510, 1791}

\bibitem[\protect\citeauthoryear{{Nolan} et~al.,}{{Nolan}
  et~al.}{2012}]{2012ApJS..199...31N}
{Nolan} P.~L.,  et~al., 2012, \mn@doi [\apjs] {10.1088/0067-0049/199/2/31},
  \href {https://ui.adsabs.harvard.edu/abs/2012ApJS..199...31N} {199, 31}

\bibitem[\protect\citeauthoryear{{Padovani} \& {Giommi}}{{Padovani} \&
  {Giommi}}{1995}]{1995ApJ...444..567P}
{Padovani} P.,  {Giommi} P.,  1995, \mn@doi [\apj] {10.1086/175631}, \href
  {https://ui.adsabs.harvard.edu/abs/1995ApJ...444..567P} {444, 567}

\bibitem[\protect\citeauthoryear{{Paliya}, {Sahayanathan}  \&
  {Stalin}}{{Paliya} et~al.}{2015}]{2015ApJ...803...15P}
{Paliya} V.~S.,  {Sahayanathan} S.,   {Stalin} C.~S.,  2015, \mn@doi [\apj]
  {10.1088/0004-637X/803/1/15}, \href
  {https://ui.adsabs.harvard.edu/abs/2015ApJ...803...15P} {803, 15}

\bibitem[\protect\citeauthoryear{{Paliya}, {Diltz}, {B{\"o}ttcher}, {Stalin}
  \& {Buckley}}{{Paliya} et~al.}{2016}]{2016ApJ...817...61P}
{Paliya} V.~S.,  {Diltz} C.,  {B{\"o}ttcher} M.,  {Stalin} C.~S.,   {Buckley}
  D.,  2016, \mn@doi [\apj] {10.3847/0004-637X/817/1/61}, \href
  {https://ui.adsabs.harvard.edu/abs/2016ApJ...817...61P} {817, 61}

\bibitem[\protect\citeauthoryear{{Paliya}, {Marcotulli}, {Ajello}, {Joshi},
  {Sahayanathan}, {Rao}  \& {Hartmann}}{{Paliya}
  et~al.}{2017}]{2017ApJ...851...33P}
{Paliya} V.~S.,  {Marcotulli} L.,  {Ajello} M.,  {Joshi} M.,  {Sahayanathan}
  S.,  {Rao} A.~R.,   {Hartmann} D.,  2017, \mn@doi [\apj]
  {10.3847/1538-4357/aa98e1}, \href
  {https://ui.adsabs.harvard.edu/abs/2017ApJ...851...33P} {851, 33}

\bibitem[\protect\citeauthoryear{{Patel} \& {Chitnis}}{{Patel} \&
  {Chitnis}}{2020}]{2020MNRAS.492...72P}
{Patel} S.~R.,  {Chitnis} V.~R.,  2020, \mn@doi [\mnras]
  {10.1093/mnras/stz3490}, \href
  {https://ui.adsabs.harvard.edu/abs/2020MNRAS.492...72P} {492, 72}

\bibitem[\protect\citeauthoryear{{Patel}, {Chitnis}, {Shukla}, {Rao}  \&
  {Nagare}}{{Patel} et~al.}{2018}]{2018ApJ...866..102P}
{Patel} S.~R.,  {Chitnis} V.~R.,  {Shukla} A.,  {Rao} A.~R.,   {Nagare} B.~J.,
  2018, \mn@doi [\apj] {10.3847/1538-4357/aae1fc}, \href
  {https://ui.adsabs.harvard.edu/abs/2018ApJ...866..102P} {866, 102}

\bibitem[\protect\citeauthoryear{{Poutanen} \& {Stern}}{{Poutanen} \&
  {Stern}}{2010}]{2010ApJ...717L.118P}
{Poutanen} J.,  {Stern} B.,  2010, \mn@doi [\apjl]
  {10.1088/2041-8205/717/2/L118}, \href
  {https://ui.adsabs.harvard.edu/abs/2010ApJ...717L.118P} {717, L118}

\bibitem[\protect\citeauthoryear{{Prince}}{{Prince}}{2019}]{2019ApJ...871..101P}
{Prince} R.,  2019, \mn@doi [\apj] {10.3847/1538-4357/aaf475}, \href
  {https://ui.adsabs.harvard.edu/abs/2019ApJ...871..101P} {871, 101}

\bibitem[\protect\citeauthoryear{{Prince}}{{Prince}}{2023}]{2023ATel15854....1P}
{Prince} R.,  2023, The Astronomer's Telegram, \href
  {https://ui.adsabs.harvard.edu/abs/2023ATel15854....1P} {15854, 1}

\bibitem[\protect\citeauthoryear{{Pushkarev}, {Kovalev}, {Lister}  \&
  {Savolainen}}{{Pushkarev} et~al.}{2009}]{2009A&A...507L..33P}
{Pushkarev} A.~B.,  {Kovalev} Y.~Y.,  {Lister} M.~L.,   {Savolainen} T.,  2009,
  \mn@doi [\aap] {10.1051/0004-6361/200913422}, \href
  {https://ui.adsabs.harvard.edu/abs/2009A&A...507L..33P} {507, L33}

\bibitem[\protect\citeauthoryear{{Rajput} \& {Pandey}}{{Rajput} \&
  {Pandey}}{2021}]{2021Galax...9..118R}
{Rajput} B.,  {Pandey} A.,  2021, \mn@doi [Galaxies] {10.3390/galaxies9040118},
  \href {https://ui.adsabs.harvard.edu/abs/2021Galax...9..118R} {9, 118}

\bibitem[\protect\citeauthoryear{{Rajput}, {Stalin}, {Sahayanathan}, {Rakshit}
  \& {Mandal}}{{Rajput} et~al.}{2019}]{2019MNRAS.486.1781R}
{Rajput} B.,  {Stalin} C.~S.,  {Sahayanathan} S.,  {Rakshit} S.,   {Mandal}
  A.~K.,  2019, \mn@doi [\mnras] {10.1093/mnras/stz941}, \href
  {https://ui.adsabs.harvard.edu/abs/2019MNRAS.486.1781R} {486, 1781}

\bibitem[\protect\citeauthoryear{{Rajput}, {Stalin}  \&
  {Sahayanathan}}{{Rajput} et~al.}{2020}]{2020MNRAS.498.5128R}
{Rajput} B.,  {Stalin} C.~S.,   {Sahayanathan} S.,  2020, \mn@doi [\mnras]
  {10.1093/mnras/staa2708}, \href
  {https://ui.adsabs.harvard.edu/abs/2020MNRAS.498.5128R} {498, 5128}

\bibitem[\protect\citeauthoryear{{Safna}, {Stalin}, {Rakshit}  \&
  {Mathew}}{{Safna} et~al.}{2020}]{2020MNRAS.498.3578S}
{Safna} P.~Z.,  {Stalin} C.~S.,  {Rakshit} S.,   {Mathew} B.,  2020, \mn@doi
  [\mnras] {10.1093/mnras/staa2622}, \href
  {https://ui.adsabs.harvard.edu/abs/2020MNRAS.498.3578S} {498, 3578}

\bibitem[\protect\citeauthoryear{{Sahakyan}}{{Sahakyan}}{2020}]{2020A&A...635A..25S}
{Sahakyan} N.,  2020, \mn@doi [\aap] {10.1051/0004-6361/201936715}, \href
  {https://ui.adsabs.harvard.edu/abs/2020A&A...635A..25S} {635, A25}

\bibitem[\protect\citeauthoryear{{Sbarrato}, {Ghisellini}, {Maraschi}  \&
  {Colpi}}{{Sbarrato} et~al.}{2012}]{2012MNRAS.421.1764S}
{Sbarrato} T.,  {Ghisellini} G.,  {Maraschi} L.,   {Colpi} M.,  2012, \mn@doi
  [\mnras] {10.1111/j.1365-2966.2012.20442.x}, \href
  {https://ui.adsabs.harvard.edu/abs/2012MNRAS.421.1764S} {421, 1764}

\bibitem[\protect\citeauthoryear{{Scarpa} \& {Falomo}}{{Scarpa} \&
  {Falomo}}{1997}]{1997A&A...325..109S}
{Scarpa} R.,  {Falomo} R.,  1997, \aap, \href
  {https://ui.adsabs.harvard.edu/abs/1997A&A...325..109S} {325, 109}

\bibitem[\protect\citeauthoryear{{Shah}, {Sahayanathan}, {Mankuzhiyil},
  {Kushwaha}, {Misra}  \& {Iqbal}}{{Shah} et~al.}{2017}]{2017MNRAS.470.3283S}
{Shah} Z.,  {Sahayanathan} S.,  {Mankuzhiyil} N.,  {Kushwaha} P.,  {Misra} R.,
   {Iqbal} N.,  2017, \mn@doi [\mnras] {10.1093/mnras/stx1194}, \href
  {https://ui.adsabs.harvard.edu/abs/2017MNRAS.470.3283S} {470, 3283}

\bibitem[\protect\citeauthoryear{{Shakura} \& {Sunyaev}}{{Shakura} \&
  {Sunyaev}}{1973}]{1973A&A....24..337S}
{Shakura} N.~I.,  {Sunyaev} R.~A.,  1973, \aap, \href
  {http://adsabs.harvard.edu/abs/1973A%26A....24..337S} {24, 337}

\bibitem[\protect\citeauthoryear{{Shappee} et~al.,}{{Shappee}
  et~al.}{2014}]{2014ApJ...788...48S}
{Shappee} B.~J.,  et~al., 2014, \mn@doi [\apj] {10.1088/0004-637X/788/1/48},
  \href {https://ui.adsabs.harvard.edu/abs/2014ApJ...788...48S} {788, 48}

\bibitem[\protect\citeauthoryear{{Shaw} et~al.,}{{Shaw}
  et~al.}{2012}]{2012ApJ...748...49S}
{Shaw} M.~S.,  et~al., 2012, \mn@doi [\apj] {10.1088/0004-637X/748/1/49}, \href
  {https://ui.adsabs.harvard.edu/abs/2012ApJ...748...49S} {748, 49}

\bibitem[\protect\citeauthoryear{{Sikora}, {Moderski}  \& {Madejski}}{{Sikora}
  et~al.}{2008}]{2008ApJ...675...71S}
{Sikora} M.,  {Moderski} R.,   {Madejski} G.~M.,  2008, \mn@doi [\apj]
  {10.1086/526419}, \href
  {https://ui.adsabs.harvard.edu/abs/2008ApJ...675...71S} {675, 71}

\bibitem[\protect\citeauthoryear{{Stocke}, {Morris}, {Gioia}, {Maccacaro},
  {Schild}, {Wolter}, {Fleming}  \& {Henry}}{{Stocke}
  et~al.}{1991}]{1991ApJS...76..813S}
{Stocke} J.~T.,  {Morris} S.~L.,  {Gioia} I.~M.,  {Maccacaro} T.,  {Schild} R.,
   {Wolter} A.,  {Fleming} T.~A.,   {Henry} J.~P.,  1991, \mn@doi [\apjs]
  {10.1086/191582}, \href
  {https://ui.adsabs.harvard.edu/abs/1991ApJS...76..813S} {76, 813}

\bibitem[\protect\citeauthoryear{{Thompson} et~al.,}{{Thompson}
  et~al.}{1995}]{1995ApJS..101..259T}
{Thompson} D.~J.,  et~al., 1995, \mn@doi [\apjs] {10.1086/192240}, \href
  {https://ui.adsabs.harvard.edu/abs/1995ApJS..101..259T} {101, 259}

\bibitem[\protect\citeauthoryear{{Tramacere}}{{Tramacere}}{2020}]{2020ascl.soft09001T}
{Tramacere} A.,  2020, {JetSeT: Numerical modeling and SED fitting tool for
  relativistic jets}, Astrophysics Source Code Library, record ascl:2009.001
  (\mn@eprint {ascl} {2009.001})

\bibitem[\protect\citeauthoryear{{Tramacere}, {Giommi}, {Perri}, {Verrecchia}
  \& {Tosti}}{{Tramacere} et~al.}{2009}]{2009A&A...501..879T}
{Tramacere} A.,  {Giommi} P.,  {Perri} M.,  {Verrecchia} F.,   {Tosti} G.,
  2009, \mn@doi [\aap] {10.1051/0004-6361/200810865}, \href
  {https://ui.adsabs.harvard.edu/abs/2009A&A...501..879T} {501, 879}

\bibitem[\protect\citeauthoryear{{Tramacere}, {Massaro}  \&
  {Taylor}}{{Tramacere} et~al.}{2011}]{2011ApJ...739...66T}
{Tramacere} A.,  {Massaro} E.,   {Taylor} A.~M.,  2011, \mn@doi [\apj]
  {10.1088/0004-637X/739/2/66}, \href
  {https://ui.adsabs.harvard.edu/abs/2011ApJ...739...66T} {739, 66}

\bibitem[\protect\citeauthoryear{{Tripathi}, {Devanand}, {Gupta}, {Kishore},
  {Dogra}, {Krishna Mohana}  \& {Dhiman}}{{Tripathi}
  et~al.}{2023}]{2023ATel15875....1T}
{Tripathi} T.,  {Devanand} P.~U.,  {Gupta} A.~C.,  {Kishore} S.,  {Dogra} K.,
  {Krishna Mohana} A.,   {Dhiman} V.,  2023, The Astronomer's Telegram, \href
  {https://ui.adsabs.harvard.edu/abs/2023ATel15875....1T} {15875, 1}

\bibitem[\protect\citeauthoryear{{Urry} \& {Mushotzky}}{{Urry} \&
  {Mushotzky}}{1982}]{1982ApJ...253...38U}
{Urry} C.~M.,  {Mushotzky} R.~F.,  1982, \mn@doi [\apj] {10.1086/159607}, \href
  {https://ui.adsabs.harvard.edu/abs/1982ApJ...253...38U} {253, 38}

\bibitem[\protect\citeauthoryear{{Urry} \& {Padovani}}{{Urry} \&
  {Padovani}}{1995}]{1995PASP..107..803U}
{Urry} C.~M.,  {Padovani} P.,  1995, \mn@doi [\pasp] {10.1086/133630}, \href
  {http://cdsads.u-strasbg.fr/abs/1995PASP..107..803U} {107, 803}

\bibitem[\protect\citeauthoryear{{Urry}, {Scarpa}, {O'Dowd}, {Falomo}, {Pesce}
  \& {Treves}}{{Urry} et~al.}{2000}]{2000ApJ...532..816U}
{Urry} C.~M.,  {Scarpa} R.,  {O'Dowd} M.,  {Falomo} R.,  {Pesce} J.~E.,
  {Treves} A.,  2000, \mn@doi [\apj] {10.1086/308616}, \href
  {https://ui.adsabs.harvard.edu/abs/2000ApJ...532..816U} {532, 816}

\bibitem[\protect\citeauthoryear{{Vermeulen}, {Ogle}, {Tran}, {Browne},
  {Cohen}, {Readhead}, {Taylor}  \& {Goodrich}}{{Vermeulen}
  et~al.}{1995}]{1995ApJ...452L...5V}
{Vermeulen} R.~C.,  {Ogle} P.~M.,  {Tran} H.~D.,  {Browne} I.~W.~A.,  {Cohen}
  M.~H.,  {Readhead} A.~C.~S.,  {Taylor} G.~B.,   {Goodrich} R.~W.,  1995,
  \mn@doi [\apjl] {10.1086/309716}, \href
  {https://ui.adsabs.harvard.edu/abs/1995ApJ...452L...5V} {452, L5}

\bibitem[\protect\citeauthoryear{{Wu}, {Zhou}, {Ma}  \& {Jiang}}{{Wu}
  et~al.}{2011}]{2011MNRAS.418.1640W}
{Wu} J.,  {Zhou} X.,  {Ma} J.,   {Jiang} Z.,  2011, \mn@doi [\mnras]
  {10.1111/j.1365-2966.2011.19565.x}, \href
  {https://ui.adsabs.harvard.edu/abs/2011MNRAS.418.1640W} {418, 1640}

\bibitem[\protect\citeauthoryear{{Zhang}, {Zhou}, {Zhao}  \& {Dai}}{{Zhang}
  et~al.}{2015}]{2015RAA....15.1784Z}
{Zhang} B.-K.,  {Zhou} X.-S.,  {Zhao} X.-Y.,   {Dai} B.-Z.,  2015, \mn@doi
  [Research in Astronomy and Astrophysics] {10.1088/1674-4527/15/11/002}, \href
  {https://ui.adsabs.harvard.edu/abs/2015RAA....15.1784Z} {15, 1784}

\bibitem[\protect\citeauthoryear{{de Jaeger} et~al.,}{{de Jaeger}
  et~al.}{2023}]{2023MNRAS.519.6349D}
{de Jaeger} T.,  et~al., 2023, \mn@doi [\mnras] {10.1093/mnras/stad060}, \href
  {https://ui.adsabs.harvard.edu/abs/2023MNRAS.519.6349D} {519, 6349}

\makeatother
\end{thebibliography}





\bsp	
\label{lastpage}
\end{document}